\documentclass[reprint,
 amsmath,amssymb,
 aps,
 prb,
 superscriptaddress
]{revtex4-2}
\usepackage{physics}
\usepackage{dsfont}
\usepackage{color}
\usepackage{amsmath}
\usepackage{graphicx}
\usepackage{dcolumn}
\usepackage{bm}
\usepackage[mode=buildnew]{standalone}
\usepackage{tikz}

\begin{document}

\title{Variational manifolds for ground states and scarred dynamics of blockade-constrained spin models on two and three dimensional lattices}

\author{Joey Li}
\affiliation{Institute for Theoretical Physics, University of Innsbruck, 6020 Innsbruck, Austria}
\affiliation{Institute for Quantum Optics and Quantum Information of the Austrian Academy of Sciences, 6020 Innsbruck, Austria}
\author{Giuliano Giudici}
\affiliation{Institute for Theoretical Physics, University of Innsbruck, 6020 Innsbruck, Austria}
\affiliation{Institute for Quantum Optics and Quantum Information of the Austrian Academy of Sciences, 6020 Innsbruck, Austria}
\author{Hannes Pichler}
\affiliation{Institute for Theoretical Physics, University of Innsbruck, 6020 Innsbruck, Austria}
\affiliation{Institute for Quantum Optics and Quantum Information of the Austrian Academy of Sciences, 6020 Innsbruck, Austria}

\date{\today}

\begin{abstract}
 We introduce a variational manifold of simple tensor network states for the study of a family of constrained models that describe spin-1/2 systems as realized by Rydberg atom arrays. Our manifold permits  analytical calculation via perturbative expansion of one- and two-point functions in arbitrary spatial dimensions and allows for efficient computation of the matrix elements required for variational energy minimization and variational time evolution in up to three dimensions. We apply this framework to the PXP model on the hypercubic lattice in 1D, 2D, and 3D, and show that, in each case, it exhibits quantum phase transitions breaking the sub-lattice symmetry in equilibrium, and hosts quantum many body scars out of equilibrium. We demonstrate that our variational ansatz qualitatively captures all these phenomena and predicts key quantities with an accuracy that increases with the dimensionality of the lattice, and conclude that our method  can be interpreted as a generalization of mean-field theory to constrained spin models.
\end{abstract}

 \maketitle


\section{Introduction}
Arrays of neutral atoms trapped in optical tweezers~\cite{Weimer2010,RydbergReview0,RydbergReview,RydbergReview2,RydbergReview3} have recently emerged as a promising platform for the quantum simulation of many-body spin models.  These Rydberg atom quantum simulators have been used for the experimental study of a variety of quantum phenomena in equilibrium, including zero temperature quantum phase transitions between trivially disordered states and states that break a variety of spatial symmetries~\cite{KeeslingKibbleZurek,256atom,Browaeys2021hundredsofRydbergs,ContinuousSymmetryBreaking}, as well as spin liquids with topological  order \cite{SemeghiniQSL,PredictionToricCodeVishwanath,giulianoQSL,sahay2023quantum,QSLbootstrap,QSLtrimer} and symmetry protected topological phases~\cite{RydbergSPT,RydbergSPTBrowaeys}. They can also be used to study thermalization dynamics \cite{RydbergKAIST,MicrowaveEngineering} and were instrumental in the experimental discovery \cite{51atom, drivenRydberg} of the out-of-equilibrium phenomenon now known as quantum many-body scars (QMBS) \cite{QscarsReview}, which have sparked interest as an example of non-thermalizing behaviour in quantum many-body systems, violating the eigenstate thermalization hypothesis~\cite{ETH1,ETH2}. Remarkably, often the physics of these systems simply emerges from the Rydberg blockade mechanism~\cite{blockade1,blockade2,blockade3} that forbids neighbouring Rydberg excitations, introducing a constraint between otherwise freely rotating spins. It is this constraint that renders the problems in general non-trivial and thus underlies the difficulty of describing these systems in a many-body setting.

\begin{figure}[b]
    \centering
    \includegraphics[width = 7cm]{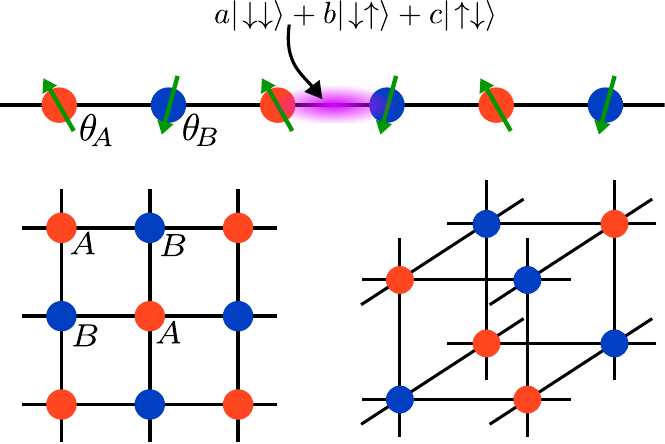}
 
    \caption{Schematic depiction of the setups considered in this work, consisting of PXP models in 1D, 2D and 3D hyper-cubic lattices. The blockade-constraint in these systems does not allow two neighboring spins to be simultaneously in the Rydberg state. After imposing unit cell translation invariance, the system is parameterized by two variational parameters $\theta_A$ and $\theta_B$, corresponding to the two sub-lattices.}
    \label{fig:lattices}
\end{figure}

Nevertheless, we expect that it should be possible to capture some of the above phenomena by an appropriately constructed mean-field theory for these constrained systems. 
This includes in particular phase transitions breaking the sublattice symmetry, as well as the dynamics associated with QMBS. On a technical level such an attempt faces the challenge of formulating a proper variational manifold of states that satisfies the blockade constraint, and at the same time 
allows for an efficient calculation of physical observables. In this work we address this challenge and introduce a manifold of constraint-satisfying states that can be elegantly represented as minimally entangled tensor networks. Our ansatz can be seen as a generalization of minimal matrix-product wavefunctions that have been employed for 1D chains~\cite{periodicOrbits,OptimalSteering} and the related tree networks in Ref.~\cite{slowQuantThermSerbyn}.

Going beyond these previous results, we show that our variational ansatz allows for efficient calculation of expectation values, in the relevant parameter regimes, in 2D square and 3D cubic lattices directly in the thermodynamic limit. In particular, we recover the quantum phase transition between a disordered and a symmetry broken phase in the ground states of generalized PXP models, as well as the periodic dynamics of special initial states associated to quantum scars in PXP models, via a time-dependent version of the variational principle~\cite{TDVP1,TDVP2}. In 1D, the phase diagram has been studied analytically~\cite{FendleySachdev2004}, and in 1D and 2D, these phenomena have been observed experimentally \cite{51atom,256atom,drivenRydberg} and studied numerically with density matrix renormalization group techniques \cite{PichlerSachdev2020}. Here our method provides a complementary viewpoint and adds the benefit of an analytical description. In addition, it also treats the so far unexplored case of 3D lattices, and allows for quantitative predictions, such as the location of the quantum phase transition point in equilibrium, or the period of revival in scarred dynamics.

The paper is structured as follows. We first introduce the class of PXP models analyzed in this work in Sec.~\ref{sec:PXPmodel} and review the basic phenomena that this model is expected to display. In Sec.~\ref{sec:variationalansatz} we introduce our variational manifold and we present a method that allow us to perform tensor network contractions in two and three dimensions in the form of a power series. In Sec.~\ref{sec:ground state} we discuss the ground state phase diagram of our model in 1D, 2D and 3D using our variational ansatz, and compare its predictions with exact diagonalization results. In Sec.~\ref{sec:dynamics} we discuss time evolution obtained via the TDVP and contrast its predictions with the exact dynamics of the PXP model.

\section{Model}\label{sec:PXPmodel}

\subsection{Constrained Hilbert space}
In this work we are interested in constrained spin models that are often referred to as PXP models. Specifically, we consider $N$ spin-1/2 particles, each with basis states $\ket{\downarrow}$ and $\ket{\uparrow}$, that are defined on the sites of a $D$-dimensional lattice, as depicted in Fig.~\ref{fig:lattices}. We consider states of this spin systems that satisfy a constraint that prevents two spins on neighbouring lattice sites to be simultaneously in the $\ket{\uparrow}$ state. In physical realizations this constraint arises dynamically from strong state-dependent nearest neighbour interactions. Specifically, this constraint  captures the essence of the Rydberg blockade mechanism in Rydberg atom arrays. The spin configurations that satisfy this constraint form a subspace of the $2^N$ dimensional Hilbert space. For large system sizes, the dimension of this constrained Hilbert space grows as $x^N$, where the value of $x<2$ depends on the lattice. For a 1D chain, $x=(1+\sqrt{5})/2\simeq 1.618$ \footnote{The Hilbert space dimension of $N$ spins $d_H(N)$ satisfies the equation $d_H(N+1) = d_H (N) + d_H(N-1)$, which implies $d_H \sim \phi^N$ for large $N$, where $\phi$ is the golden ratio.}. In higher dimensions the larger coordination number increases the effect of the constraint, leading to a decrease of $x$ with higher $D$. 

In the following we denote the projector onto the constraint-satisfying subspace by $\mathcal{P}$. It can be composed from local projectors $\mathcal{P}_{ij}$ that act on pairs of neighboring spins $i$ and $j$, i.e., $\mathcal{P} = \prod_{\langle ij\rangle}\mathcal{P}_{ij}$. Here $\mathcal{P}_{ij}=\ket{\downarrow\downarrow}\bra{\downarrow \downarrow}+\ket{\downarrow \uparrow}\bra{\downarrow \uparrow}+\ket{\uparrow \downarrow}\bra{\uparrow \downarrow}$ simply annihilates components that do not satisfy the constraint on the link between $i$ and $j$.

\subsection{Hamiltonian}
Let us now write down the Hamiltonian of a spin model that lives in this constrained Hilbert space; the simplest example is the PXP model
\begin{align}\label{eq:hamiltonian}
    H_\text{PXP}=  \Omega\sum_i\mathcal{P}\sigma^x_i\mathcal{P},
\end{align}
where $\sigma^x=\ket{\uparrow}\bra{\downarrow}+\ket{\downarrow}\bra{\uparrow}$. The Hamiltonian \eqref{eq:hamiltonian} can be interpreted as the projection of a non-interacting model of independently rotating spins into the constraint-satisfying subspace. It induces a conditional dynamics, where a spin only precesses if all its neighbours are in the $\ket{\downarrow}$ state. Crucially this projected Hamiltonian is interacting, and in general non-integrable~\cite{PXPintegable,Signaturesofintegrability}. Note that the geometry of the lattice enters in \eqref{eq:hamiltonian} implicitly via the projection operation, which manifestly depends on the lattice structure. 
 
 In this work we study the physics of what we shall call generalized PXP models, the family of Hamiltonians of the PXP model \eqref{eq:hamiltonian} with the addition of a transverse-field term and a next-nearest-neighbour interaction:

\begin{align}\label{eq:hamiltonianNNN}
    H = \sum_i \mathcal{P}(\Omega\sigma^x_i-\Delta n_i) \mathcal{P} + V\sum_{\langle\!\langle ij\rangle\!\rangle}\mathcal{P}n_i n_j\mathcal{P},
\end{align}
where $n =\ket{\uparrow}\bra{\uparrow}$.

 This Hamiltonian may be interpreted as a toy model for an array of coherently driven Rydberg atoms. In these systems the state $\ket{\downarrow}$ represents an internal electronic ground state of an atom and the state $\ket{\uparrow}$ represents a highly excited Rydberg state. When driven by a laser that couples these two states, the system is described by a Hamiltonian of the form
 \begin{align}\label{eq:hamiltonianExp}
    H_\text{Rydberg}= \sum_i (\Omega\sigma^x_i-\Delta n_i) + \sum_{ij}V_{ij}n_i n_j.
\end{align}
Here $\Omega$ and $\Delta$ denote the Rabi frequency and the laser detuning, and $V_{ij} = C/|i-j|^6$ is the van der Waals interaction between two atoms $i$ and $j$ that are both in the Rydberg state. 
The strength of the interaction depends on the geometric distance between the atoms, with a length scale set by the blockade radius $R_b = (C/\Omega)^{1/6}$. In the so-called blockade approximation one considers the limit $V_{ij}\rightarrow\infty$ if two atoms are less than one blockade radius apart, and $V_{ij}\rightarrow 0$ otherwise. In particular, when $R_b$ equals one lattice spacing $a$, the model reduces to Eq.~\eqref{eq:hamiltonianNNN} with $V=0$, which becomes the PXP model for $\Delta=0$. If one sets $V_{ij}=V$ for next-nearest-neighbour sites, i.e. for $|i-j|=2 a$ in 1D or $|i-j|=\sqrt{2} a$ in $D>1$, then one recovers the Hamiltonian in Eq.~\eqref{eq:hamiltonianNNN}. While it is natural to have $V>0$, it is also theoretically and experimentally~\cite{EndresRydberg2023} interesting to consider the case $V<0$, i.e. with attractive next nearest neighbour interactions.  In what follows, we will set $\Omega=1$.

\subsection{Physics of the PXP model on bipartite lattices}
Hamiltonians of the form~\eqref{eq:hamiltonianNNN} host a remarkably rich variety of physical phenomena. For instance, depending on the choice of the lattice, which selects the form of $\mathcal{P}$, the ground state of~\eqref{eq:hamiltonianNNN} can host a plethora of symmetry broken ordered phases~\cite{FendleySachdev2004,PichlerSachdev2020,Samajdar2021} as well as topologically ordered phases~\cite{Verresen2021}. In this work we focus on bipartite lattices, in particular on hypercubic lattices, and denote the two sublattices by $A$ and $B$ respectively. For such bipartite lattices we now briefly review the expected qualitative features of the system in and out of equilibrium.  

\subsubsection{Equilibrium}
\begin{figure}
    \centering
    \includegraphics[width=5cm]{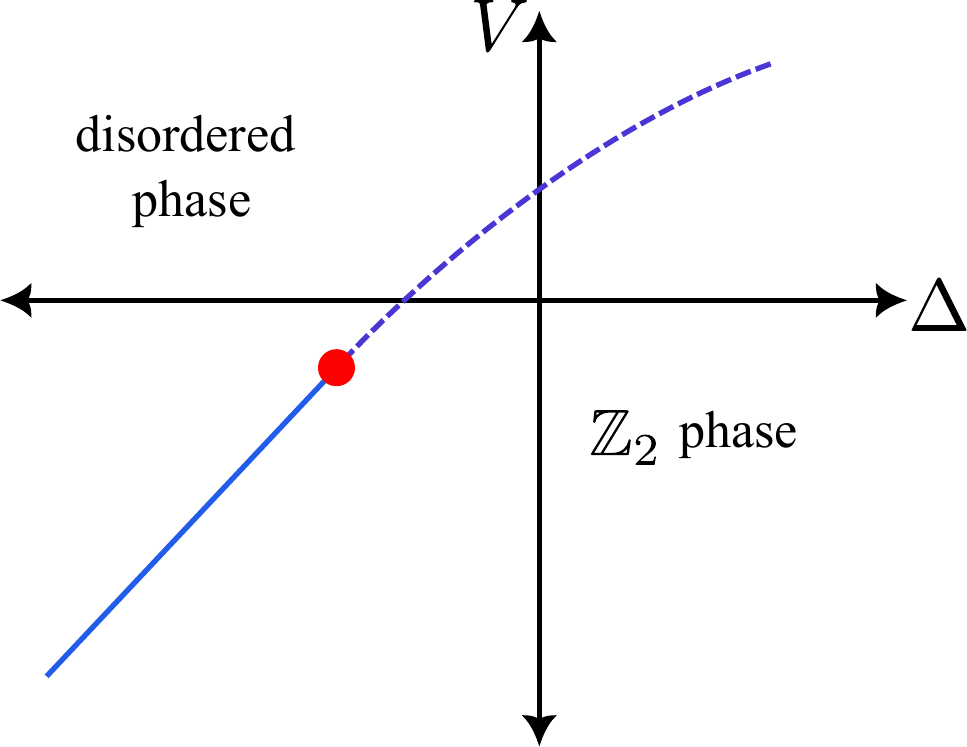}
    \caption{Sketch of the phase diagram for the generalized PXP model with next-nearest-neighbour interaction Eq.~\eqref{eq:hamiltonianNNN}. A disordered phase and a $\mathbb{Z}_2$ ordered phase are separated by second-order (dotted line) and first order (solid line) transitions, which meet at the tricritical point (red dot). Quantitative versions of this for 1D, 2D, and 3D are shown in Fig. \ref{fig:phasediagrams}. }
    \label{fig:phasediagramcartoon}
\end{figure}

The ground state phase diagram of the Hamiltonian in Eq.~\eqref{eq:hamiltonianNNN}, at fixed finite values of $V$, depends only on $\Delta$. For $\Delta\to -\infty$, the ground state is unique and given by $\ket{\downarrow}^{\otimes N}$ independent of the lattice geometry. For $\Delta \to +\infty$ and any finite $V$, the ground state maximizes the number of spins in the $\ket{\uparrow}$ state that is consistent with the constraints. For a bipartite lattice this gives a twofold degenerate ground state, which spontaneously breaks the sublattice symmetry: the two ground states are given by configurations where all spins on sublattice $A$ are in the $\ket{\uparrow}$ state and all spins on sublattice $B$ are in the $\ket{\downarrow}$ state, or viceversa. We denote these states by $\ket{\mathbb{Z}_2}=\ket{\uparrow}^{\otimes A}\ket{\downarrow}^{\otimes B}$ and  $\ket{\mathbb{Z}_2^\prime}=\ket{\downarrow}^{\otimes A}\ket{\uparrow}^{\otimes B}$, respectively. 
In the limit $\Delta,V \to + \infty$, with constant $\Delta/V = c$, the ground state can be different. In 1D for $c > 3$ a period-3 density-wave state with $1/3$ density of up spins is favored. In 2D and 3D the states $\ket{\mathbb{Z}_2},\ket{\mathbb{Z}_2^\prime}$ cease to be the classical ground states for $c>4$ and $c>6$, respectively, in favor of states with density $1/4$, leading to a four-fold degenerate striated phase which arises perturbatively due to quantum fluctuations~\cite{PichlerSachdev2020}. The latter can stabilize these non-$\mathbb{Z}_2$-ordered phases even at finite $V$. In fact, in 1D, the interplay between $\mathbb{Z}_2$ and $\mathbb{Z}_3$ order gives rise to several interesting phenomena such as floating phases and chiral transitions~\cite{chepiga,PichlerSachdev2018,giudici2019,rader2019floating}. In $D>1$, the precise nature of the phase transition along all phase boundaries of~\eqref{eq:hamiltonianNNN} is less clear~\cite{PichlerSachdev2020}.
Here, we focus on the regime where $V$, when positive, is sufficiently small, such that
as $\Delta$ is tuned from $-\infty$ to $+\infty$ the system undergoes a direct quantum phase transition from the gapped, disordered phase to the gapped, $\mathbb{Z}_2$ ordered phase. When $D=1$ and for $V\geq 0$, the phase transition is second order. For sufficiently strong attractive next-nearest-neighbour interaction $V<0$, the phase transition becomes discontinuous and first order. A tricritical point for $V<0$ separates these two regimes. The most accurate numerical study of the entire transition line was performed in Ref.~\cite{chepiga}. The location of the tricritical point is known analytically, since it lies on the integrable line $\Delta = V - 1/V$ for $V = -( \sqrt{5} + 1) / 2 )^{5/2}$~\cite{FendleySachdev2004}. 
A sketch of the $(\Delta,V)$ phase diagram is shown in Fig.~\ref{fig:phasediagramcartoon}.
We expect a similar phase diagram to hold in $D>1$ and, as we show below, our variational method predicts that this is indeed the case; in particular, we will use it to locate the phase boundary and the tricritical point in $D=1,2,3$ and we will compare the outcome with exact diagonalization results on 2D and 3D lattices with up to $48$ and $64$ sites, respectively.
We note that, whereas in 1D the transition is always surrounded by a disordered regime for any finite $\Delta,V$, in 2D and for large enough $\Delta$ the system undergoes a direct second order transition from the $\mathbb{Z}_2$ phase to the striated phase, with $V$ as driving parameter~\cite{PichlerSachdev2020,kalinowski}. Since our ansatz only captures $\mathbb{Z}_2$ order, we will restrict our discussion to the regime where the effect of the $1/4$-density ordered phase is negligible. 

\subsubsection{Quantum many-body scars}
When $\Delta=V=0$, the Hamiltonian in Eq.~\eqref{eq:hamiltonianNNN} reduces to the PXP model which is known to host quantum many-body scars. These are special eigenstates that do not satisfy the eigenstate thermalization hypothesis~\cite{ETH1,ETH2} and have been proposed in Ref.~\cite{Turner2018} as the explanation of the anomalous dynamics observed in a Rydberg atom simulator initialized in one of the two $\ket{\mathbb{Z}_2}$ states~\cite{51atom}.
The anomaly consists in the fact that despite the Hamiltonian~\eqref{eq:hamiltonian} being non-integrable, and the initial $\ket{\mathbb{Z}_2}$ state having the energy density of an infinite temperature ensemble, the system does not rapidly thermalize, but rather exhibits (approximate) periodic dynamics with alternating revivals of the two $\ket{\mathbb{Z}_2}$ states. 
It is now believed that this phenomenon arises in the PXP model in any dimension whenever the lattice is bipartite~\cite{stabilizing}, although numerical studies in $D>2$ are absent in the literature.
The name quantum scars comes from the analogy, put forward in Ref.~\cite{Turner2018}, with single-particle physics, where the quantum system possesses special eigenstates that are ``scars'' of unstable periodic orbits of their, otherwise chaotic, classical counterpart~\cite{Heller1984}.
In this context, the one-dimensional version of the variational ansatz which we will introduce in Sec.~\ref{sec:variationalansatz} was employed in Ref.~\cite{periodicOrbits} to map the constrained dynamics of the PXP model into classical equations of motion where the $\ket{\mathbb{Z}_2}$ states indeed lie on unstable periodic orbits.
Below we generalize these findings to $D>1$ and demonstrate the predictive power of our variational ansatz in two and three dimensions.
A characteristic quantity that we can compute in our framework is the revival time, which we find to increase with the coordination number of the lattice. By comparing the result with the exact many-body quantum dynamics on small systems, we will show that our method predicts the period with an accuracy increasing with the lattice dimension. 

\begin{figure}
    \centering
    \includegraphics[width=8cm]{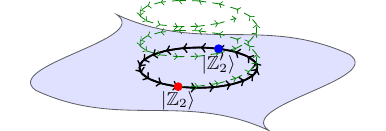}
    \caption{Schematic illustration of the periodic revivals in quantum scarred systems, and how they are captured by the variational method. The 2D surface represents the variational manifold, and the loop on it represents the exact periodic orbits in the variational dynamics connecting the two $\ket{\mathbb{Z}_2}$ states. The dotted line represents the true time evolved state $e^{-iH_\text{PXP}t}\ket{\mathbb{Z}_2}$.}
    \label{fig:manifold}
\end{figure}

\section{Variational ansatz}\label{sec:variationalansatz}
We now introduce a variational manifold of states that is designed to capture some of the physics of PXP-type models on bipartite lattices. The manifold of states $|\psi(\vec{\theta},\vec{\phi})\rangle$ can be represented as tensor network states with bond dimension $\chi=2$ on the network given by the lattice of interest. The key feature of our variational manifold is that the states satisfy the blockade constraint on the corresponding lattice by construction, i.e., they do not contain configurations where two neighboring spins are both in the $\ket{\uparrow}$ state. The variational manifold is parameterized by two angles, $\theta_i\in [-\pi,\pi)$ and $\phi_i\in [0,2\pi)$, for each lattice site $i$.  We emphasize that for the applications presented in this paper, we only work with two-site unit-cell translationally invariant systems (i.e. systems invariant under a translation by an even number of sites in any direction) with parameter space $(\theta_A,\theta_B,\phi_A,\phi_B)$, although in the discussion here we will refer to the general case $(\vec{\theta},\vec{\phi})$ when it presents no additional difficulty.

 The variational state $|\psi(\vec{\theta},\vec{\phi})\rangle$ is a projected entangled pair state (PEPS)\cite{OrusTNreview} with a PEPS tensor $M(\theta_i,\phi_i)$ defined at each lattice site $i$.   In 1D, the state is a matrix product state (MPS), and the MPS tensor is  
\begin{align}\label{eq:1dMPStensor}
	&\quad\vcenter{\hbox{\includegraphics[width=2.1cm]{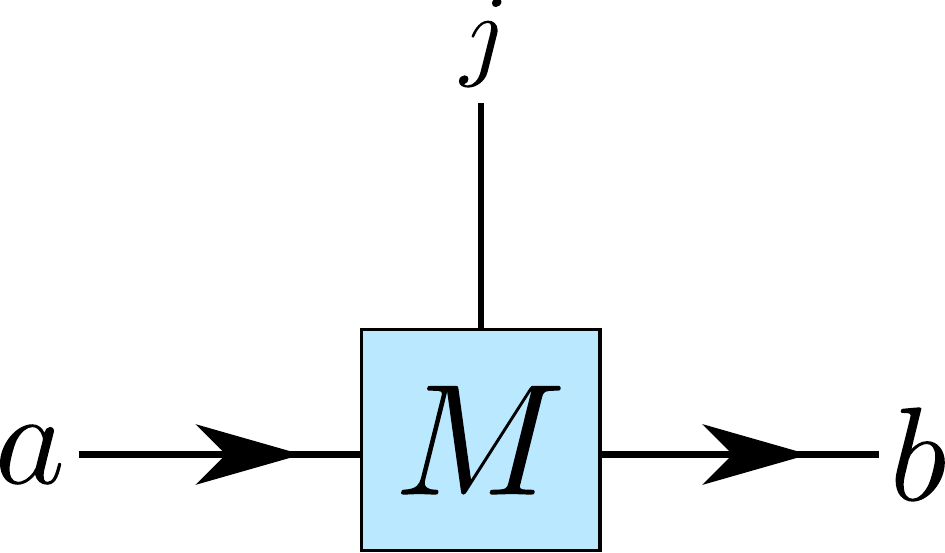}}}=M(\theta,\phi)^{(j)}_{ab} \nonumber\\=& 
		\begin{pmatrix}
		\cos(\frac{\theta}{2})\ket{\downarrow}_j & -ie^{i\phi}\sin(\frac{\theta}{2})\ket{\uparrow}_j\\
		\ket{\downarrow}_j & 0 
	\end{pmatrix}_{ab} 
\end{align}
where $\ket{\downarrow}=(1\;\;0)^T$ and $\ket{\uparrow}=(0\;\;1)^T$, and where $j$ is the physical index and $a$ and $b$ are the virtual indices. In the graphical notation above we assigned a direction to the virtual legs; the usefulness of this should be readily apparent when we consider higher dimensional generalizations.  

\begin{figure}[t]
    \centering
    \includegraphics[width = 5cm]{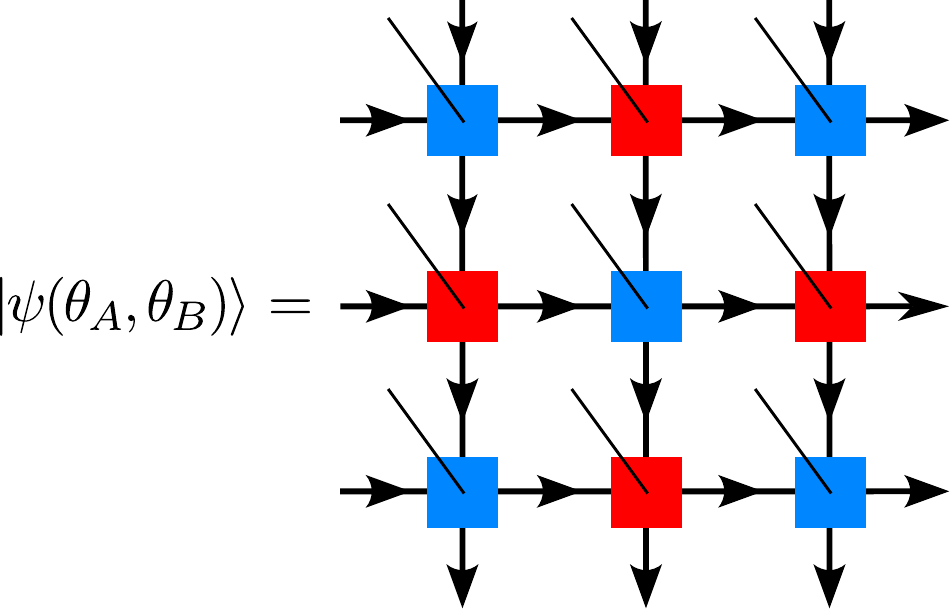}
    \hspace{0.3cm}
    \includegraphics[width=2.7 cm]{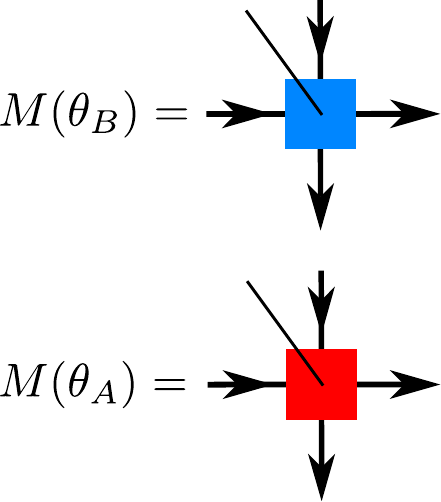}
    \caption{PEPS for the 2D square lattice}
    \label{fig:2dPEPS}
\end{figure}

We generalize this ansatz as follows. On a 2D square lattice, the PEPS tensors take the form
\begin{align}\label{eq:2DPEPStensor}
	&\vcenter{\hbox{\includegraphics[width=1.9cm]{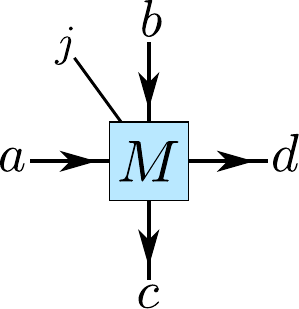}}}\;=\; M(\theta,\phi)^{(j)}_{(ab)(cd)} \nonumber\\=&\,
	\begin{pmatrix}
		\cos(\frac{\theta}{2})\ket{\downarrow}_j & \;0\;\; & \;\;0\; & -ie^{i\phi}\sin(\frac{\theta}{2})\ket{\uparrow}_j\\
		\ket{\downarrow}_j & \;0\;\; & \;\;0\; & 0 \\
		\ket{\downarrow}_j & \;0\;\; & \;\;0\; & 0 \\
		\ket{\downarrow}_j & \;0\;\; & \;\;0\; & 0  
	\end{pmatrix} _{(ab)(cd)},
\end{align}
where the matrix is written in the basis $(00,01,10,11)$. Fig.~\ref{fig:2dPEPS} depicts the 2D PEPS for a system with a two-site unit cell translation symmetry imposed. 

The above construction suggests a generalization that can be applied to arbitrary dimensional lattices: in general each tensor will have some ``in" virtual indices, and some ``out" virtual indices. For a hypercubic lattice in $D$ dimensions, each PEPS tensor will have $2D+1$ total legs, with $D$ incoming virtual indices and $D$ outgoing virtual indices.  Diagrammatically, we write 
\begin{align}\label{eq:MgeneralD}
	  \vcenter{\hbox{\includegraphics[width=2.8cm]{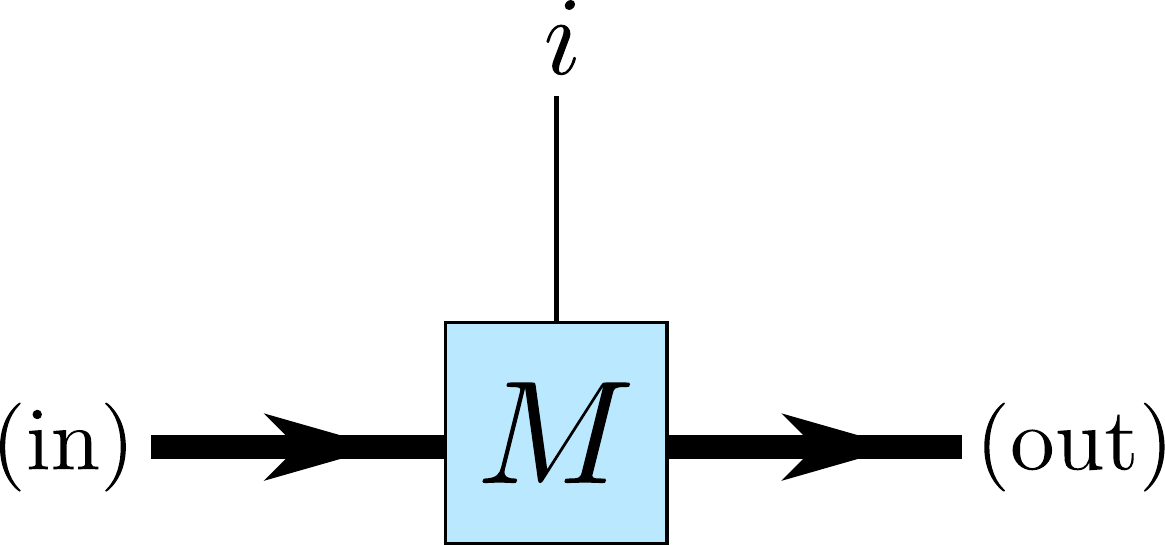}}}=M(\theta,\phi)^{(i)}_{(\text{in})(\text{out})},
\end{align}
where a thick line collectively represents all incoming or outgoing legs. Introducing a bra-ket notation on the virtual indices, the tensor $M$ reads
\begin{align}
M = \sum_{i=\downarrow,\uparrow}\sum_{\text{in, out}}M^{(i)}_{(\text{in})(\text{out})}\ket{i}\ket{\text{in}}\bra{\text{out}}.     
\end{align}
Let $I\in \{0,1\}^D$ be a binary string of length $D$, let $\sum_I$ be the sum over all $2^D$ such strings, and denote by $\ket{I}$ the corresponding product state in the virtual Hilbert space.  We further use the short-hand notation  $\Vec{0}=(0,0,0,\dots)$ and $\Vec{1}=(1,1,1,\dots)$, and define $\ket{\alpha}\equiv\sum_I\ket{I}$ and $\ket{\beta}\equiv\sum_{I\neq\Vec{0}}\ket{I}=\ket{\alpha}-|\vec{0}\rangle$.
Using this notation, the tensor $M$ can be written in general dimensions as
\begin{align}\label{eq:Mtensor}
    M(\theta,\phi) = \ket{\downarrow}\big(\cos(\theta/2)|\Vec{0}\rangle\langle\Vec{0}|+\ket{\beta}\langle\vec{0}|\big)\nonumber\\
    -ie^{i\phi}\sin(\theta/2)\ket{\uparrow}|\vec{0}\rangle\langle\vec{1}|.
\end{align}
It is straightforward to see that this reproduces Eqs.~\eqref{eq:1dMPStensor} and~\eqref{eq:2DPEPStensor} for $D=1, 2$, respectively.

 The variational ansatz $\ket{\psi}$ has several interesting properties. As already noted, it satisfies the blockade constraint. Remarkably, it is always normalized, i.e. $\langle\psi(\Vec{\theta},\Vec{\phi})|\psi(\Vec{\theta},\Vec{\phi})\rangle = 1$, in the thermodynamic limit; this follows from a simple but rather technical calculation which we present in Appendix~\ref{appendix:methods}. An important consequence of the normalization is that it allows one to calculate also expectation values and time evolution of states using the TDVP.
 For bipartite lattices, the variational manifold contains the important state $\ket{\mathbb{Z}_2}=\ket{\downarrow}^{\otimes A}\ket{\uparrow}^{\otimes B}$; this can be checked by setting $\theta_i=\pm\pi, \theta_j=0$ for all sites $i\in A, j\in B$. This is in fact a special case of a more general property: product states with all spins on one sublattice in the state $\ket{\downarrow}$, and with all spins on the other sublattice being any state on the Bloch sphere, are contained in the variational manifold. 
 In $D=1$ the variational manifold is known to be equivalent to the manifold of all product states projected into the constraint-satisfying Hilbert space~\cite{periodicOrbits}. This is not the case in $D>1$. In fact, the projected product state has a different tensor network representation that is not gauge-equivalent to Eq.~\eqref{eq:Mtensor} and does not yield a normalized state in the thermodynamic limit. Since this property provides remarkable simplifications in the following variational calculations, we will stick to the ansatz in Eq.~\eqref{eq:Mtensor} and provide a more in-depth discussion about its relation to the projected product state manifold in Appendix~\ref{appendix:projprod}.

\subsubsection{$(\theta_A,\theta_B)$ space for systems on bipartite lattices}

\begin{figure}
    \centering
    \includegraphics[width=8.5cm]{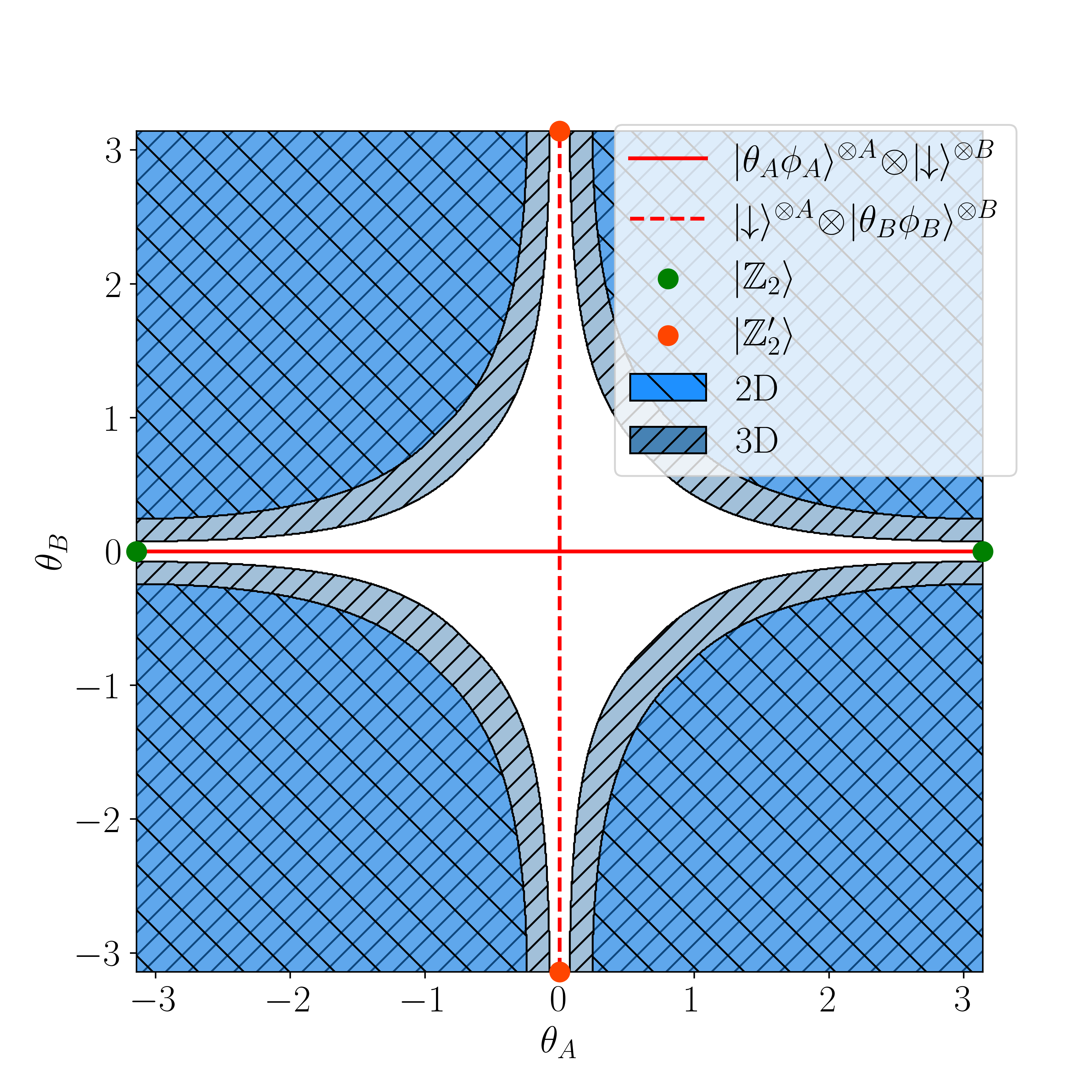}
    \caption{Depiction of the $(\theta_A,\theta_B)$ space that our variational state $\ket{\psi(\theta_A,\theta_B)}$ is in. 
 Points along the axes are all product states, where we have defined $\ket{\theta\phi}=\cos(\theta/2)\ket{\downarrow} -ie^{i\phi}\sin(\theta/2)\ket{\uparrow}$, and $\ket{\mathbb{Z}_2}=\ket{\uparrow}^{\otimes A}\ket{\downarrow}^{\otimes B}$ and $\ket{\mathbb{Z}_2^\prime}=\ket{\downarrow}^{\otimes A}\ket{\uparrow}^{\otimes B}$. (For the applications here $\phi$ is irrelevant and can be fixed to $0$ or $\pm\pi/2$.) The shaded regions correspond to the non-perturbative regime (which is slightly larger in 3D than in 2D). In 1D and 2D we can access the entire space via explicit contraction, while in 3D we are limited to the unshaded region. However, we find that the perturbative regime captures all of the physics discussed in this work.}
    \label{fig:blankplot}
\end{figure}

For the applications that follow, we only consider two-site unit-cell translational invariant states $\ket{\psi(\theta_A,\theta_B,\phi_A,\phi_B)}$.  The $\phi$ variables are \textit{a priori} relevant, but we will show that for the variational dynamics, the trajectories contain only states with $\phi_A,\phi_B = 0$, and that the variational ground states have $\phi_A,\phi_B =\pm\pi/2$.  Thus, the value of  $\phi_A$ and $\phi_B$ can be considered fixed, and we have a two dimensional parameter space $\theta_A, \theta_B \in [-\pi,\pi)$, which contains the states relevant for variational energy minimization, and in which the variational time evolution takes place.  This space is shown in Fig.~\ref{fig:blankplot}, where we label the $\ket{\mathbb{Z}_2}$ states and the product states along the $\theta_A$ and $\theta_B$ axes, as well the non-perturbative and perturbative regimes for the series expansion calculation of expectation values (discussed below).

\subsection{Calculation of expectation values}\label{sec:asymptoticseries}
The structure of the variational state $\ket{\psi} = |\psi(\Vec{\theta},\Vec{\phi})\rangle$ allows us to efficiently calculate quantities of the form $\langle\psi|\hat{\mathcal{O}}|\psi\rangle$, for product operators $\hat{\mathcal{O}}$ in up to three dimensions. This includes in particular local observables but also $n$-point functions, although in this work we only consider one- and two-point functions.  In this section we give an outline of the methods we use; a full discussion is in Appendix~\ref{appendix:methods}.

In general, for a bond-dimension $\chi$ tensor network state $\ket{\Psi}$, evaluating quantities like $\langle\Psi|\hat{\mathcal{O}}|\Psi\rangle$ requires the contraction of a tensor network of bond dimension $\chi^2$. However, even though the tensor networks states introduced above have bond dimension $\chi=2$, it turns out that the evaluation of quantities of the form $\langle\psi|\hat{\mathcal{O}}|\psi\rangle$ reduces to the contraction of a tensor network, whose bond dimension is not 4 but instead only 2.
For $D=1$ the tensor network can be contracted analytically for infinte systems and for $D= 2$, one can calculate $\langle\psi|\hat{\mathcal{O}}|\psi\rangle$ via explicit tensor network contraction on infinite cylinders with finite circumference. The reduced bond dimension allows one to perform the calculation for up to relatively large circumference.  We performed our numerical calculations with cylinders of circumference $L=10$ sites, but we verified that compared to larger cylinders the results are unchanged to several significant figures.  

\subsubsection{Perturbative expansion}\label{sec:PertExp_main}
For $D\geq 2$, we introduce a method to express the tensor network contraction as a perturbative expansion. This perturbative method, while powerful, is quite technical and does not have a simple intuitive interpretation.  In this section, we give a high-level overview and then simply state the result, while a full derivation of the method can be found in Appendix~\ref{sec:perturbativeexp}.  

The main idea is to decompose each tensor in the network into a sum of two tensors. This seems counterproductive at first, since it expresses the quantity $\bra{\psi}\hat{\mathcal{O}}\ket{\psi}$ not as a contraction of a single tensor network, but instead as sum over contractions of $2^N$  tensor networks. However, with a clever choice of the decomposition, one can show that only few of those contractions give a nonzero contribution. Moreover, those contractions that give a finite contribution, can be easily evaluated. These terms can be counted and collected into a series expansion with $\sin^2(\theta_A/2)$ and $\sin^2(\theta_B/2)$ as the expansion parameter.  

Let us illustrate this. For simplicity, we focus on the case of a local operator  $\hat{\mathcal{O}}$ acting on a single site in a 2D lattice. An important object in the tensor network representing the quantity $\bra{\psi}\hat{\mathcal{O}}\ket{\psi}$ is the tensor $T$ 
formed by contracting the PEPS tensor $M$ with its conjugate:
\begin{align}\label{eq:Tdef}
T= \;\vcenter{\hbox{\includegraphics[width=1.9cm]{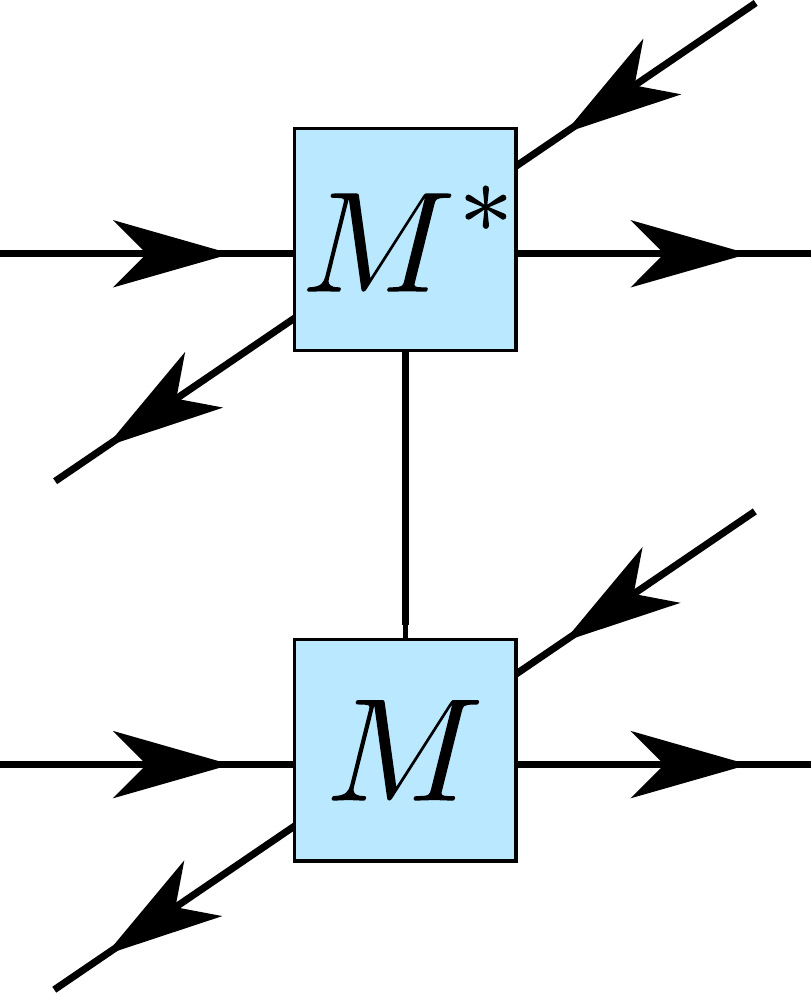}}}\; =\;\; \vcenter{\hbox{\includegraphics[width=1.8cm]{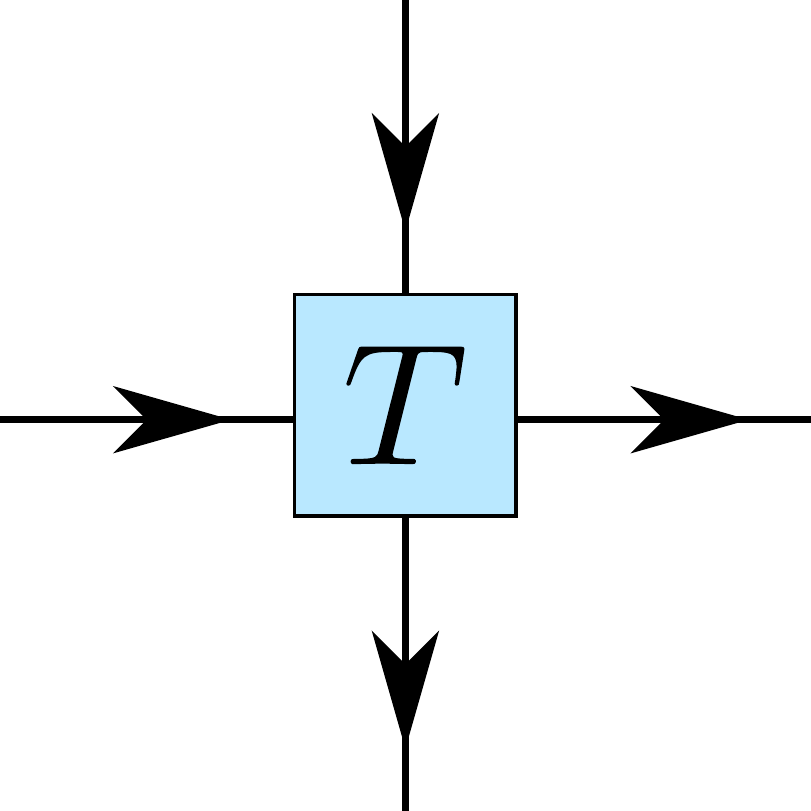}}}\;.
\end{align}
As mentioned above, it is useful to express $T$ as a sum of two tensors:
\begin{align}\label{eq:Tdecomp}
T =\, 
\begin{tikzpicture}[scale=0.7,baseline=-0.5ex]
\draw[thick] (-0.5,0) -- (0.5,0);
\draw[thick] (0,-0.5) -- (0,0.5);
\fill[black] (0,0) circle (0.1cm);
\end{tikzpicture} \;+\;
\begin{tikzpicture}[scale=0.7,baseline=-0.5ex]
\draw[thick] (-0.5,0) -- (0.5,0);
\draw[thick] (0,-0.5) -- (0,0.5);
\fill[blue] (0,0) circle (0.2cm);
\end{tikzpicture}.
\end{align}
The black dot is a constant tensor (i.e. independent of the variational parameters), while the blue dot carries a factor of  $\sin^2(\theta/2)$. In the notation introduced in Appendix~\ref{sec:properties}, the black dot is $p$ and the blue dot is $-\sin^2(\theta/2)q$. These tensors have the crucial property that the following contraction is zero:
\begin{align}\label{eq:pqZero2D}
    \begin{tikzpicture}[scale=0.7,baseline=-2ex]
\draw[thick] (-1.5,-1.5) grid (0.5,0.5);
\fill[black] (0,0) circle (0.1cm);
\fill[black] (-1,-1) circle (0.1cm);
\fill[blue] (-1,0) circle (0.2cm);
\def\x{0.55}
\fill[white] (-\x,-1-\x) rectangle (0+\x,-1+\x);
\end{tikzpicture} \,=0
\end{align}
Using the decomposition \eqref{eq:Tdecomp} and the property \eqref{eq:pqZero2D}, it is easy to see  $\bra{\psi}\hat{\mathcal{O}}\ket{\psi}$ can be expressed as a sum of tensor network contractions of the following form:
\begin{align}\label{eq:expansionSchematic}
\bra{\psi}\hat{\mathcal{O}}\ket{\psi} =\quad
\begin{tikzpicture}[scale=0.5,baseline={(0,0.1)}]
\draw[thick] (-2.5,-1.5) grid (1.5,2.5);
\foreach \x in {-2,...,1}
    \foreach \y in {-1,...,2}
    {
        \fill[black] (\x,\y) circle (0.1cm);
    }
\fill[red] (0,0) circle (0.3cm);
\end{tikzpicture}
\quad&+\quad
\begin{tikzpicture}[scale=0.5,baseline={(0,0.1)}]
\draw[thick] (-2.5,-1.5) grid (1.5,2.5);
\foreach \x in {-2,...,1}
    \foreach \y in {-1,...,2}
    {
        \fill[black] (\x,\y) circle (0.1cm);
    }
\fill[red] (0,0) circle (0.3cm);
\fill[blue] (-1,0) circle (0.2cm);
\end{tikzpicture} 
\nonumber\\[.5cm] +\quad
\begin{tikzpicture}[scale=0.5,baseline={(0,0.1)}]
\draw[thick] (-2.5,-1.5) grid (1.5,2.5);
\foreach \x in {-2,...,1}
    \foreach \y in {-1,...,2}
    {
        \fill[black] (\x,\y) circle (0.1cm);
    }
\fill[red] (0,0) circle (0.3cm);
\fill[blue] (-1,0) circle (0.2cm);
\fill[blue] (0,1) circle (0.2cm);
\end{tikzpicture}
\quad&+\quad
\begin{tikzpicture}[scale=0.5,baseline={(0,0.1)}]
\draw[thick] (-2.5,-1.5) grid (1.5,2.5);
\foreach \x in {-2,...,1}
    \foreach \y in {-1,...,2}
    {
        \fill[black] (\x,\y) circle (0.1cm);
    }
\fill[red] (0,0) circle (0.3cm);
\fill[blue] (-1,0) circle (0.2cm);
\fill[blue] (-2,1) circle (0.2cm);
\fill[blue] (-1,1) circle (0.2cm);
\end{tikzpicture}
\nonumber\\[.5cm] +\quad
\begin{tikzpicture}[scale=0.5,baseline={(0,0.1)}]
\draw[thick] (-2.5,-1.5) grid (1.5,2.5);
\foreach \x in {-2,...,1}
    \foreach \y in {-1,...,2}
    {
        \fill[black] (\x,\y) circle (0.1cm);
    }
\fill[red] (0,0) circle (0.3cm);
\fill[blue] (-1,1) circle (0.2cm);
\fill[blue] (0,1) circle (0.2cm);
\fill[blue] (-2,1) circle (0.2cm);
\fill[blue] (-2,2) circle (0.2cm);
\end{tikzpicture}
\quad &+ \quad\cdots
\end{align}
Here the red tensor is obtained by inserting the local operator $\hat{\mathcal{O}}$ between the tensors $M$ and $M^\ast$ in eq.~\eqref{eq:Tdef}. Eq.~\eqref{eq:expansionSchematic} shows a subset of diagrams that give non-zero contribution at various order of $\sin(\theta/2)$. To obtain the perturbative expansion, we collect all terms that contribute in the same order of $\sin(\theta/2)$. For the unit cell translation invariant states, $\ket{\psi(\theta_A,\theta_B,\phi_A,\phi_B)}$, that we consider below, the quantity $\bra{\psi}\hat{\mathcal{O}}\ket{\psi}$ can thus be expressed as 
\begin{align}\label{eq:expansion1}
    \bra{\psi}\hat{\mathcal{O}}\ket{\psi} &= h(\theta_A,\theta_B,\phi_A,\phi_B)\Sigma(f,\theta_A,\theta_B),\nonumber\\
    \Sigma(f,\theta_A,\theta_B) =&\sum_{n,m=0}^\infty\! (-1)^{n+m}f_{n,m}\sin^{2n}\!\!\Big(\frac{\theta_A}{2}\Big)\sin^{2m}\!\!\Big(\frac{\theta_B}{2}\Big),
\end{align}
where the prefactor $h(\theta_A,\theta_B,\phi_A,\phi_B)$ is a function that depends on the operator $\hat{\mathcal{O}}$, while $f$ is a simple matrix of ``counting factors", that also depends on $\hat{\mathcal{O}}$, but is independent of the variational parameters. 

\begin{figure*}[t]
    \centering
    \includegraphics[width=5.8cm]{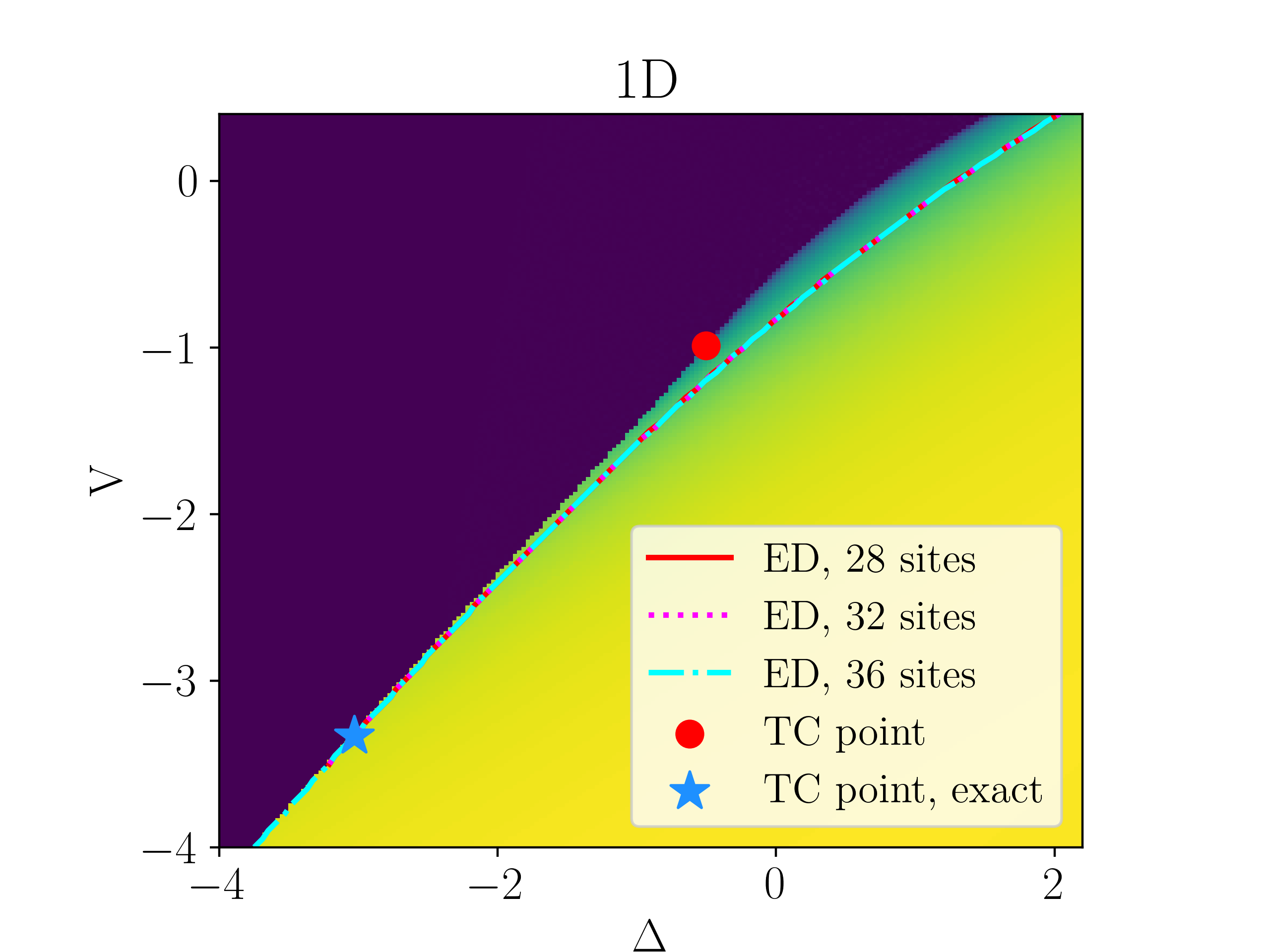}
    \hspace{-1cm}
    \includegraphics[width=5.8cm]{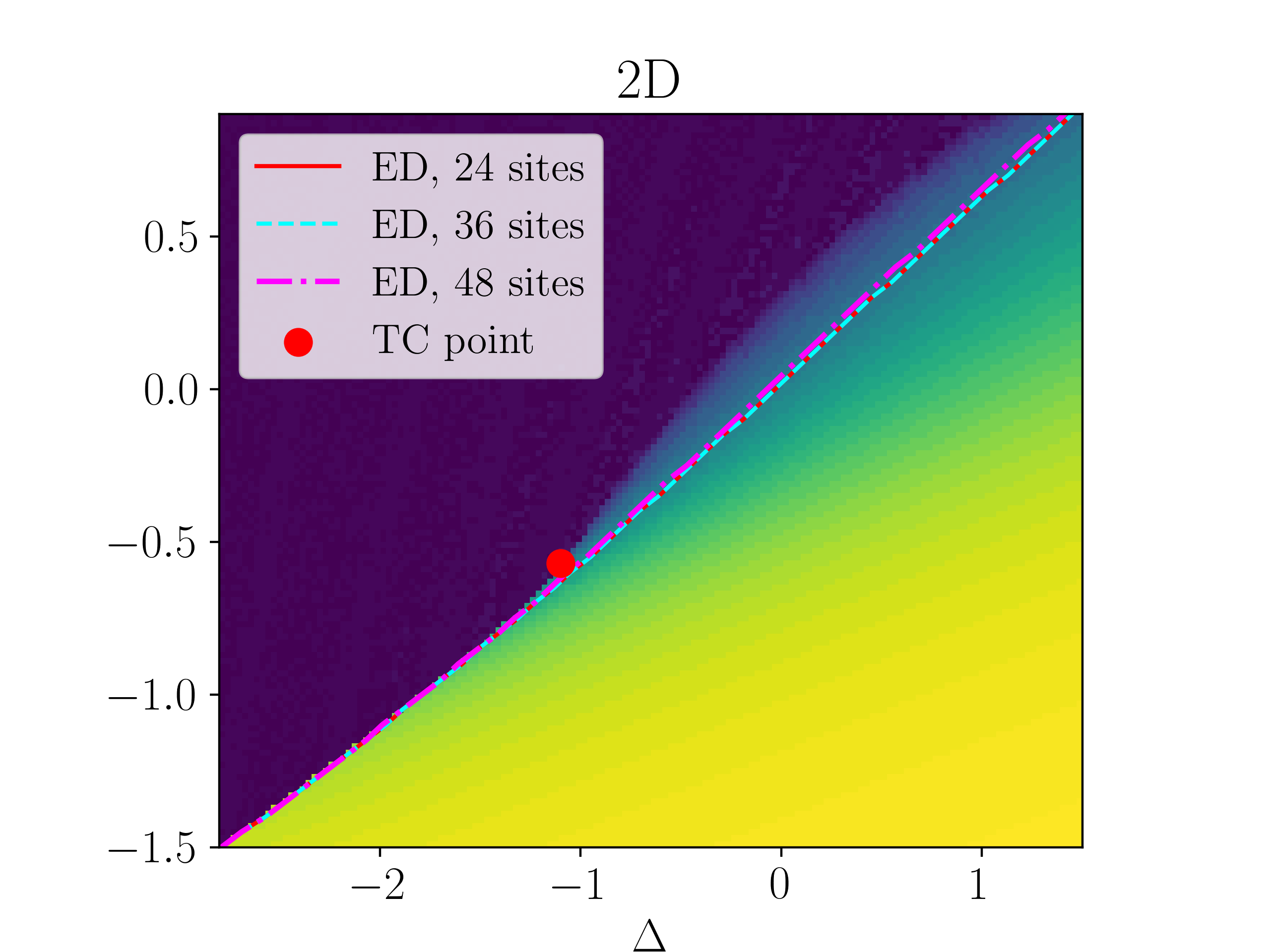}
    \hspace{-.7cm}
    \includegraphics[width=5.9cm]{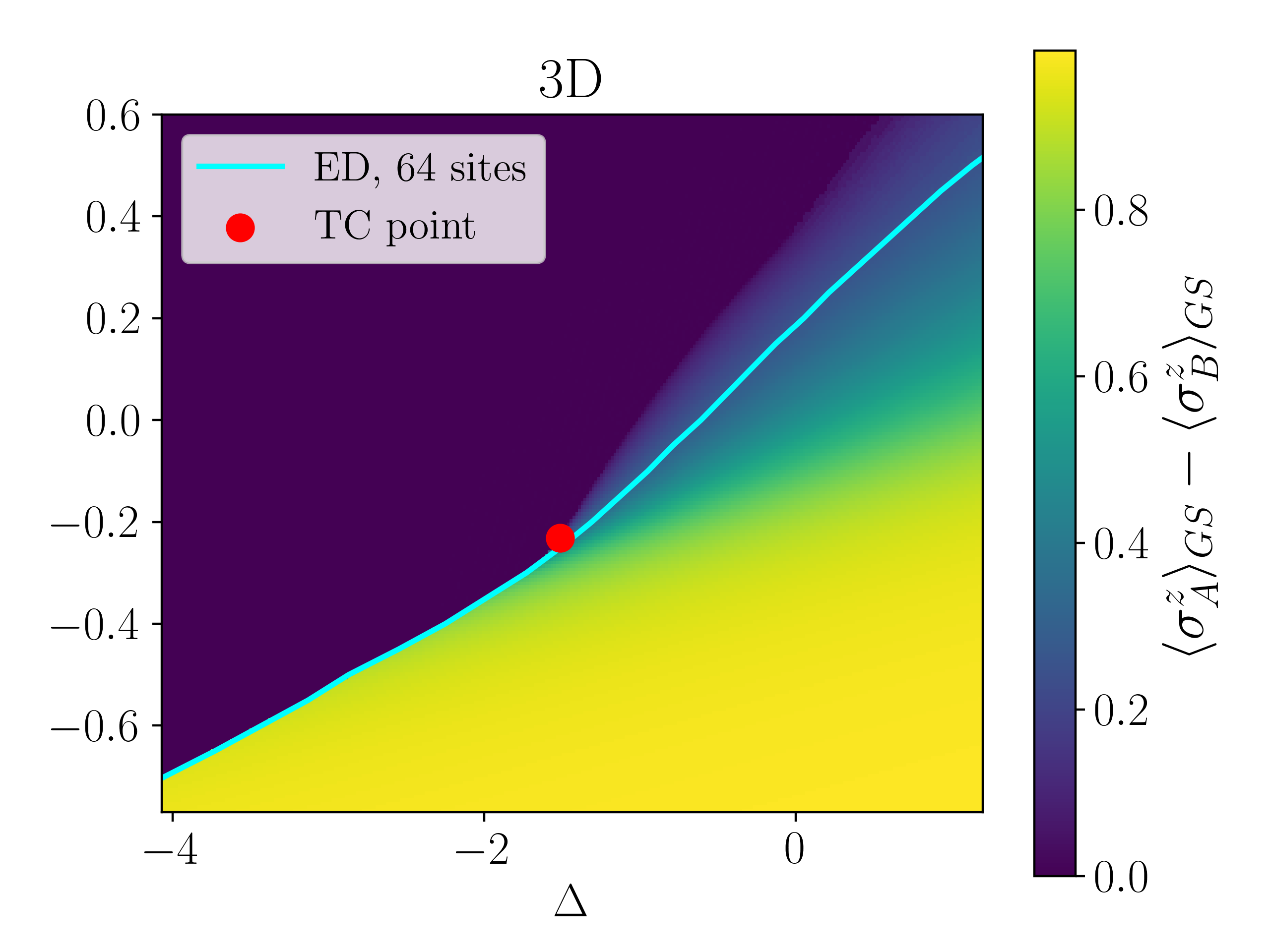}    \caption{$(\Delta,V)$ phase diagrams for the generalized PXP model~\eqref{eq:hamiltonianNNN} with the magnetization $\langle\sigma^z_A\rangle_{GS}-\langle\sigma^z_B\rangle_{GS}$ as the order parameter.  The red dots denote the tricritical points (as found from the variational method) while the lines show, for comparison, the phase boundaries found via exact diagonalization on various system sizes. Finite-size critical points are extracted from the peaks in the ground state fidelity susceptibility (see Appendix~\ref{appendix:ED} for details). In 1D, we also show the analytically known tricritical point.}
    \label{fig:phasediagrams}
\end{figure*} 

We note that Eq.~\eqref{eq:expansion1} should be understood as an asymptotic series. 
This is because the counting factors $f_{n,m}$ scale super exponentially (that is, there do not exist any $a,b\in\mathbb{R}$ such that $f_{n,m}<a^n b^m$ for all $n,m\in \mathbb{N}$), causing the series to eventually diverge at high enough order.  However, we find that in regions of the $(\theta_A,\theta_B)$ plane where the effective expansion parameters $\sin^2(\theta_A/2)$ and $\sin^2(\theta_B/2)$ are small enough, i.e. a star shaped region around the $x$ and $y$ axes, the terms of the series decrease quickly, and accurate results are obtained at finite order.  The perturbative regimes in Fig.~\ref{fig:blankplot} were defined as the region where the expansion is accurate to an error of $10^{-3}$.
We verified that this perturbative method is valid for the two applications we work with (energy minimization and time evolution): for the former, we verify that our variational ground states lie inside the perturbative region, and for the latter, we considered time-evolved trajectories from different orders of the expansion, and verified that they converge. 

Finally, we note that the perturbative expansion can be interpreted as an expansion around product states.  For states in the variational manifold that are product states (i.e. $\theta_A=0$ or $\theta_B=0$), the sum \eqref{eq:expansion1} contains only a finite number of terms. As seen in Fig.~\ref{fig:blankplot}, the perturbative regime is located in the region around the set of product states, with the non-perturbative region corresponding to the states in the manifold with higher entanglement.  We also note that states with increasing correlation length require higher order expansion terms. For example, if one considers the equivalent of Eq.~\eqref{eq:expansionSchematic} for a two-point function $\bra{\psi}\hat{\mathcal{O}}_i\hat{\mathcal{O}}_j\ket{\psi}$, one would have to go to at least order $|i-j|$ to obtain nontrivial results.  

\section{Ground state phase diagram}\label{sec:ground state}

As a first application of the variational manifold, we calculate the energy $E = \bra{\psi}H\ket{\psi}$ and study the properties of the variational ground state.  We consider the generalized PXP model including the detuning term and a next-nearest-neighbour (NNN) term Eq.~\eqref{eq:hamiltonianNNN}.

 We use our variational manifold as a minimal ansatz beyond mean-field theory to study the ground state of this Hamiltonian and to reproduce the phase diagram.  With translation invariance we have a four dimensional parameter space with variational states $\ket{\psi(\theta_A,\theta_B,\phi_A,\phi_B)}$. However, we can restrict the manifold to real states by setting $\phi_A = \phi_B = \pi/2$, as the Hamiltonian is real. We consider the energy function $E(\theta_A,\theta_B)$ on the variational space $\theta_A, \theta_B \in [-\pi,\pi)$ and find the variational ground state by minimizing $E(\theta_A,\theta_B)$. Fig.~\ref{fig:phasediagrams} shows the resulting phase diagrams.

Let us first qualitatively describe the landscape of the optimal variational parameters $\theta_A^{GS},\theta_B^{GS}$ as the Hamiltonian is tuned. Suppose we fix $V$ and allow $\Delta$ to vary: when $\Delta=-\infty$, the ground state is the $|\Vec{0}\rangle$ state, and when $\Delta$ remains large and negative, we find that $\theta_A^{GS} =\theta_B^{GS}$ at the minimum energy state of the variational manifold. As $\Delta$ reaches a critical value $\Delta_c$, the variational ground state changes from the symmetric phase to a symmetry-broken phase where $\theta_A \ne \theta_B$. When $\Delta>\Delta_c$, the variational ground state becomes twofold degenerate, with nonzero $\theta_A^{GS}-\theta_B^{GS}$. For $V=0$, we find that the transition occurs at $\Delta_c^{1D}\simeq0.77$ in 1D, at $\Delta_c^{2D}\simeq-0.45$ in 2D, and at $\Delta_c^{3D}\simeq-1.0$ in 3D. In 1D it is known from exact diagonalization studies that $\Delta_c\simeq 1.31$~\cite{ssg}. In 2D, exact diagonalization on small systems sizes combined with finite-size scaling indicate that $\Delta_c\simeq-0.2$ (see Appendix~\ref{appendix:ED}). In 3D, the critical point for 64 sites is at $\Delta_c\simeq-0.61$. Thus we see that in all cases the variational method approximates, but underestimates the transition point.

\begin{figure}[t]
    \centering
    \includegraphics[width=2.9cm]{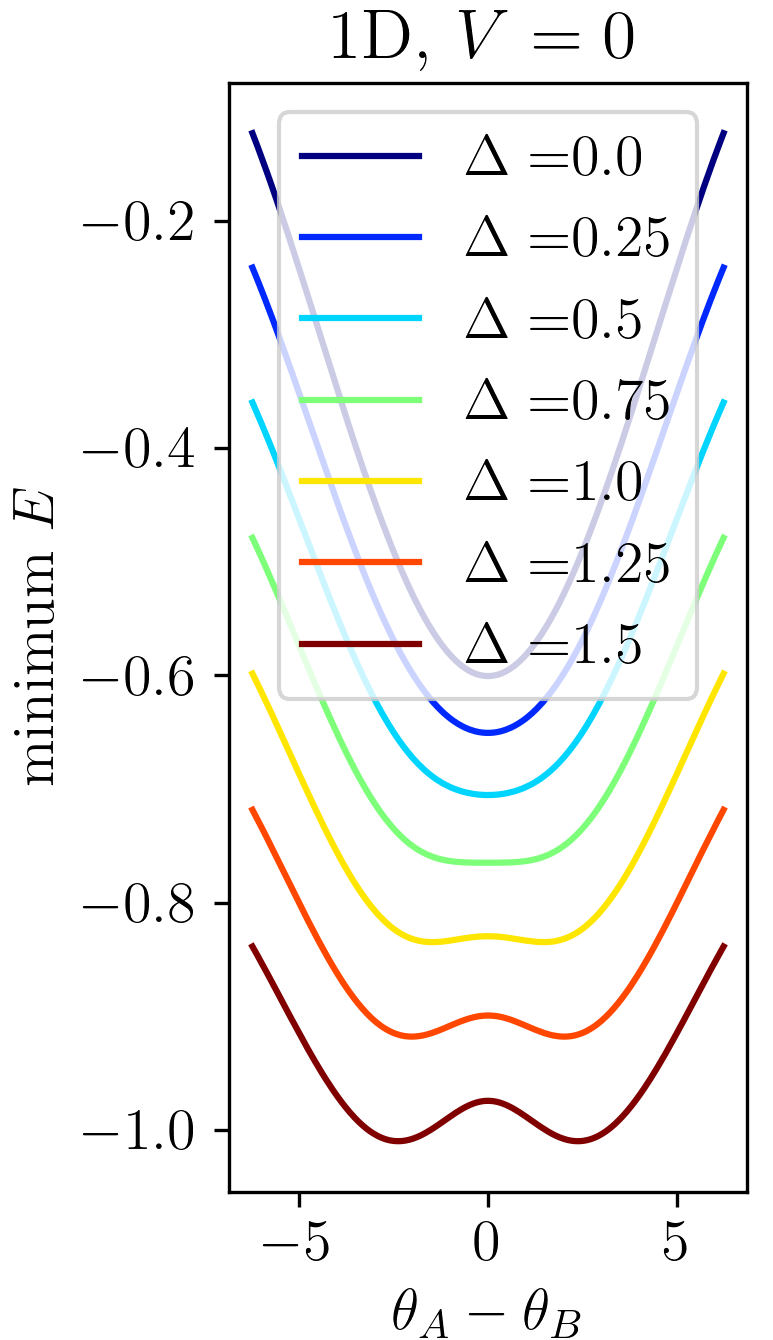} 
    \includegraphics[width=2.7cm]{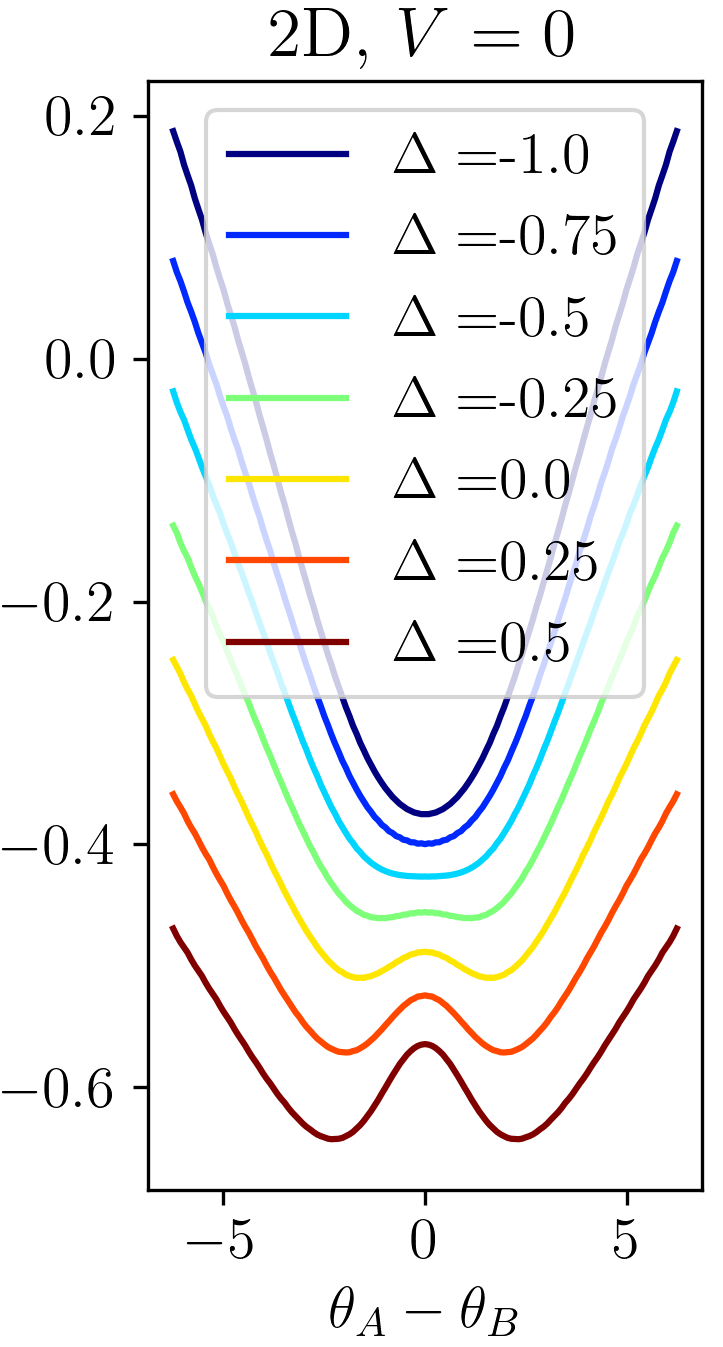}
    \includegraphics[width=2.66cm]{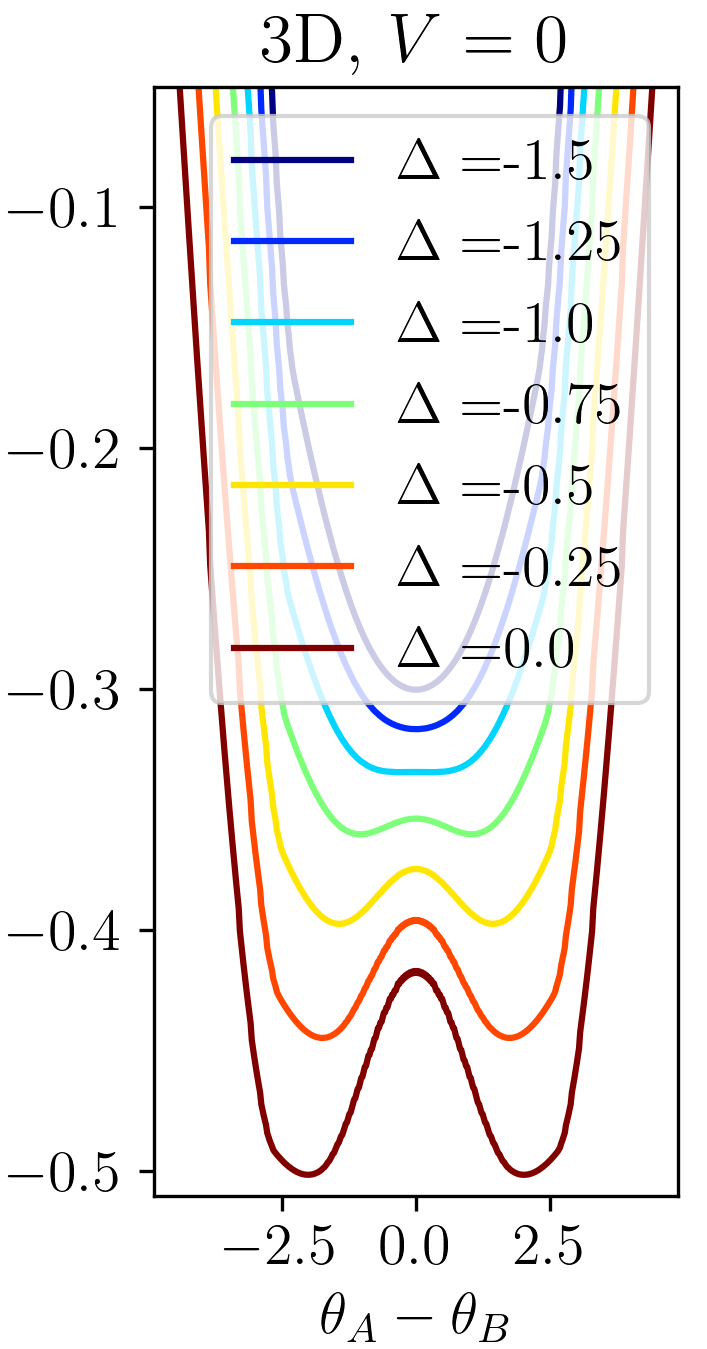}
    \includegraphics[width=2.9cm]{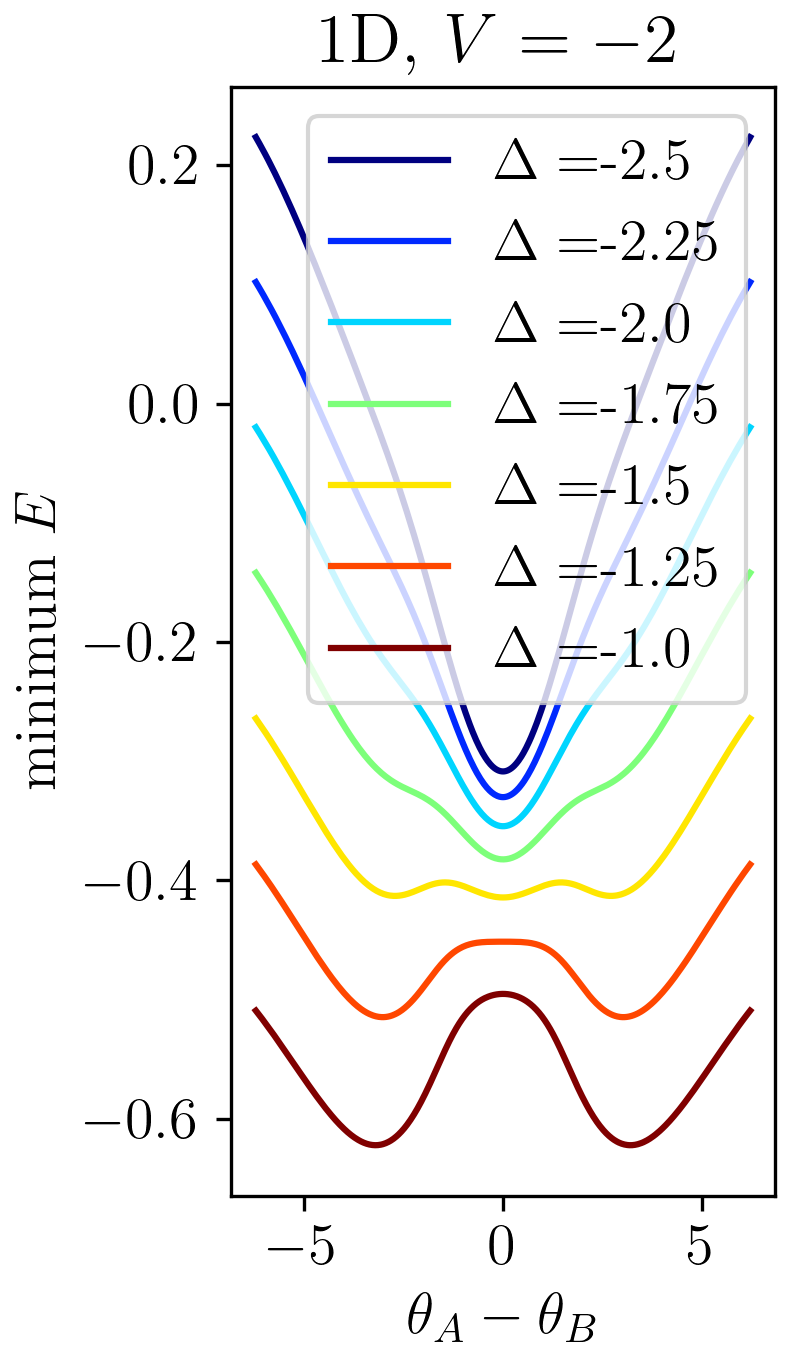} 
    \includegraphics[width=2.62cm]{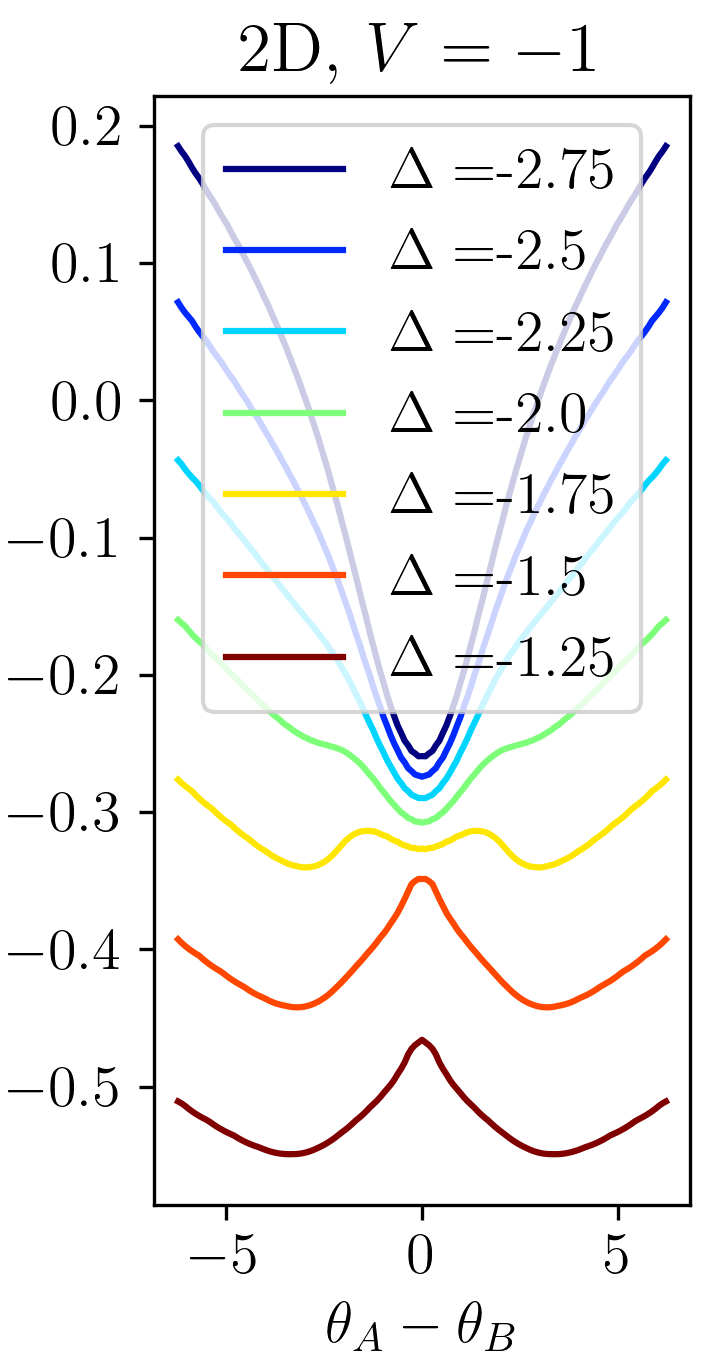}
    \includegraphics[width=2.6cm]{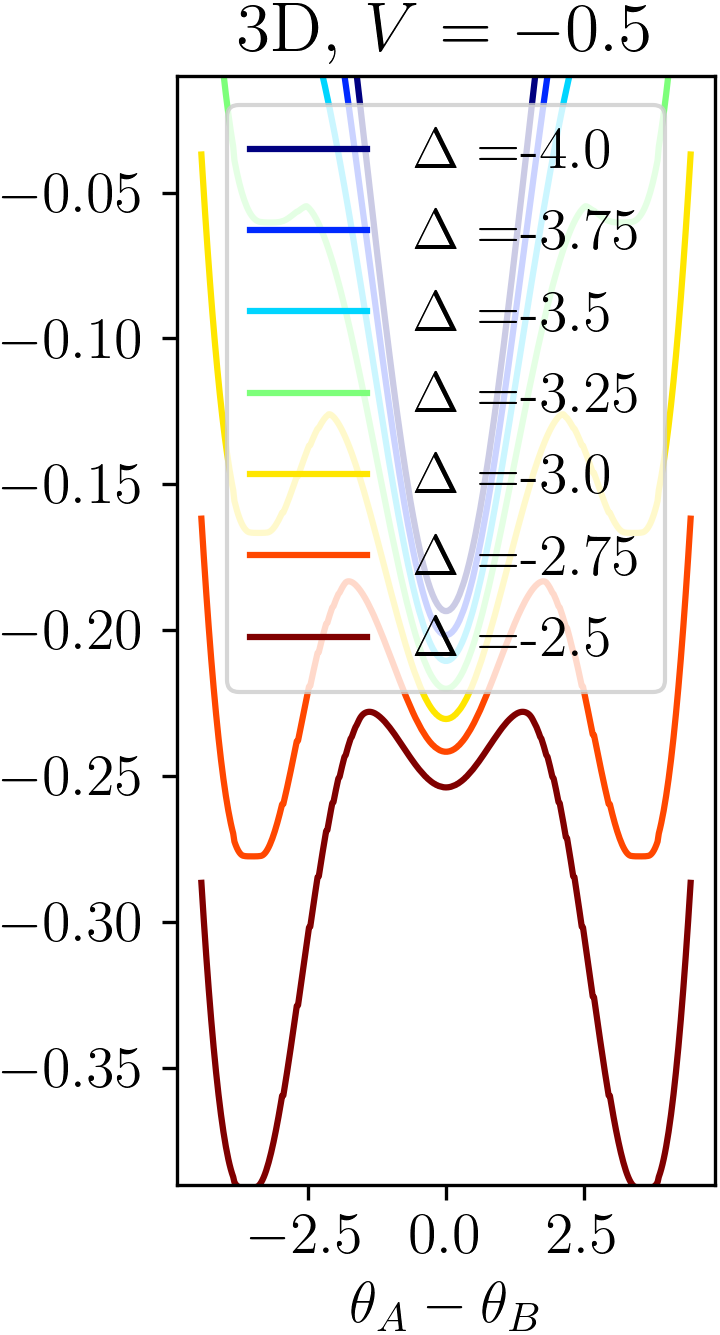}
    \caption{Plots of the minimum energy (per site) along lines of constant $\theta_A-\theta_B$ for $D=1,2,3$, showing the second order transition (top row) and first order transition (bottom row).}
    \label{fig:diagplots}
\end{figure}

When considering different values of the NNN interaction $V$, we find that in all three cases ($D = 1,2,3$) the variational method predicts a change from a second order transition to a first order transition as $V$ is decreased. For $V>V_c$ the variational ground state $(\theta_A^{GS},\theta_B^{GS})$ evolves smoothly with $\Delta$, while for $V<V_c$ it changes abruptly at the transition point.  We find that $V_c\simeq -1.0, -0.6,$ and $-0.25$ in 1D, 2D, and 3D, respectively.  In 1D, this significantly overestimates the tricritical point compared with the known analytical value of $V_c \simeq -3.33$~\cite{FendleySachdev2004}, although it is still notable that the effect is captured by the variational method.  For $\Delta$, we find that $\Delta_\text{tricritical}\simeq -0.51, -1.1,$ and $-1.5$ in 1D, 2D, and 3D, respectively.

\begin{figure}
    \centering
    \begin{minipage}{0.45\linewidth}
        \includegraphics[width=\linewidth]{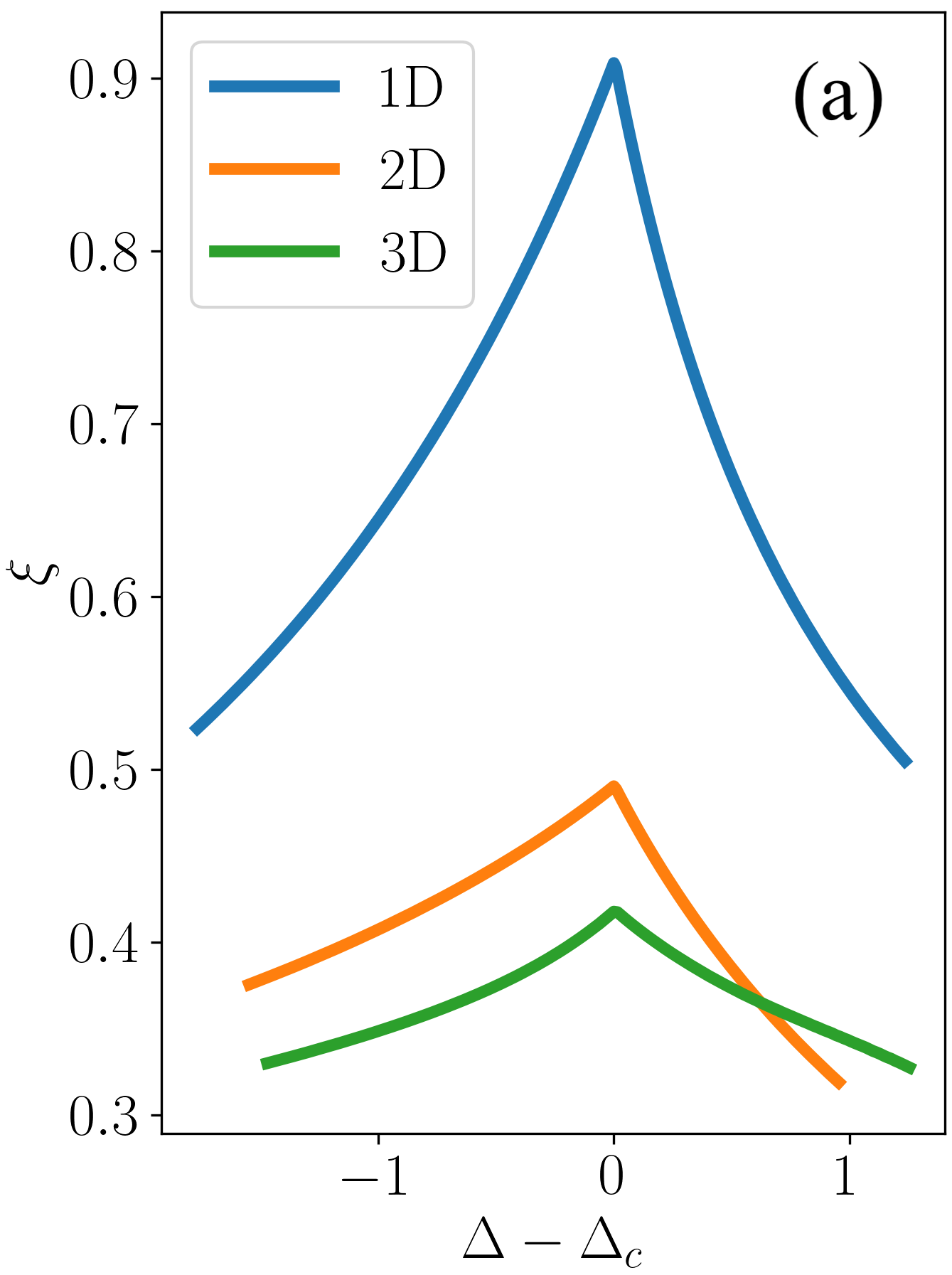}
    \end{minipage}
    \begin{minipage}{0.45\linewidth}
        \includegraphics[width=\linewidth]{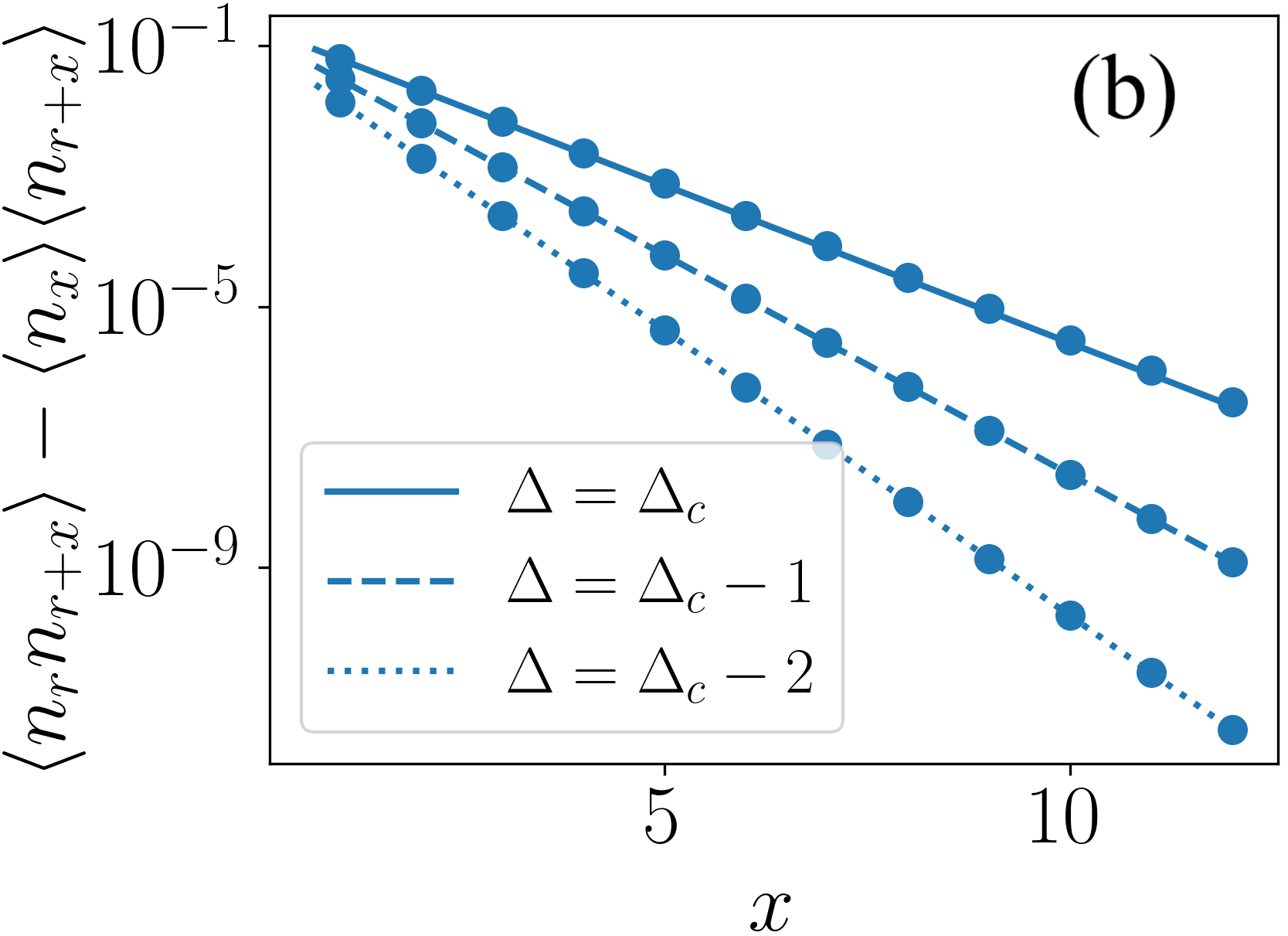}  
        \includegraphics[width=.95\linewidth]{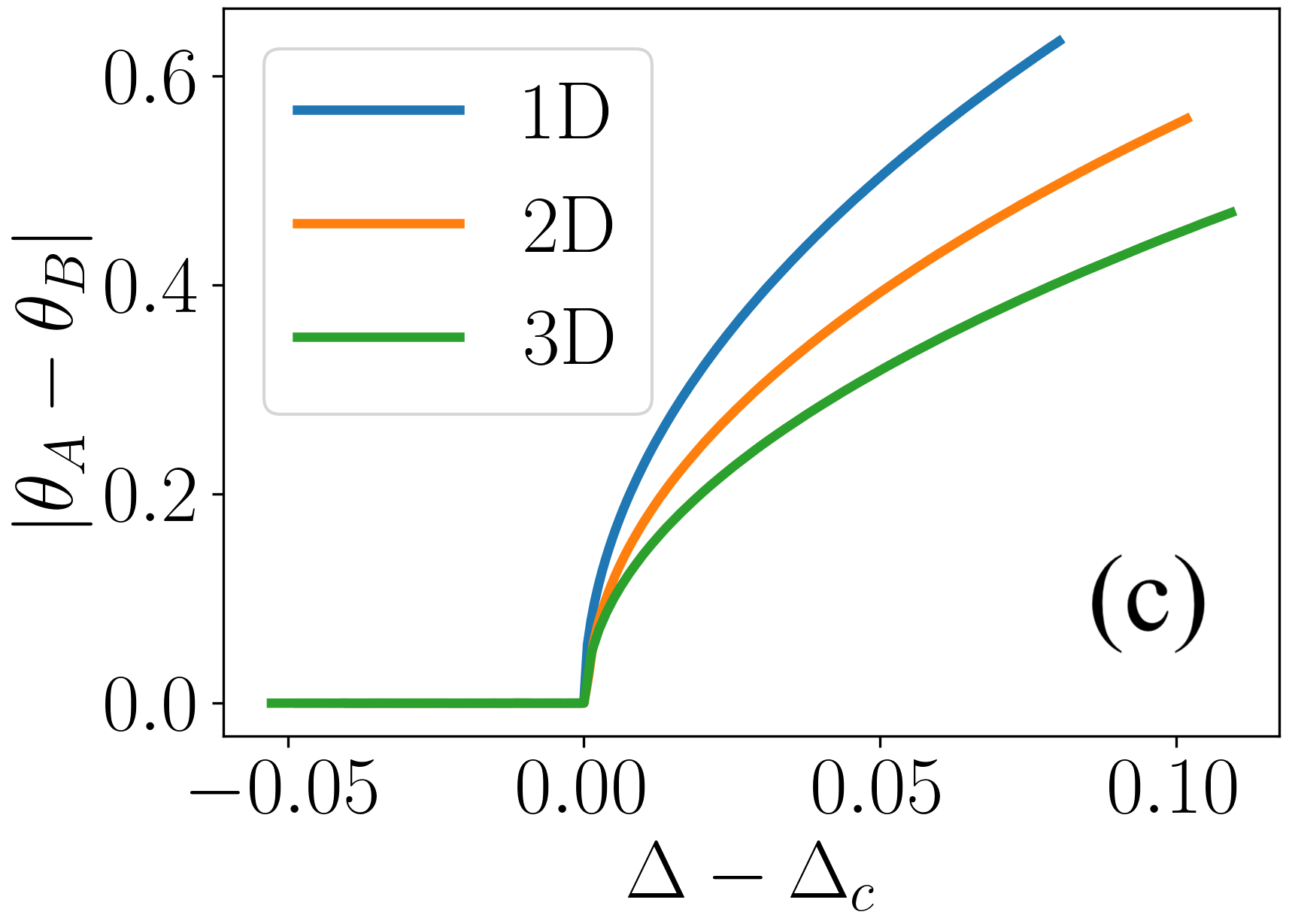}
      
    \end{minipage}
    \caption{(a) Correlation lengths $\xi$ as function of $\Delta-\Delta_c$ of the variational ground state, for $V=0$. $\Delta_c = 0.77, -0.45, -1.0$ for 1D, 2D, and 3D, respectively. (b) Illustration of how $\xi$ is computed: plots of the correlation function at different values of $\Delta$, in 1D.   (c) Plot of $|\theta_A^{GS}-\theta_B^{GS}|$ as a function of $\Delta-\Delta_c$, showing scaling with the critical exponent $\beta=1/2$.}
    \label{fig:xiDelta}
\end{figure}

We present the data here in two series of plots. In Fig.~\ref{fig:phasediagrams} we plot the $(\Delta,V)$ phase diagrams, using the order parameter $\langle\sigma^z_A\rangle_{GS}-\langle\sigma^z_B\rangle_{GS}$. In Fig.~\ref{fig:diagplots} we present the one-dimensional plots of minimum energies along lines of constant $\theta_A-\theta_B$, i.e. $E_\text{minimum}(\zeta=\theta_A - \theta_B)$ where $E_\text{minimum}(\zeta) = \min_{\theta_A}[E(\theta_A,\theta_A-\zeta)]$ , for various values of $\Delta$ at both $V=0$ and $V<V_c$. In these plots one can clearly observe the continuous and discontinuous change of the ground state in the second-order and first-order transition regimes, respectively. Moreover, we extract the critical exponent of the order parameter by fitting the numerical data close to the transition point at $V=0$, where the transition is continuous, and find that it agrees with the Ising mean-field critical exponent $\beta = 1/2$ (see Fig.~\ref{fig:xiDelta}(c)).

\subsection{Correlation length from two-point functions}
The energy optimization presented above only requires the calculation of one-point functions. However, as described in Appendix~\ref{appendix:expansion}, the perturbative expansion can be employed to compute also two-point functions, and can be in principle extended to the calculation of any $n$-point function. Here, we focus on the density-density connected two-point functions $f(|i-j|) = \langle n_i n_j\rangle -\langle n_i \rangle\langle n_j\rangle$, which we compute on the optimal variational ground state for various values of $\Delta$ and $V=0$. We plot a few examples of $f(|i-j|)$ for 1D in Fig.~\ref{fig:xiDelta} (b). By fitting the exponential decay via $f(|i-j|) = A \exp(-|i-j|/\xi)$ we can extract the correlation length $\xi$, which we plot as a function of $\Delta$ (for the case $V=0$) in Fig.~\ref{fig:xiDelta} (a).
As expected from a mean-field ansatz, the correlation length is always finite. Note however, that our ansatz is a tensor network with bond dimension 2; such states can be gapless in $D>1$~\cite{gapless}, but we find that our variational manifold only includes gapped states.

\section{Variational dynamics}\label{sec:dynamics}

We now use our variational manifold combined with the time-dependent variational principle (TDVP)~\cite{TDVP1,TDVP2} to study the time evolution of the PXP model Eq.~\eqref{eq:hamiltonian} with $\Omega=1$,
and show that it captures the periodic revivals characteristic of quantum scarred systems. The 1D results were already obtained in Ref.~\cite{periodicOrbits} and are presented here for comparison with the $D>1$ case. 

\subsection{TDVP equations of motion}
In general, the variational manifold is parameterized by the variables $\Vec{\theta}$ and $\Vec{\phi}$. However, since we are mainly concerned with time evolution from the initial states $\ket{\mathbb{Z}_2}=\ket{\uparrow}^{\otimes A}\ket{\downarrow}^{\otimes B}$ and $\ket{\mathbb{Z}_2^\prime}=\ket{\downarrow}^{\otimes A}\ket{\uparrow}^{\otimes B}$, it suffices to consider unit cell translationally invariant states. Also, since the TDVP is energy conserving~\cite{TDVP1,TDVP2}, it can be shown that $d\Vec{\phi}/dt=0$ when starting from the $\ket{\mathbb{Z}_2}$ states, and we can set all $\Vec{\phi} = 0$ (see Ref.~\cite{periodicOrbits} and Appendix~\ref{sec:Scalculation}). Thus our variational manifold is two dimensional, parameterized by the position vector $\Vec{\theta}=(\theta_A,\theta_B)$. We calculate the dynamics within this variational space by minimizing the leakage rate, or ``quantum leakage", given by $\Gamma^2 = \langle\delta|\delta\rangle$, where $\ket{\delta}$ is the component of the exact time evolution that is orthogonal to the variational space.

The application of the TDVP to this variational manifold (described in Appendix~\ref{appendix:TDVP}), results in  equations of motion $\dot{\Vec{\theta}}=(\dot{\theta}_A,\dot{\theta}_B)$ that, due to symmetry,  may be expressed in the form $\dot{\theta}_A = f(\theta_A,\theta_B)$ and $\dot{\theta}_B = f(\theta_B,\theta_A)$, where $f(\theta_A,\theta_B)$ can be calculated by exact tensor network contraction in $D\leq2$ and via perturbative expansion in any $D$.

Along the axes of the $(\theta_A,\theta_B)$ space, the equations of motion take on the following, particularly simple form (derived in Appendix~\ref{appendix:leakageprodstates})

\begin{subequations}\label{eq:EOMaxes}
\begin{align}
\dot{\Vec{\theta}}(\theta_A,0)&=2 (\hat{\theta}_A+\cos^D\!(\theta_A/2)\hat{\theta}_B)\\\dot{\Vec{\theta}}(0,\theta_B)&=2 (\cos^D\!(\theta_B/2)\hat{\theta}_A+\hat{\theta}_B),
\end{align}
\end{subequations}
where $\hat{\theta}_A$ and $\hat{\theta}_B$ are the unit vectors. In 1D there also exists a closed-form expression for the equations of motion in general; it was shown~\cite{periodicOrbits} that these tensor network calculations can be performed analytically to show that $f(\theta_A,\theta_B) = 2 (\cos(\theta_B/2)+\sin(\theta_A/2)\cos(\theta_A/2)\tan(\theta_B/2))$.

\begin{figure}[t]
    \centering
    \includegraphics[height=3.95cm]{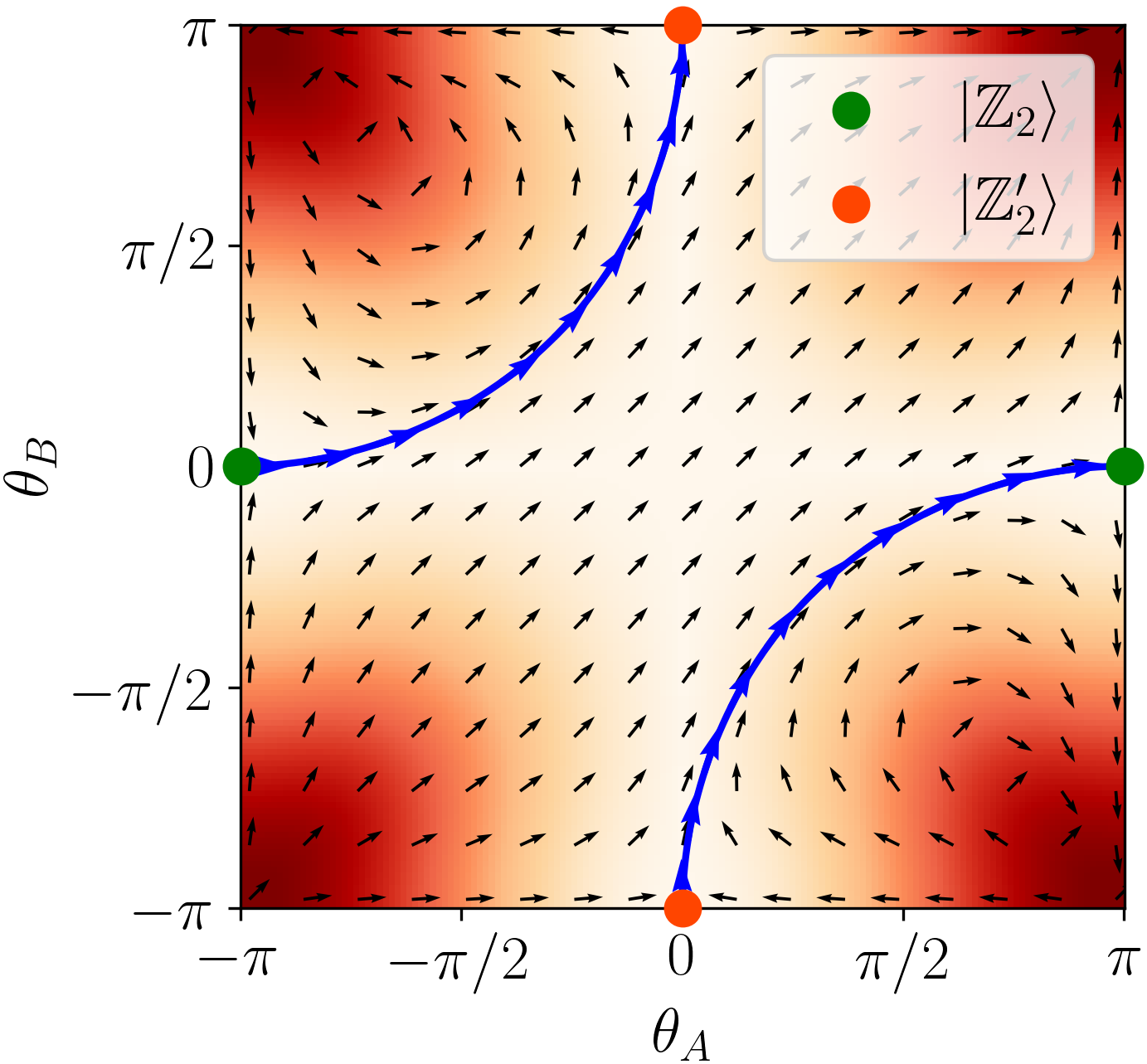}
    \includegraphics[height=3.95cm]{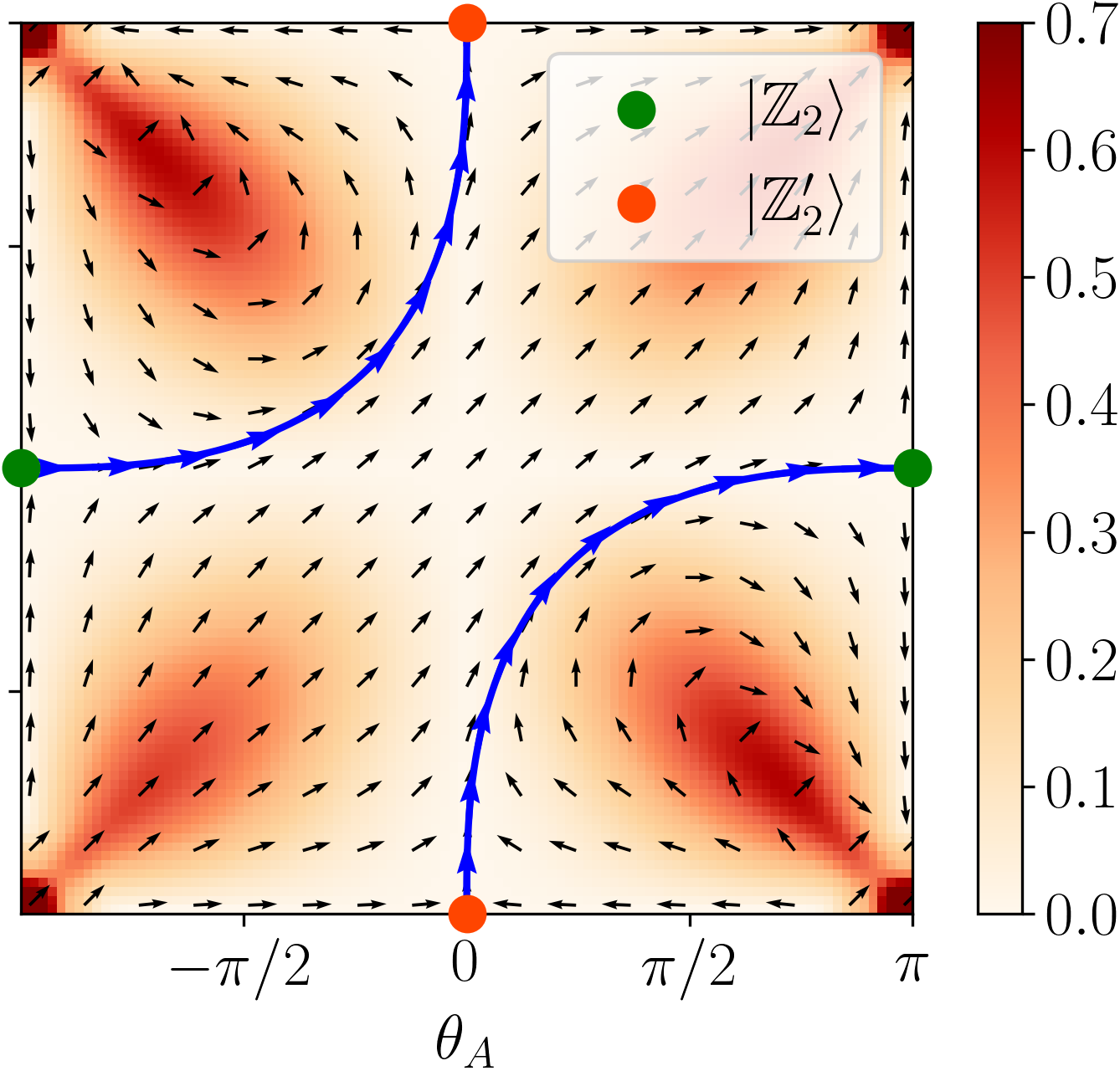}
    \caption{Flow diagrams and paths for 1D (left) and 2D (right). The heatmap is the leakage rate $\gamma = \sqrt{\braket{\delta}/N}$; note that the leakage rate is exactly zero along the $x$ and $y$ axes.}
    \label{fig:flows1d2d}
\end{figure}

\subsection{Periodic trajectories}
\begin{figure}[t]
    \centering
    \includegraphics[height=3.95cm]{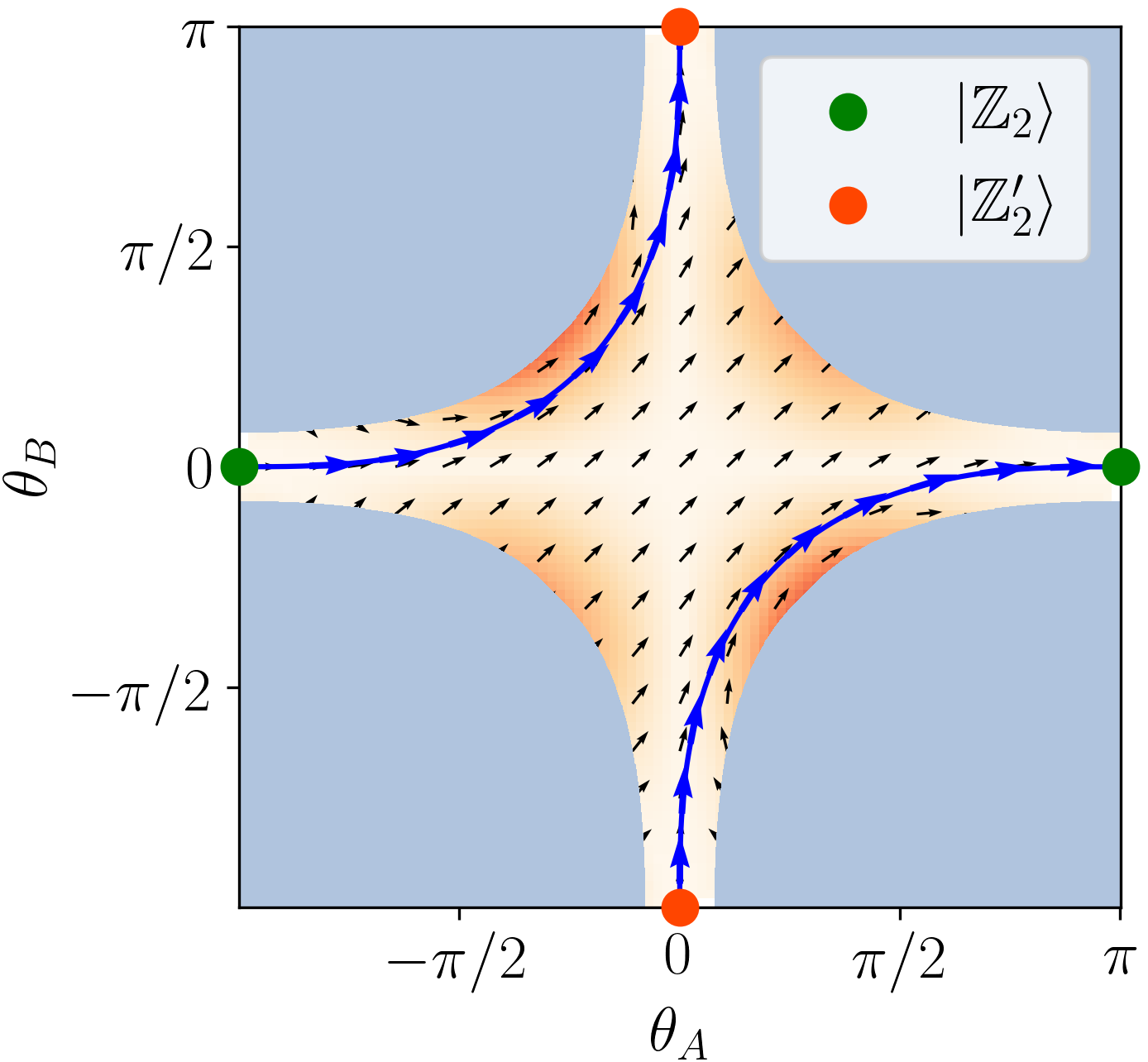}
    \includegraphics[height=3.95cm]{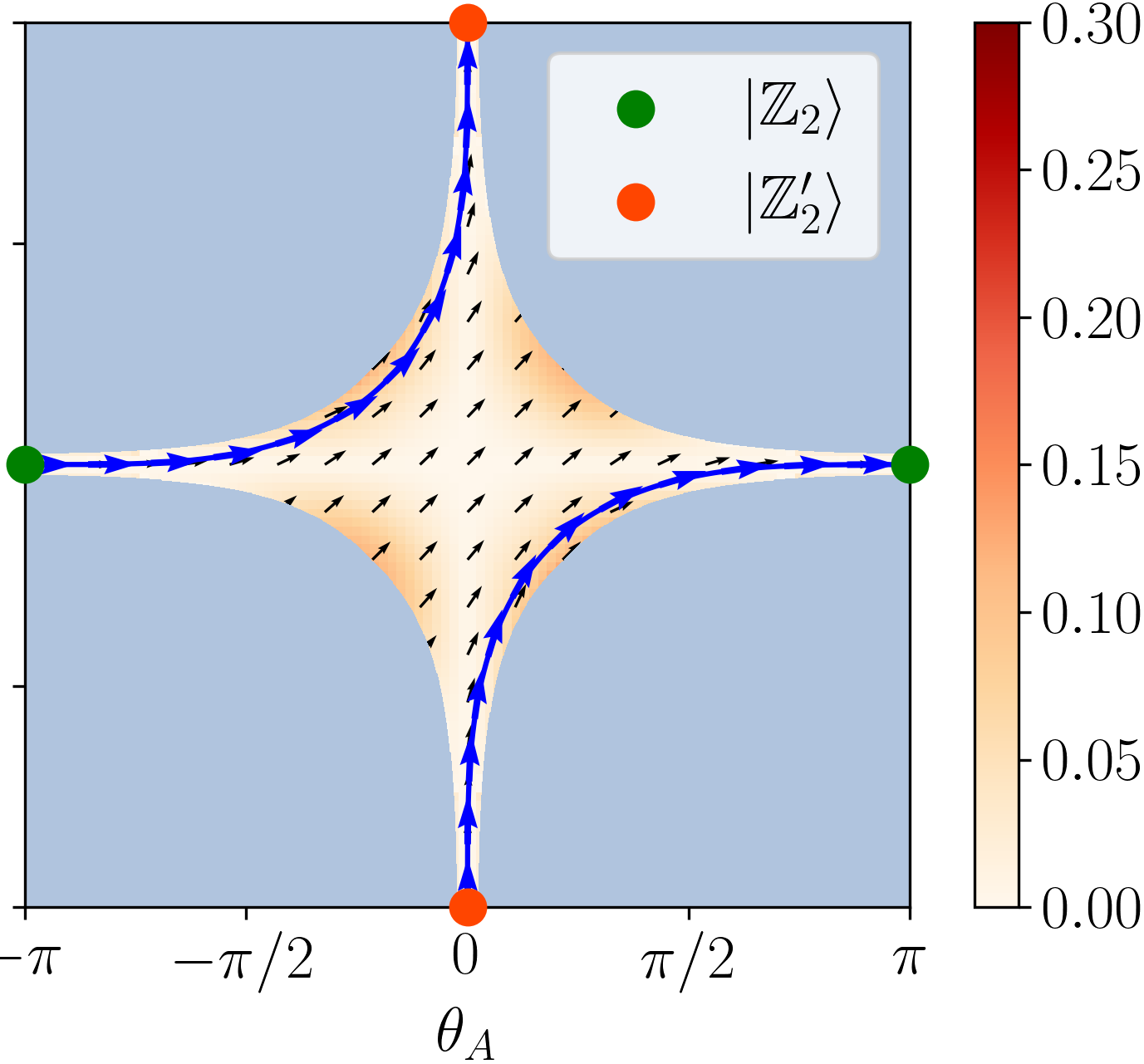}
    \caption{Flow diagrams and paths for 2D (left) and 3D (right), from the perturbative method. Note that the corners have been removed: in these regions the perturbative expansion behavior is uncontrolled at the order $(\mathcal{N} = 12$ (2D) and $(\mathcal{N} = 9$ (3D) at which the calculation is performed. The heatmap is the leakage rate $\gamma = \sqrt{\langle\delta|\delta\rangle/N}$.  }
    \label{fig:flowsPert}
\end{figure}

In the TDVP framework, the interacting spins problem is approximated by a two dimensional nonlinear dynamical system with equations of motion $d\Vec{\theta}/dt = \dot{\Vec{\theta}}(\theta_A,\theta_B)$. These equations can be integrated numerically to obtain trajectories starting from any initial condition, and the vector field $\dot{\Vec{\theta}}(\theta_A,\theta_B)$ can be straightforwardly analyzed to infer the stability of these trajectories. In Figs.~\ref{fig:flows1d2d}-\ref{fig:flowsPert} we plot $\dot{\Vec{\theta}}(\theta_A,\theta_B)$ and the leakage rate in the $(\theta_A,\theta_B)$ plane of the exact equations of motion obtained analytically in Ref.~\cite{periodicOrbits} in 1D, on finite cylinders via exact contraction of the tensor network in 2D (Fig.~\ref{fig:flows1d2d}), and the equation of motions obtained from the perturbative expansion in 2D and 3D (Fig.~\ref{fig:flowsPert}).
We have marked the $\ket{\mathbb{Z}_2}$ and $\ket{\mathbb{Z}_2^\prime}$ states; the all spin down state $\ket{\downarrow}^{\otimes N}$ is at the origin. Note that all points on the $x$ and $y$ axes correspond to product states. For example, points on the $x$ axis are states of the form $\bigotimes_{i\in A}(\cos(\theta_A/2)\ket{\downarrow}_i-i\sin(\theta_A/2)\ket{\uparrow}_i)\otimes(\bigotimes_{j\in B}\ket{\downarrow}_j)$.

In each case (1D, 2D, and 3D), we observe a periodic path connecting the $\ket{\mathbb{Z}_2}$ and $\ket{\mathbb{Z}_2^\prime}$ states; these non-ergodic trajectories correspond to the revival behaviour observed in the quantum spin system and it is easy to see from the vector field plot of $\dot{\Vec{\theta}}(\theta_A,\theta_B)$ that they are unstable orbits alike the one-dimensional case. The path periods, i.e. the time it takes to go $\ket{\mathbb{Z}_2}\rightarrow\ket{\mathbb{Z}_2^\prime}\rightarrow\ket{\mathbb{Z}_2}$, are $T_\mathrm{TDVP} = 4.820, 5.168, 5.345$ for 1D, 2D, 3D, respectively. These have to be compared to the revival times $T_{\mathrm{exact}} = 4.786, 5.154, 5.340$ that we obtained from the exact calculation of the many-body dynamics on periodic systems of up to $36$, $48$, and $64$ sites in 1D, 2D, and 3D, respectively. Despite these numbers being extracted from small systems, they exhibit no noticeable finite-size effects (see Appendix~\ref{appendix:ED} for details) and they demonstrate that the TDVP combined with our variational ansatz provide periods that improve with increasing $D$ ($T_\mathrm{TDVP} - T_\mathrm{Exact} = 0.035,0.014,0.005$ for $D=1,2,3$).

We also plot the leakage rate $\gamma = \sqrt{\braket{\delta}/N},$ normalized to be an intensive quantity. In regions where the leakage rate is small, the variational dynamics closely approximates the true evolution of the quantum system. We observe that the periodic paths lie within regions of low leakage, thus giving validity to the results. We calculated the integrated leakage rates $\oint\gamma dt\simeq0.17$ in 1D, $\oint\gamma dt\simeq0.16$ in 2D, and $\oint\gamma dt\simeq0.13$ in 3D. The leakage rate is exactly zero along the lines $\theta_A=0$ and $\theta_B=0$; we show this analytically (in any dimension) in Appendix~\ref{appendix:leakageprodstates}.

It is nontrivial that the variational dynamics remains accurate along the entire path. In general, a starting point in an area of low leakage may travel to an area of high leakage, given our variational manifold of low-entanglement states. A simple example is the path starting at the origin (i.e. with initial state $\ket{\downarrow}^{\otimes N}$) and ending at $(\pi,\pi)$, which has leakage $\oint\gamma dt\simeq 1.28$ and $0.46$ in 1D and 2D, respectively. 

Observing the trend from one to three dimensions, we conjecture that as one goes to higher dimensional hypercubic lattices, the path period will monotonically increase, asymptotically approaching $2\pi$ in the infinite dimensional limit, with the trajectory becoming more and more square shaped, i.e. more closely following the $x$ and $y$ axes.
We can give an intuitive argument for this. Consider starting in the state $\ket{\mathbb{Z}_2}=\ket{\uparrow}^{\otimes A}\ket{\downarrow}^{\otimes B}$. Initially the $A$ spins will rotate freely while the $B$ spins are frozen. As the dimension increases, so does the number of neighbours of each site, thus increasing the effect of the Rydberg blockade. In the infinite dimensional limit, the $B$ spins will remain nearly frozen until the $A$ spins are nearly pointing down, so we expect the time of half a periodic path to approach $\pi$. 

This is corroborated by the variational equations of motion Eq.~\eqref{eq:EOMaxes}: in the limit $D\rightarrow\infty$, the velocity is $2 \hat{\theta}_A$ along the $x$ axis and $2\hat{\theta}_B$ along the $y$ axis. An alternative picture of the higher-dimensional limit can be given via the related idea of PXP models on complete bipartite graphs~\cite{squeezingQMBS}.

\section{Conclusion}
We have introduced a manifold of low-entanglement states that accurately captures much of the essential physics of the Rydberg arrays in the blockade regime on hypercubic lattices. Our approach provides a general way to study such constrained systems by a variational mean-field method, which is otherwise challenging due to the blockade, that precludes direct application of product state ansatzes.

We studied the equilibrium properties of the generalized PXP model with the variational ansatz in cubic lattices up to three dimensions, and found that it predicts phase boundaries and transitions in good agreement with exact diagonalization and previously known results.  

We applied the TDVP to calculate the approximate time evolution of the PXP model by considering only the component that lies within the variational manifold. We not only obtained important evidence of the non-ergodic behavior that originate from the $\mathbb{Z}_2$ initial state in $D>1$ PXP models, but also demonstrated that TDVP on this manifold of states yields quantitative predictions that improve with increasing lattice dimensionality, as validated by leakage rate and accuracy of the revivals period $T$. Moreover, our results allowed us to infer the infinite-dimensional limit of the PXP non-ergodic dynamics, that consists of a periodic orbit exactly supported on the cartesian axes in the parameter space of our variational manifold, with period $T=2 \pi$.

The are several potential directions for future work. At equilibrium, our ansatz can be generalized to larger blockade radii, to encode lower density symmetry broken phases of Rydberg arrays. Moreover, thanks to its tensor network representation, it can be employed to approximate low-energy excitations of Rydberg atom Hamiltonians~\cite{tn_excitations}. Out of equilibrium, the TDVP can expended to calculate the variational dynamics for constrained systems with a time-dependent Hamiltonian, which is especially relevant in experimental platforms~\cite{drivenRydberg}.

\begin{acknowledgments}
This research was funded in whole or in part by the Austrian Science Fund (FWF) [grant DOI 10.55776/COE1]. For open access purposes, the author has applied a CC BY public copyright license to any author accepted manuscript version arising from this submission. This work is also supported by the European Union’s Horizon Europe research and innovation program under Grant Agreement No. 101113690 (PASQuanS2.1), the ERC Starting grant QARA (Grant No.~101041435), the EU-QUANTERA project TNiSQ (N-6001). G.G. acknowledges support from the European Union’s Horizon Europe program under the Marie Sklodowska Curie Action TOPORYD (Grant No. 101106005).

\end{acknowledgments}

\appendix
\section{Methods}\label{appendix:methods}
The simple structure of the variational ansatz allows one to perform tensor network calculations relatively easily.  Here we provide the details of how these calculations are done.

Specifically, we discuss methods to calculate $\langle\psi|\hat{\mathcal{O}}|\psi\rangle$, where $\ket{\psi} = |\psi(\Vec{\theta},\Vec{\phi})\rangle$ is the variational state and $\hat{\mathcal{O}}$ is a local operator. We will show that it is possible to calculate $\langle\psi|\hat{\mathcal{O}}|\psi\rangle$ by contracting a bond dimension 2 tensor network. This allows one to perform explicit tensor network contractions for up to relatively large quasi-2D cylinders.  We also introduce, for $D\geq 2$, a method to express the tensor network contraction as a perturbative expansion. As a corollary, we will show that the variational state is always normalized, i.e. $\braket{\psi}=1$.   

\subsection{Properties of the variational ansatz}\label{sec:properties}
 For clarity, let us temporarily work in two dimensions. We define the tensor $T$ as a contraction of the PEPS tensor $M$ with its adjoint: 

\begin{align}
T= \;\vcenter{\hbox{\includegraphics[width=1.9cm]{figs/MM2D.pdf}}}\; =\;\; \vcenter{\hbox{\includegraphics[width=1.8cm]{figs/T2D.pdf}}}\;,
\end{align}

where parallel in and out legs have been combined, so the legs of $T$ have dimension $4$. The quantity $\braket{\psi}$ is thus a contraction of an infinite network of $T$s.  

It turns out that the network of $T$s is equal to a network of bond-dimension-$2$ tensors $t$.  That is, for the purposes of calculating $\braket{\psi}$ or $\langle\psi|\hat{\mathcal{O}}|\psi\rangle$, we can perform a ``bond dimension reduction" operation

\begin{align}
\vcenter{\hbox{\includegraphics[width=1.8cm]{figs/T2D.pdf}}}\quad\longrightarrow\quad \vcenter{\hbox{\includegraphics[width=1.6cm]{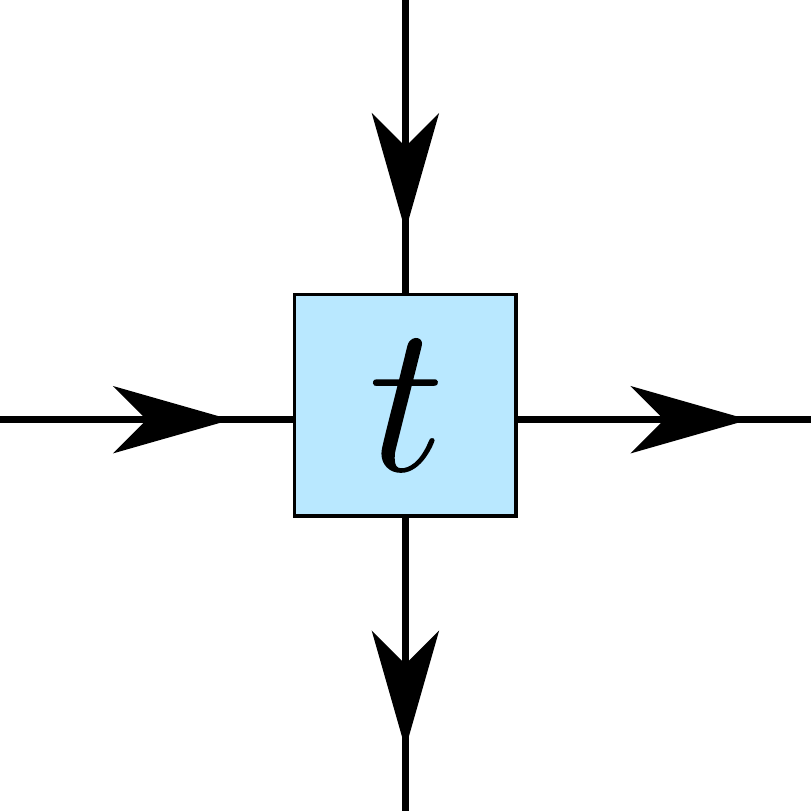}}}\, ,
\end{align}
where $t_{abcd} = T_{(aa)(bb)(cc)(dd)}$ for $a,b,c,d\in\{0,1\}$, and all other elements of $T$ are discarded.
This works because for all the nonzero elements of $T$, the outgoing legs are $(00)$ or $(11)$.  Thus when considering the entire network contraction, we can ignore the indices $(01)$ and $(10)$.

Let us now consider the general ($D$ dimensional) case.  Using the thick-line notation introduced in Eq.~\eqref{eq:MgeneralD}, we write the tensor $T$ as

\begin{align}
  T = \vcenter{\hbox{\includegraphics[width=1.9cm]{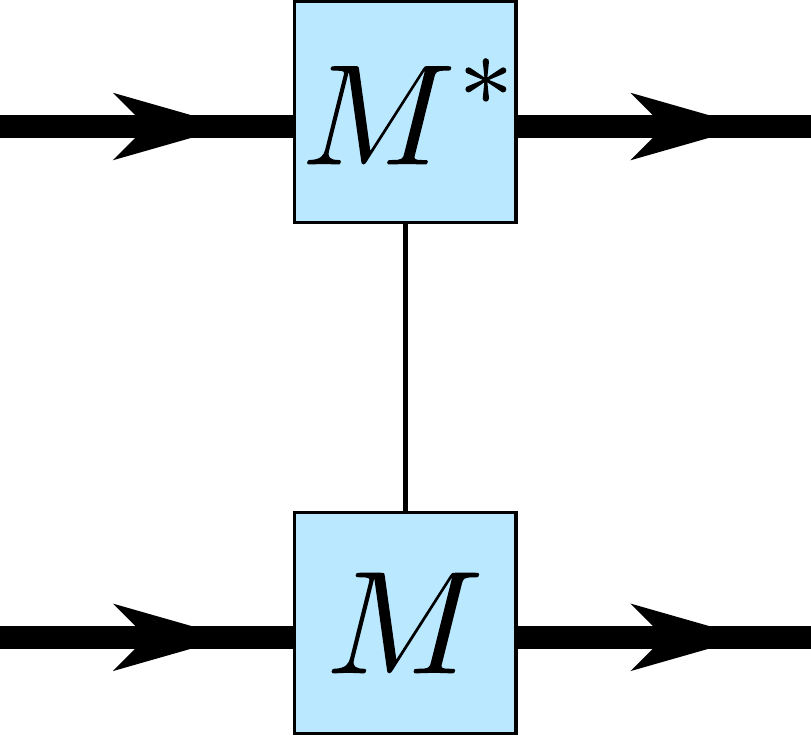}}}\;.
\end{align}

Using the matrix elements of $M$ from Eq.~\eqref{eq:Mtensor}, we can find the matrix elements of $T = \braket{M}$, which is given by

\begin{align}\label{eq:T_tensor}
    T = \cos^2(\theta/2)|\Vec{0}\rangle\langle\Vec{0}|\otimes|\Vec{0}\rangle\langle\Vec{0}|
    &+\sin^2(\theta/2)|\Vec{0}\rangle\langle\Vec{1}|\otimes|\Vec{0}\rangle\langle\Vec{1}|\nonumber\\
    +\cos(\theta/2)\big( |\beta\rangle\langle\Vec{0}|\otimes|\Vec{0}\rangle\langle\Vec{0}| &+ |\Vec{0}\rangle\langle\Vec{0}|\otimes|\beta\rangle\langle\Vec{0}| \big) \nonumber\\&+|\beta\rangle\langle\Vec{0}|\otimes|\beta\rangle\langle\Vec{0}|,
\end{align}
where the $\ket{\alpha}, \ket{\beta}$ notation is the same as originally introduced for Eq.~\eqref{eq:Mtensor}. The exact meaning of Eq.~\eqref{eq:T_tensor} is as follows:  $T$ is a tensor with $D$ incoming legs and $D$ outgoing legs, and each leg has dimension 4. Let $a,b,c,d$ each be a string of ones and zeros of length $D$.   When the term $x|\Vec{a}\rangle\langle\Vec{b}|\otimes|\Vec{c}\rangle\langle\Vec{d}|$ shows up in the right hand side of Eq. \eqref{eq:T_tensor}, it means that the element of $T$ with incoming indices set to $(a_1c_1),(a_2c_2),\dots,(a_Dc_D)$ and outgoing indices set to $(b_1d_1),(b_2d_2),\dots,(b_Dd_D)$ has the value $x$. 

Like in the 2D case, we can perform a bond dimension reduction, and then $\braket{\psi}$ is equal to the contraction of a network of bond-dimension-2 tensors $t(\theta)$, which we express diagrammatically as

\begin{align}
  t_{(\text{in})(\text{out})}= \vcenter{\hbox{\includegraphics[width=2.8cm]{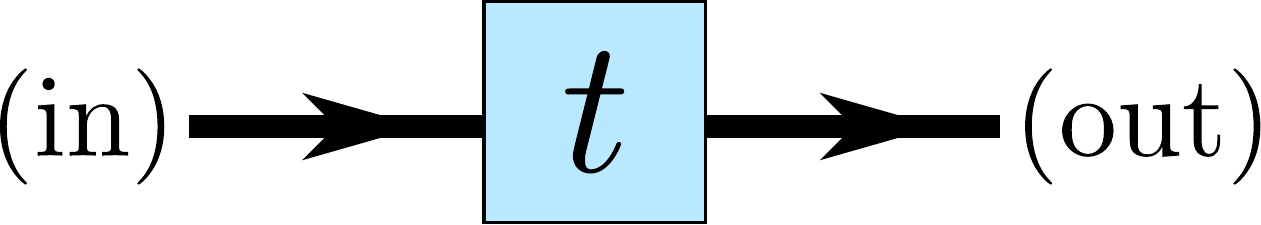}}}
\end{align}
and are given by

\begin{align}
    t=\cos^2(\theta/2)|\Vec{0}\rangle\langle\Vec{0}|
    +\sin^2(\theta/2)|\Vec{0}\rangle\langle\Vec{1}|+\ket{\beta}\langle\Vec{0}|
\end{align}

We can write $t$ as $t(\theta) = p - \sin^2(\theta/2)q$, where the constant tensors $p$ and $q$ are

\begin{subequations}
\label{eq:nq}
\begin{align}
     p = \ket{\alpha}\langle\Vec{0}|\\
    q = |\Vec{0}\rangle\langle\Vec{0}| - |\Vec{0}\rangle\langle\Vec{1}|
\end{align}
\end{subequations}
The decomposition $t = p - \sin^2(\theta/2)q$ turns out to be very useful in performing calculations. Observe that in the network of $t$s, following a directed line never results in a closed loop. (In the 2D square lattice, for example, this follows trivially if we define all the arrows to point rightwards or downwards.)

A simple but important result (which follows directly from Eq.~\eqref{eq:nq}) is the fact when $q$ is contracted with a $p$ on each of its outgoing legs, the result is a tensor with all elements equal to zero.  That is,

\begin{align}\label{eq:qnzero}
    \vcenter{\hbox{\includegraphics[width=3.3cm]{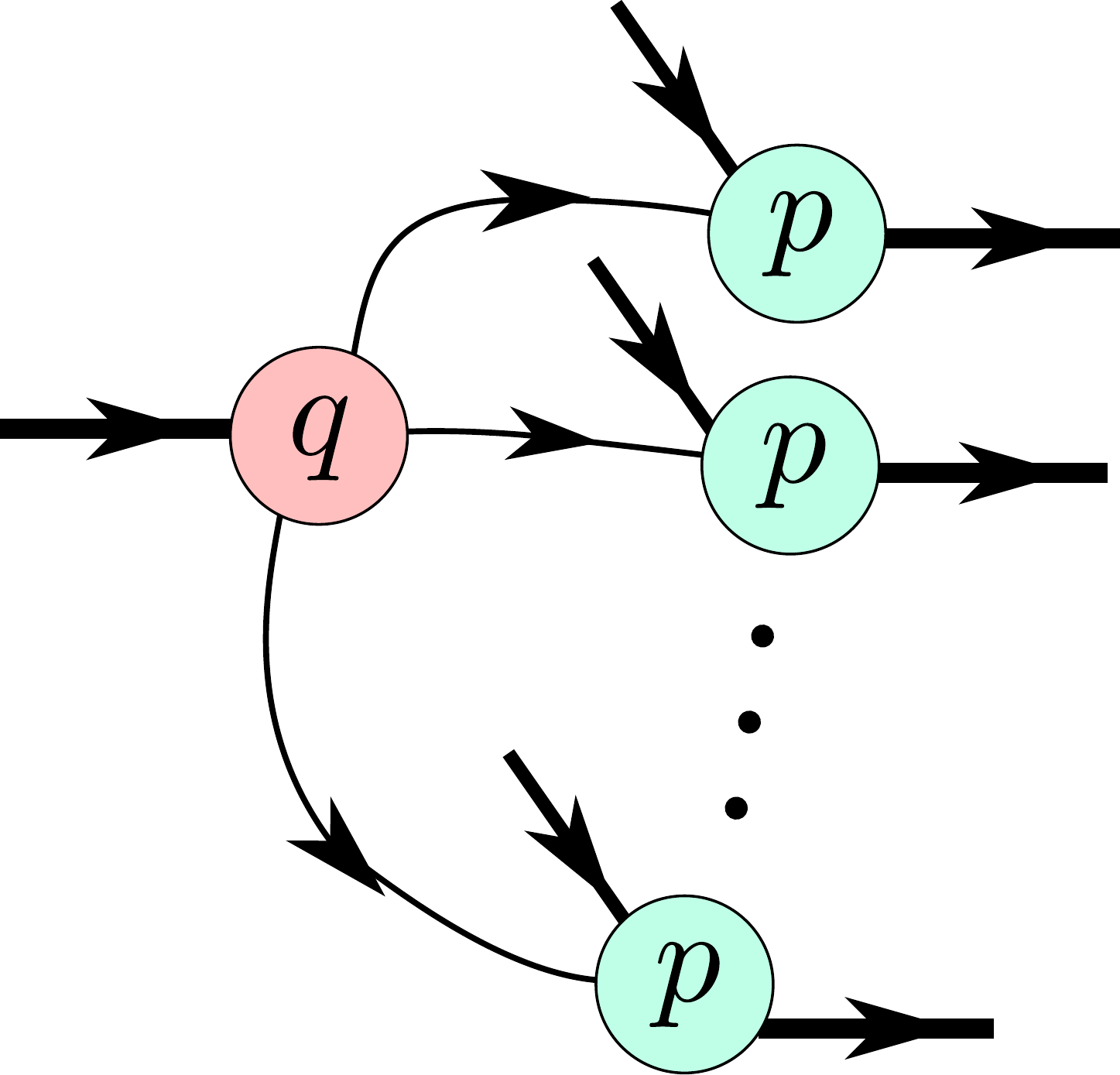}}} = 0.
\end{align}

Thus, each $q$ must be attached via an outgoing leg to at least one other $q$; otherwise the contraction of the network will be zero. Note that Eq.~\eqref{eq:pqZero2D} is the 2D version of Eq.~\eqref{eq:qnzero} here.

\subsubsection{Normalization}\label{sec:norm}
An immediate corollary is that the variational state is normalized in the thermodynamic limit, i.e. that $\braket{\psi} = 1$.  There is one additional condition that we must impose: we must assume that at most a finite number of the $\theta_i$s in $\Vec{\theta}$ are integer multiples of $\pi$.  That is, we assume that $\sin(\theta_i/2)\neq\pm 1$ for all but a finite number of  $\theta_i$s. 

In the expansion of $\braket{\psi}$ using $t = p - \sin^2(\theta/2)q$, there are no terms with a finite number of $q$s due to Eq.~\eqref{eq:qnzero}: there are either no $q$s, or at least an infinite number of them, since each $q$ must attach to at least one other $q$, ad infinitum.   But the tensor networks with an infinite number of $q$s give zero contribution, because each $q$ comes with a factor of $\sin^2(\theta/2)$.  Thus  $\braket{\psi} = (\text{network of all }p\text{s}) = 1$.

\subsubsection{Calculation by explicit contraction}
In $D\leq2$, the most straightforward way to calculate $\langle\psi|\hat{\mathcal{O}}|\psi\rangle$ is to explicitly contract the tensor network.  In 1D this can be done for arbitrary system sizes, and in 2D, we can define the system on a cylinder.  Due to the reduced effective bond dimension, we only need to work with a $2^L\times 2^L$ transfer matrix (rather than $4^L\times 4^L$), where $L$ is the circumference of the cylinder; in other words, the computational cost is equivalent to performing exact diagonalization on a 1D spin chain of length $L$.  

Due to the limited ability of the ansatz to encode long range correlations, the results do not depend strongly on the size of the cylinder.  We performed numerical calculations on $L=10$ cylinders, but we verified that increasing the circumference to $L=12$ or $L=14$ leaves all results unchanged well beyond the relevant accuracy of a few significant figures.  

\subsection{Perturbative expansion}\label{appendix:expansion}
Using the results of the previous section, we show how one can express $\bra{\psi}\hat{\mathcal{O}}\ket{\psi}$ as an infinite series. The exposition here is quite formal: for a practical introduction, the reader may skip to Sec.~\ref{sec:PertExp_example} below, which is a natural continuation of the high-level overview in Sec.~\ref{sec:PertExp_main}.

 In the interest of presenting the ideas here compactly, we write things down as algebraic equations rather than diagrams, but at the expense of using somewhat schematic notation.  Here, $\Tr(\cdots)$ means a contraction over the infinite network, and when tensors inside a $\Tr(\cdots)$ are ``multiplied", they are being placed on lattice sites and then contracted.  In the one dimensional case, the tensors are matrices, $\Tr(\cdots)$ is indeed a trace, and the equations here may be taken literally.  

Recall that $\braket{\psi}$ is the contraction of a network of $t$ tensors. Let $\Theta$ be the tensor that results when $\hat{\mathcal{O}}$ is sandwiched with the PEPS tensors: then $\bra{\psi}\hat{\mathcal{O}}\ket{\psi}$ is the contraction of a network with $\Theta$ on a local region and $t$ on all other sites. That is, 

\begin{align}
    \Theta = \Big(\prod_{i\in I}M(\theta_i)^*\Big)\hat{\mathcal{O}}\Big(\prod_{i\in I}M(\theta_i)\Big),\\
    \bra{\psi}\hat{\mathcal{O}}\ket{\psi} = \Tr\Big(\Theta \prod_{i\notin I}t(\theta_i)\Big),
\end{align}
where $I$ is the region (i.e. set of sites) that $\Theta$ covers.  Note that due to the mechanics of the bond dimension reduction operation, $I$ may be slightly larger than the region that $\hat{\mathcal{O}}$ itself acts on.  

Using $t(\theta_i) = p - \sin^2(\theta_i/2)q$, we can expand:

\begin{gather}
     \bra{\psi}\hat{\mathcal{O}}\ket{\psi}  =  \Tr\Big(\Theta_I \prod\limits_{i\notin I} (p_i - \sin^2(\theta_i/2)q_i)\Big)\\ 
     = \sum\limits_J\Tr\Big[\prod\limits_{j\in J}(-\sin^2(\theta_j/2) q_j)\Big(\prod\limits_{i\notin J\cup I} p_i \Big)\Theta_I\Big] \\  
     = \sum\limits_J\Bigg( (-1)^{|J|} \Big(\prod\limits_{j\in J}\sin^2(\theta_j/2)\Big)\Tr\Big[\Theta_I\prod\limits_{\substack{j\in J\\i\notin J\cup I}}q_j p_i \Big]\Bigg).\label{eq:expansiongeneral}
\end{gather}

We have expressed $\bra{\psi}\hat{\mathcal{O}}\ket{\psi}$ as a sum over regions $J$. Note that $p$ and $q$ are constant tensors: the subscript denotes which site they are on.  Likewise we wrote $\Theta_I$ to emphasize that it lies on the region $I$.  $|J|$ is the number of sites in $J$.

The key point now is that $\Tr(\Theta\prod q_i p_j)$ will be zero, due to Eq.~\eqref{eq:qnzero}, unless the region $J$ is ``anchored" by $\Theta$.  When $\Tr(\Theta\prod q_i p_j)$ is nonzero, it can usually be computed quickly by reading off the tensor elements of $\Theta$.  

The perturbative expansion is thus a sum over allowed regions $J$.  This is the general idea; the numerical implementation is described below.

\subsection{Numerical implementation}\label{sec:perturbativeexp}
Now let us discuss the practical implementation.  Variants of the calculation we discuss here are used for both the ground state and the TDVP calculations.

Let us now restrict ourselves to the unit cell translation invariant case, where the variational ansatz becomes $\ket{\psi(\theta_A,\theta_B,\phi_A,\phi_B)}$. Using Eq.~\eqref{eq:expansiongeneral}, we can express $\bra{\psi}\hat{\mathcal{O}}\ket{\psi}$ as 

\begin{align}
    \bra{\psi}\hat{\mathcal{O}}\ket{\psi} = h(\theta_A,\theta_B,\phi_A,\phi_B)\Sigma(f,\theta_A,\theta_B),
\end{align}
where the prefactor $h(\theta_A,\theta_B,\phi_A,\phi_B)$ is some function that comes from the aforementioned $\Tr(\Theta\prod q_i p_j)$, and is usually not hard to evaluate, and where  $\Sigma(f,\theta_A,\theta_B)$ is the perturbative expansion: 

\begin{align}\label{eq:SigmaSum}
    \Sigma(f,\theta_A,\theta_B) =\sum_{n,m=0}^\infty\! (-1)^{n+m}f_{n,m}\sin^{2n}\!\!\Big(\frac{\theta_A}{2}\Big)\sin^{2m}\!\!\Big(\frac{\theta_B}{2}\Big),
\end{align}
where $f$ is a matrix of ``counting factors" that depends on $\hat{\mathcal{O}}$ and is obtained by counting the ways that the $q_i$s may be arranged in $\Tr(\Theta\prod q_i p_j)$.  To see how exactly Eq.\eqref{eq:SigmaSum} arises, it is instructive to work out the calculation for a specific example.

\subsubsection{Example: $\bra{\psi}n_i\ket{\psi}$}\label{sec:PertExp_example}
Consider the case where $\hat{\mathcal{O}}$ is the number operator $n_i$, for some site $i\in A$.  Note that since the state $\ket{\psi}$ lies inside the constrained subspace, this is equivalent to calculating $\bra{\psi}\mathcal{P}n_i\mathcal{P}\ket{\psi}$. 
In this case, the tensor $\Theta$ occupies only a single site, and the prefactor $h$ is simply $h=h(\theta_A)=\sin^2(\theta_A/2)$.

Let us assume that we have a 2D square lattice with all arrows pointing rightwards or downwards, though the general idea is the same for other lattices.  (In particular, Eq.~\eqref{eq:SigmaSum}, as written, applies for any lattice, as long as a suitable version of $f_{n,m}$ is used.)

 We want to calculate the contraction of a tensor network with $\Theta$ on one site, and $t$ on all other sites.  We use the fact that $t = p - \sin^2(\theta/2)q$, and then expand in powers of $\sin^2(\theta/2)$.  That is, we want to express the result of the contraction of the tensor network as sum = ($\Theta$ with $p$ on all other sites) + (configurations with one $q$) + (configurations with two $q$s) + \dots.  

The allowed configurations are constrained by where the $q$s can be placed: each $q$ must be above or to the left of another $q$, or $\Theta$.  Also, note that each $q$ comes with a factor of $-\sin^2(\theta/2)$.  We can write the expansion as in Eq.~\eqref{eq:SigmaSum}, where, in each term in the sum, $n$ is the number of $q$s on sublattice $B$ and $m$ is the number of $q$s on sublattice $A$, and where $f_{n,m}$ is a coefficient that can be obtained by counting.

Specifically, $f_{n,m}$ is defined as: the number of possible configurations on a bipartite lattice (with the origin on sublattice $A$) with: $n$ $q$-tensors on $B$, $m$ $q$-tensors on $A$, a single $\Theta$ tensor at the origin, and a $p$-tensor on all the other sites; with the constraint that each $q$ must have another $q$ (or the $\Theta$) immediately ``downstream" of it (i.e. 
 to the right of and/or below). 

The $f$ matrix elements are obtained via brute force counting using a computer. For a 2D square lattice, the first few values of $f_{n,m}$ are (where the top left element of the matrix is $f_{0,0}$):
\begin{align}\label{eq:f2d}
  f^{(2D)}=  \begin{pmatrix}
        1 & 0 & 0 & 0 & 0 &0&0&\dots\;\\
         2 & 4 & 2 & 0 & 0 &0&0\\
         1 & 11 & 25 & 21 & 6 &0&0 \\
          0 & 10 & 72 & 174 & 192&100&20\\
          0&3&87&510&1281&1680&\\
          0&0&48&732&3780&&\\
          0&0&10&560&&&&\\
          &&\vdots&&&&&\ddots
    \end{pmatrix}.
\end{align}

The perturbative expansion only is valid in the region where $\theta_A$ or $\theta_B$ or both are small enough that $f_{n,m}(\sin^2(\theta_A/2))^n(\sin^2(\theta_B/2))^m$ decreases with larger values of $n$ and $m$. 

\subsubsection{Two point functions}
The method described here also allows for the calculation of two point functions, which in practice corresponds to calculating $\bra{\psi}\hat{\mathcal{O}}\ket{\psi}$ where $\hat{\mathcal{O}}$ is a non-local operator consisting of two parts.  The simplest example would be $\bra{\psi}n_i n_j\ket{\psi}$, which may be calculated by a generalization of the procedure described above for $\bra{\psi}n_i\ket{\psi}$. 

\subsection{Asymptotic nature of the expansion}\label{sec:expansion}
Here we continue the discussion on the perturbative and non-perturbative regimes from Sec.~\ref{sec:asymptoticseries}.   Since the exact form of the series expansion depends on the operator being calculated, it is difficult to discuss this in a completely general way. We use the simplest case where the operator is $\hat{\mathcal{O}}=n_i$; that is, we use $f_{n,m}=f^{(2D)}_{n,m}$ from Eq.~\eqref{eq:f2d}. However, the results are valid for any local operator $\hat{\mathcal{O}}$ as the scaling of $f_{n,m}$ with $n$ and $m$ would be similar.

\begin{figure}[t]
    \centering
    \includegraphics[height=3.8cm]{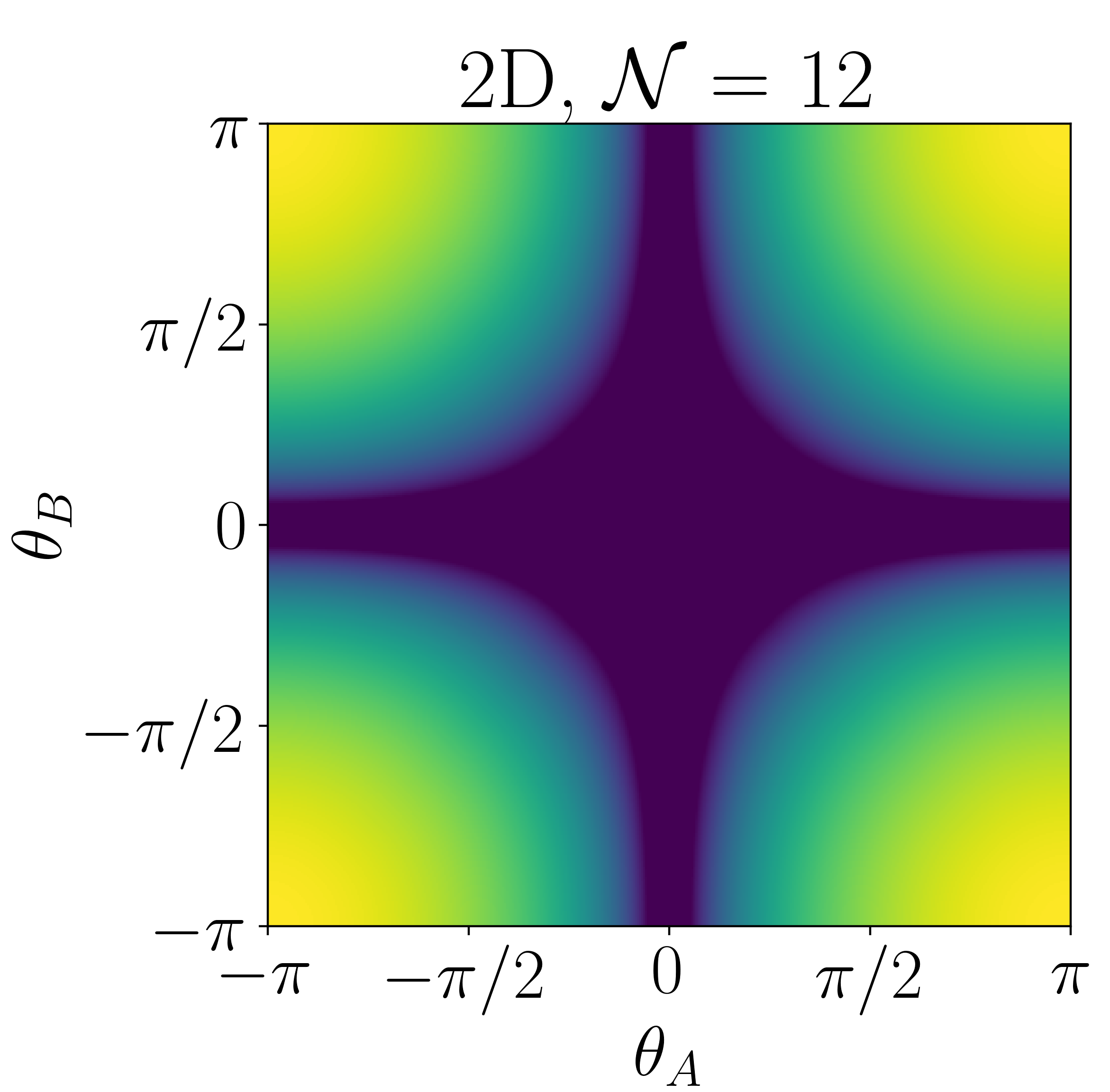}
    \includegraphics[height=3.8cm]{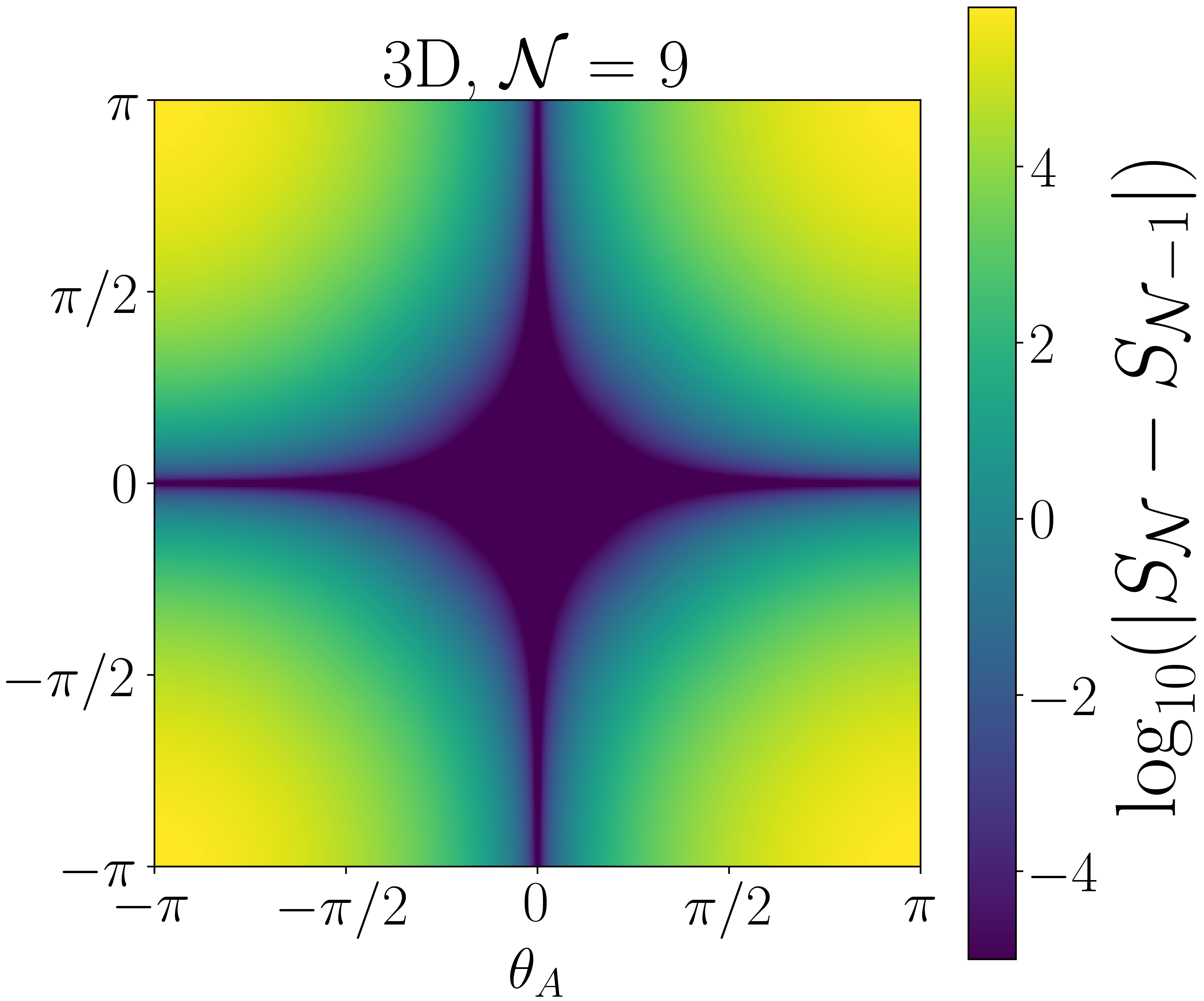}
    \caption{Plot of $|S_\mathcal{N} - S_{\mathcal{N}-1}|$ in log scale, across the parameter space, for $\mathcal{N}=12$ in 2D (left) and $\mathcal{N}=9$ in 3D (right).  Note that the colormap scale has been truncated in the sense that numbers $<10^{-5}$ have been neglected.}
    \label{fig:seriesExpansion}
\end{figure}

We now analyze the quantitative behaviour of the asymptotic series, Eq.~\eqref{eq:SigmaSum}.  We define the order of a truncated expansion to be $\mathcal{N}\equiv\max(n+m)$.  Let $s_{n,m}$ refer to a term in the sum, and let $S_\mathcal{N}\equiv \sum_{n+m\leq \mathcal{N}} s_{n,m}$ be the partial sum up to order $\mathcal{N}$.  To probe the accuracy of the expansion at a certain order, we consider the quantity $|S_\mathcal{N} - S_{\mathcal{N}-1}|$. 
 In Fig.~\ref{fig:seriesExpansion} we show $|S_\mathcal{N} - S_{\mathcal{N}-1}|$ for the highest orders of the expansion as used in the calculations in this paper, $\mathcal{N}=12$ in 2D and $\mathcal{N}=9$ in 3D.  The darker region of this plot therefore corresponds to the perturbative regime. In this work we define it as the region where $|S_\mathcal{N} - S_{\mathcal{N}-1}| < 10^{-3}$ at order $\mathcal{N}=12$ (2D) and $9$ (3D).

The boundaries of the perturbative regimes in Fig.~\ref{fig:blankplot} in the main text are the points $|S_\mathcal{N} - S_{\mathcal{N}-1}|=10^{-3}$ from the same data as in Fig.~\ref{fig:seriesExpansion}.

\subsubsection{Proof of non-convergence}
Here we prove that the series expansion Eq.~\eqref{eq:SigmaSum} divertges. In Section~\ref{sec:asymptoticseries} it was claimed that the counting factors $f_{n,m}$ scale super exponentially, i.e. that there do not exist any $a,b\in\mathbb{R}$ such that $f_{n,m}<a^n b^m$ for all $n,m\in \mathbb{N}$. This would imply that the sum \eqref{eq:SigmaSum} diverges everywhere. Here, we prove that for any $N\in\mathbb{N}$ there exists $n,m>N$ such that $f_{n,m}>\sqrt{(n+m-1)!/(n+m)}$, which is a strong enough condition to imply super-exponential scaling.  We present the proof for 2D, but the generalization to 3D is straightforward.

\begin{figure}
    \centering
    \includegraphics[width=6cm]{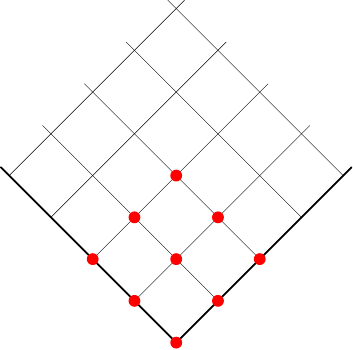}
    \caption{A visualization of the counting problem: one counts the number of ways of arranging $k$ red dots on this grid, subject only to the constraint that for each dot, at least one of the two sites immediately below it must be occupied (for sites on the edge, the condition is that the one site below it must be occupied).   Shown here is one allowed configuration for $k=9$.}
    \label{fig:grid}
\end{figure}

Instead of $f_{n,m}$, it will be easier to work with $f_k$, where $k=n+m$.  That is, define $f_k\equiv\sum_{n+m=k}f_{n,m}$, so we are counting all allowed configurations with $k$ sites occupied.  Let us state the definition of $f_k$ with the aid of a picture.  $f_k$ is the number of ways of arranging $k$ dots on the vertices of the grid shown in Fig.~\ref{fig:grid}, where the first dot is placed at the bottom point, and then each additional dot is placed with the condition that at least \textit{one} of the two sites below it must be occupied.  Now consider how $f_{k+1}$ is related to $f_k$.  The number of ways to place the $(k+1)$th dot depends on the configuration of the $k$ dots already placed: in particular, it is equal to the ``perimeter" of the region occupied by the $k$ dots.  This is clearly smallest when the $k$ dots are arranged compactly, such as in the $k=9$ example shown in Fig.~\ref{fig:grid}, and in this case the perimeter is $2\sqrt{k}$. Thus $f_{k+1}>\sqrt{k}f_k$, and by induction, $f_k>\sqrt{k!}$.

Now let us recast the result in terms of $n$ and $m$.  Here let $\Bar{f}_{n,m}$ be the \textit{largest} $f_{n,m}$ such that $n+m=k$. Clearly $k\Bar{f}_{n,m}>f_k$. Then $f_k>\sqrt{k!}$ implies that $(n+m)\Bar{f}_{n,m}>\sqrt{(n+m)!}$, which is the claimed result.

\section{Energy calculation}\label{appendix:GSdetails}
Here we elaborate on the specifics of the tensor network calculations for the variational energy $\bra{\psi}H\ket{\psi}$.  Let us consider the parts of the Hamiltonian~\eqref{eq:hamiltonianNNN},  $H = H_\text{PXP} + H_\Delta + H_\text{NNN}$ one by one.

\begin{figure}
    \centering
    \includegraphics[width=4cm]{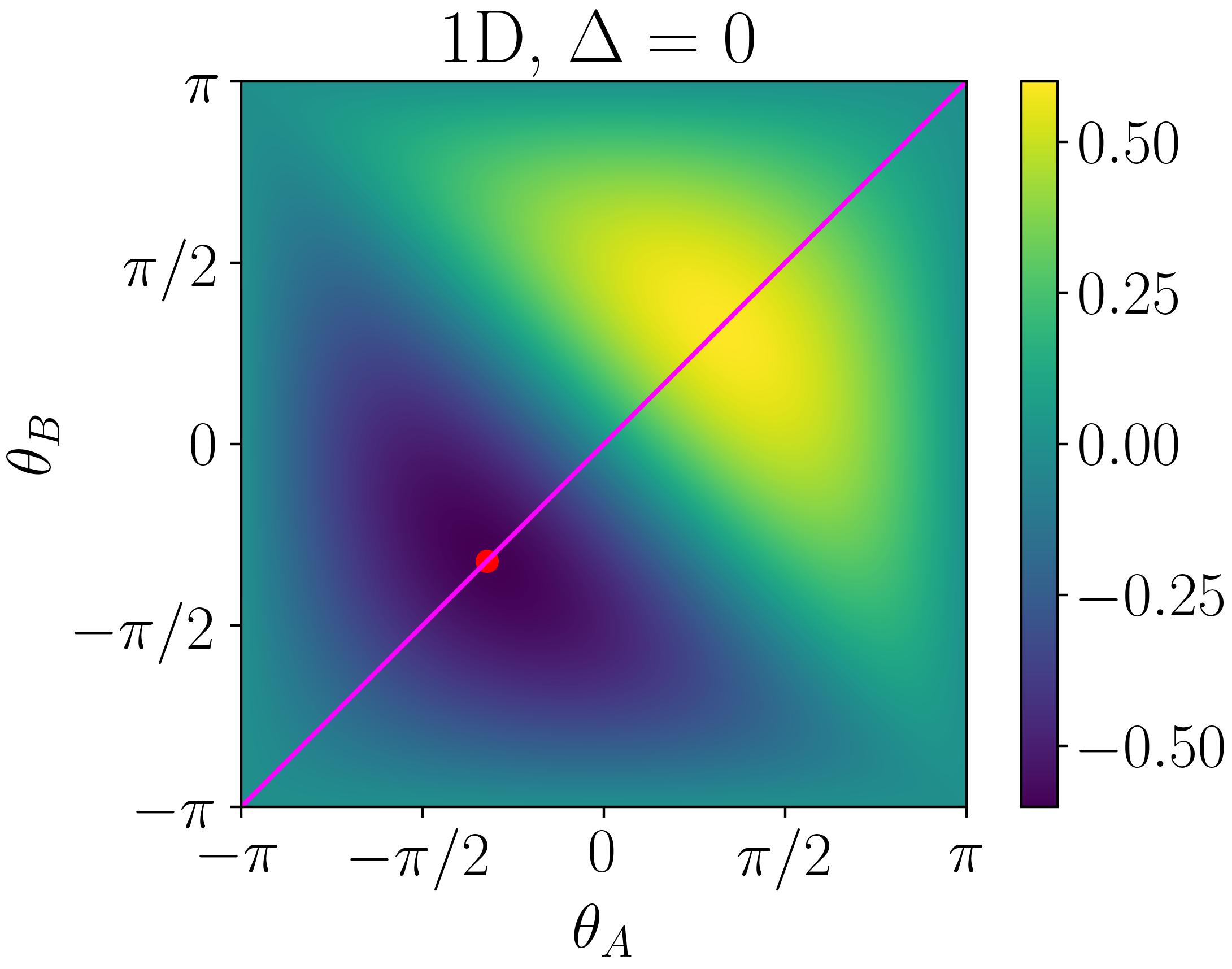}
    \includegraphics[width=4cm]{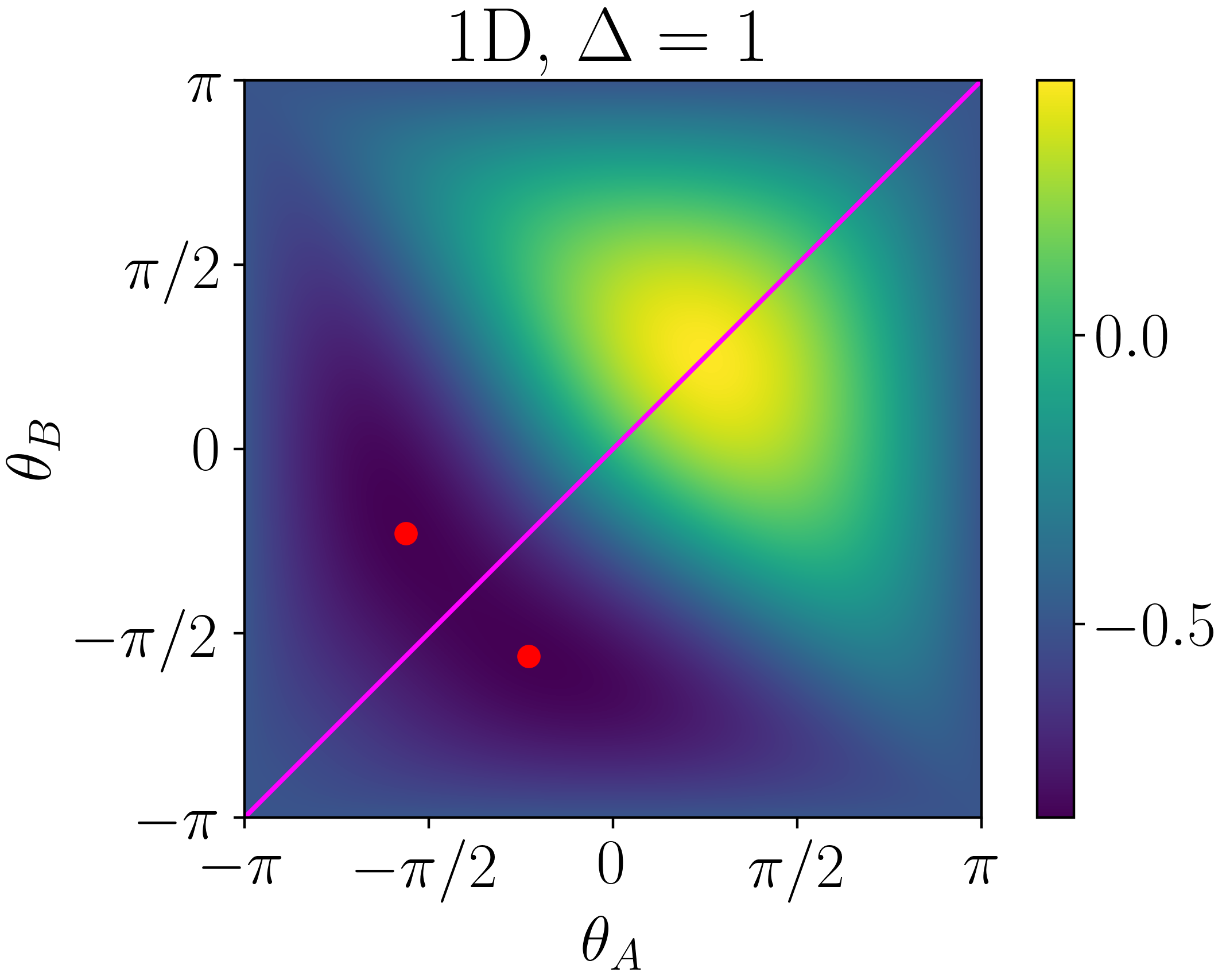}
    
    \vspace{.2cm}
    
    \includegraphics[width=4cm]{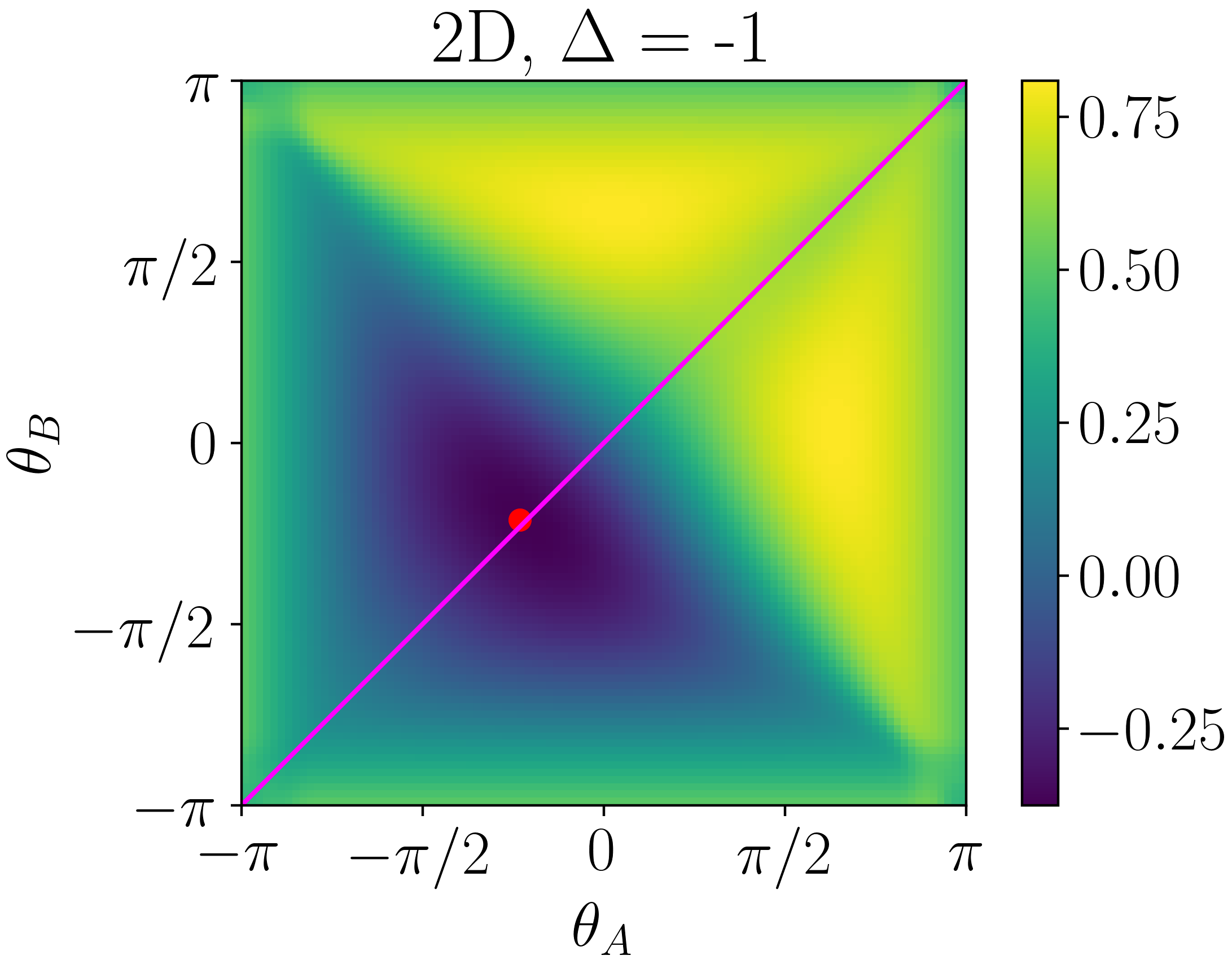}
    \includegraphics[width=4cm]{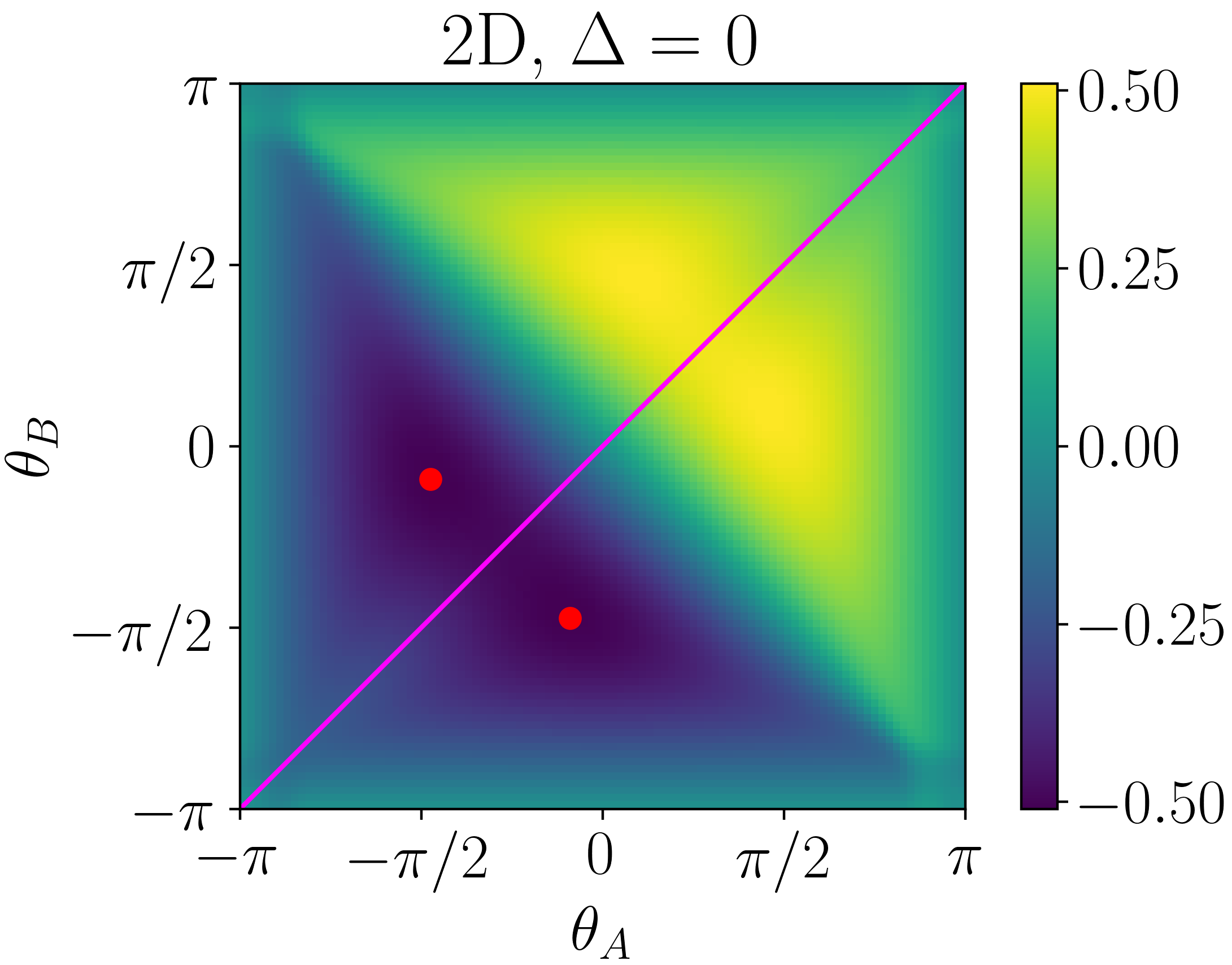}
    \caption{Plots of the energy $E(\theta_A,\theta_B)$.  Top row: 1D, at $\Delta=0$ (left) and $\Delta=1$ (right). Bottom row: 2D, at $\Delta=-1$ (left) and $\Delta=0$ (right). Thus the plots on the left are for $\Delta<\Delta_c$ and on the right for $\Delta>\Delta_c$.  In each plot, the red dot(s) indicate the minimum of $E(\theta_A,\theta_B)$.  The line $\theta_A=\theta_B$ is included as a visual aid. The red dot is at the origin at $\Delta=-\infty$.  As $\Delta$ is increased, it travels down the line, until it splits into two at $\Delta=\Delta_c$ when the variational ground state becomes degenerate.}
    \label{fig:energyplots}
\end{figure}

\begin{figure}
    \centering
    \includegraphics[width=4cm]{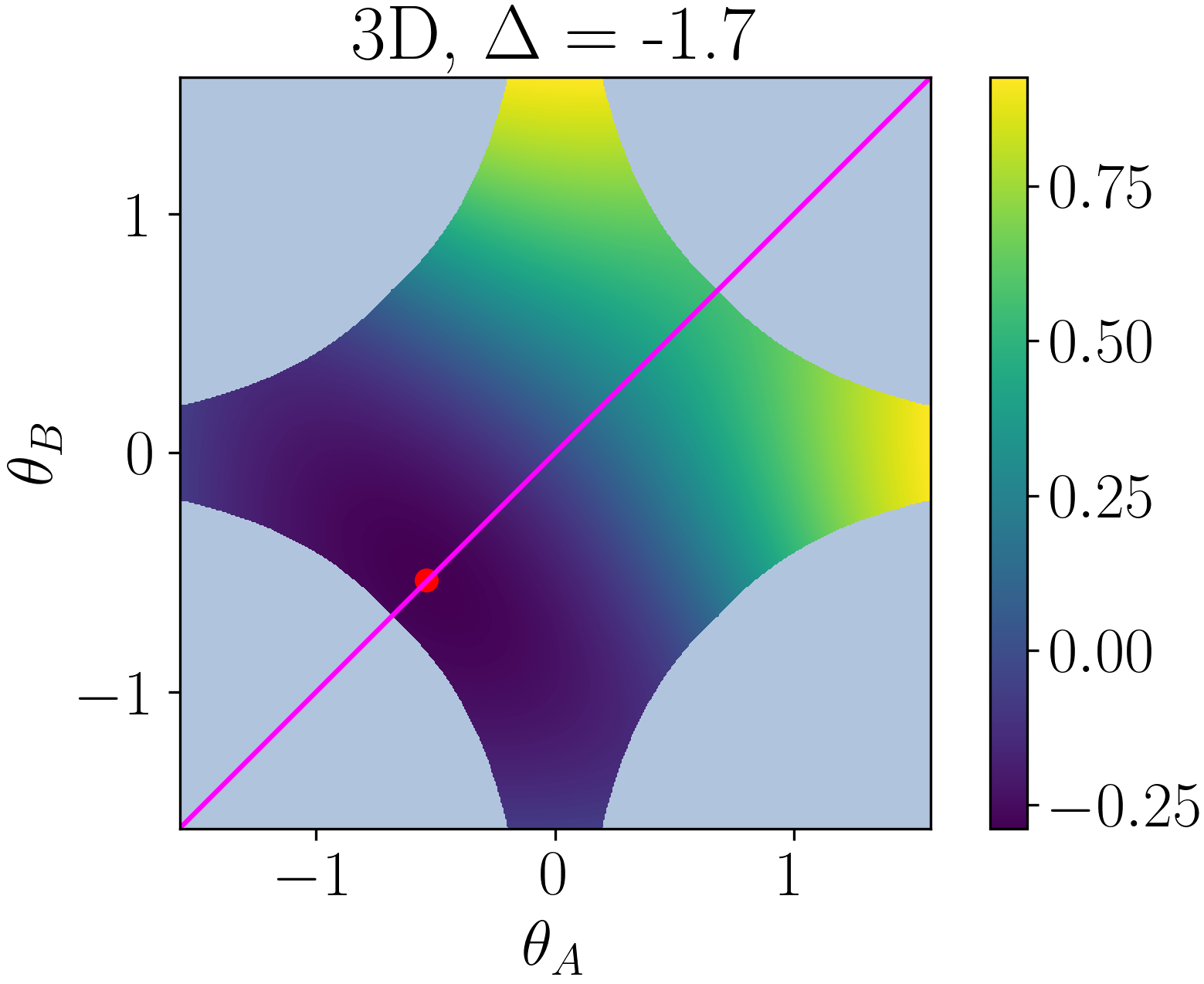}
    \includegraphics[width=4cm]{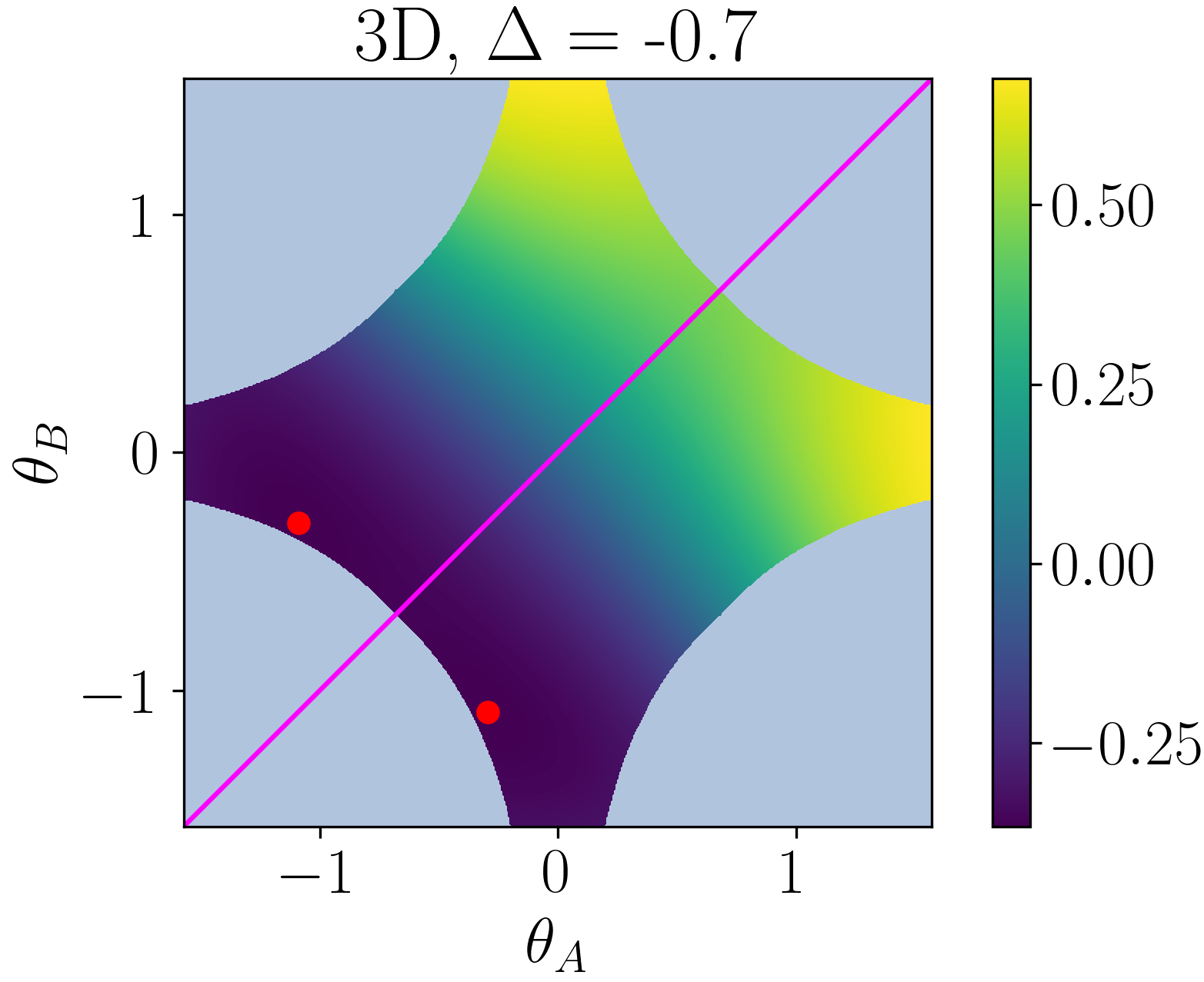}
    \caption{plots of the energy $E(\theta_A,\theta_B)$ for 3D.  at $\Delta=-1$ (left) and $\Delta=-0.6$ (right). Note that we zoomed in on a smaller window near the origin.  The red dot is the minimum.}
    \label{fig:energy3D}
\end{figure}

From these calculations we obtain the energy $E(\theta_A,\theta_B)$, plotted in Fig.~\ref{fig:energyplots} and Fig.~\ref{fig:energy3D}, and find the variational ground state by finding the $\theta_A,\theta_B$ that minimizes $E(\theta_A,\theta_B)$.

\subsection{PXP term}\label{sec:PXPenergy}

Let us first consider $E_{\text{PXP}} = \bra{\psi}H_{\text{PXP}}\ket{\psi}$. Using $\mathcal{P}\ket{\psi} = \ket{\psi}$, we have 

\begin{align}
    E_\text{PXP} = \sum_i\bra{\psi}\sigma^x_i\ket{\psi}\nonumber\\ = \frac{N}{2}\big(\bra{\psi}\sigma^x_a\ket{\psi} + \bra{\psi}\sigma^x_b\ket{\psi}\big),
\end{align}

where here (and in the sections below) we introduce the convention that $a$ and $b$ each refer to any single site on the (respectively) $A$ and $B$ sublattices. 

To calculate $\bra{\psi}\sigma^x_i\ket{\psi}$, we must consider the following tensor:

\begin{align}\label{eq:StensorDefinition}
    &\qquad\qquad\qquad S(\theta,\phi) \equiv \vcenter{\hbox{\includegraphics[width=2cm]{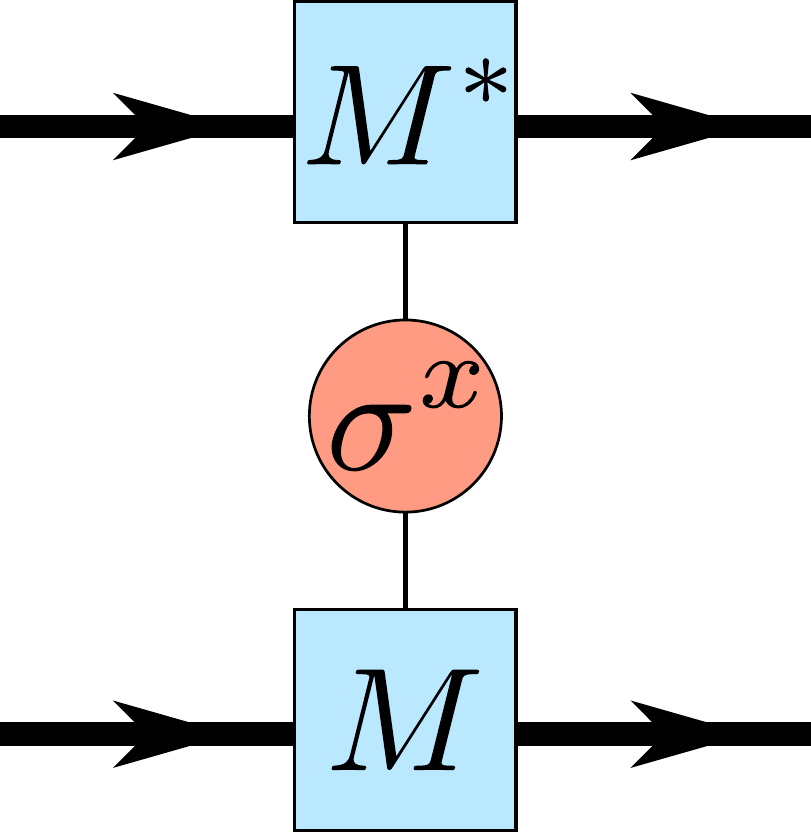}}}\nonumber\\
     &=ie^{i\phi}\sin(\theta/2)\cos(\theta/2)(|\Vec{0}\rangle\langle\Vec{0}|\otimes|\Vec{1}\rangle\langle\Vec{0}|-|\Vec{0}\rangle\langle\Vec{0}|\otimes|\Vec{0}\rangle\langle\Vec{1}|)\nonumber\\
    &=ie^{i\phi}\sin(\theta/2)\cos(\theta/2)(|\vec{0}\rangle\langle\vec{10}|-|\vec{0}\rangle\langle\vec{01}|).
\end{align}

Note that in diagrams like the above, each in (out) leg on the top is paired with an in (out) leg on the bottom, such that for a $D$ dimensional lattice we have $D$ in-pairs and $D$ out-pairs. Since this tensor $S$ has outgoing indices of the form $\Vec{01}$, we cannot perform the bond dimension reduction described above. The solution is to consider the object
 \begin{align}
  \vcenter{\hbox{\includegraphics[width=3cm]{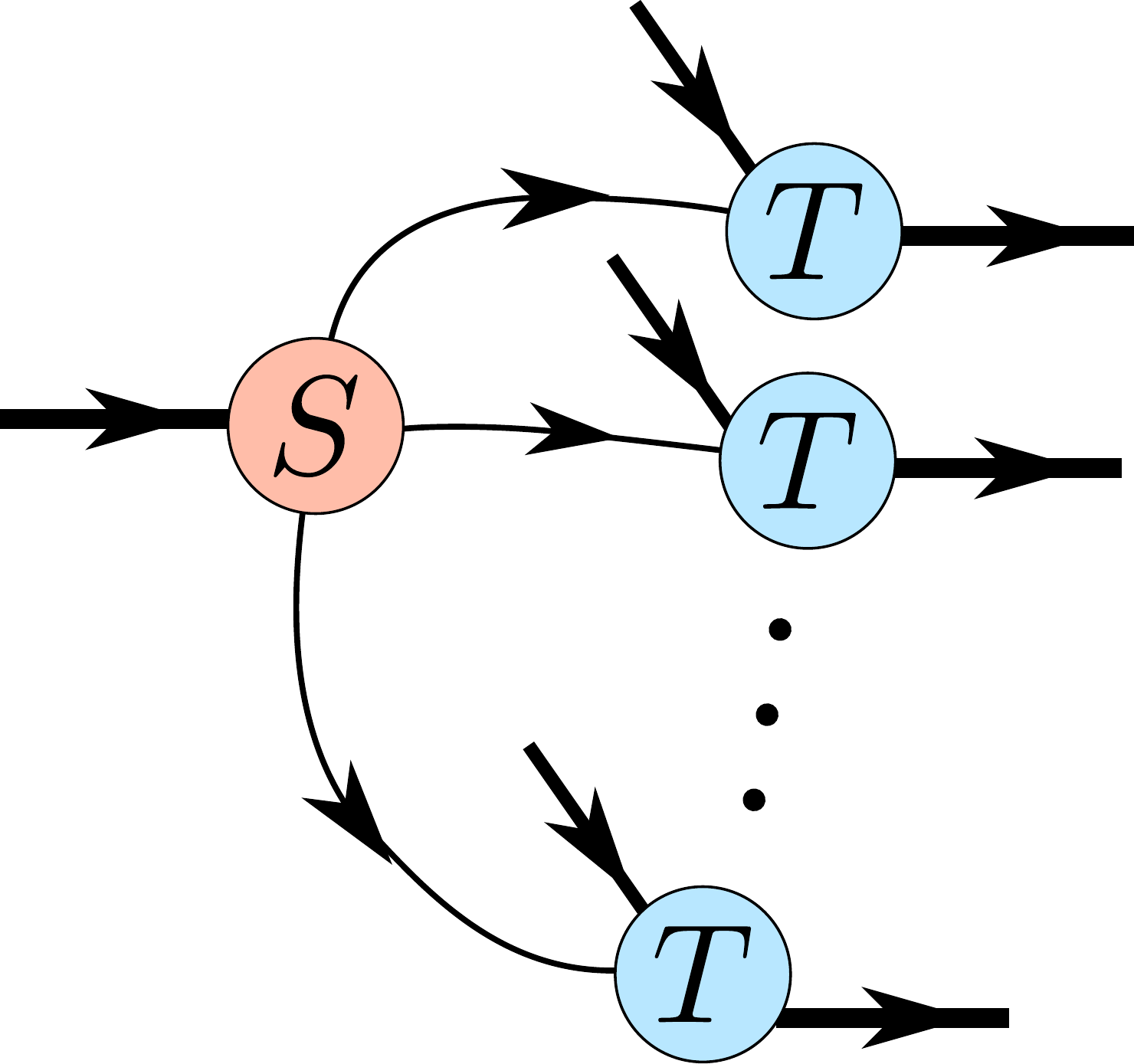}}}
 \end{align}
as a single unit, a tensor occupying $D+1$ sites, on which we can perform the bond dimension reduction.  

The quantity $\bra{\psi}\sigma^x_a\ket{\psi}$ is the contraction of the tensor network with $S(\theta_A,\phi_A)$ on a single site, and with $T(\theta_A)$ or $T(\theta_B)$ on all other sites: the calculation may be performed via either explicit contraction ($D\leq 2$) or the perturbative method ($D\geq 2$).  Upon performing this calculation, we find that the $\theta,\phi$ dependence decouples in a way that we may write $\bra{\psi}\sigma^x_a\ket{\psi} = \sin\phi_A\,\mathbb{F}(\theta_A,\theta_B),$ and thus for the energy we have 

\begin{align}
    E_{\text{PXP}}(\theta_A,\theta_B,\phi_A,\phi_B) = \\\frac{N}{2}\big(\sin\phi_A\,\mathbb{F}(\theta_A,\theta_B) + \sin\phi_B\,\mathbb{F}(\theta_B,\theta_A)\big),
\end{align}
It is easy to check that due to the symmetry of $\mathbb{F}(\theta_A,\theta_B)$, one can without loss of generality set $\phi_A=\phi_B=\pi/2$. 

\subsection{Detuning and next-nearest-neighbour terms}
For the detuning and NNN terms, we have, similar to the above:  

\begin{align}
    E_\Delta&= -\Delta\sum_i\bra{\psi}n_i\ket{\psi}\nonumber\\ &= -\frac{\Delta N}{2}\big(\bra{\psi}n_a\ket{\psi} + \bra{\psi}n_b\ket{\psi}\big)\\
    E_\text{NNN} &= V\sum_{\langle\!\langle ij\rangle\!\rangle}\bra{\psi}n_i n_j\ket{\psi}
\end{align}

We thus need to consider the tensor: 

\begin{align}
    J(\theta) \equiv \vcenter{\hbox{\includegraphics[width=2cm]{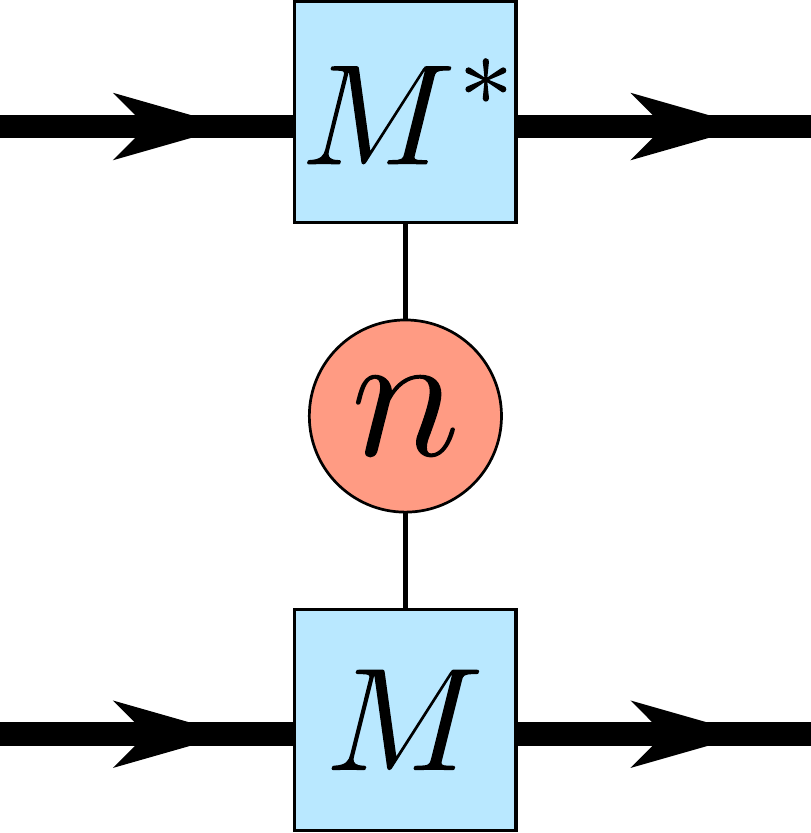}}}=\sin^2(\theta/2)|\vec{0}\rangle\langle\vec{1}|.
\end{align}

It follows that the quantity $\bra{\psi}n_i\ket{\psi}$ is the contraction of a tensor network with $J(\theta_i)$ on a single site and $T$ on all other sites, and that $\bra{\psi}n_i n_j\ket{\psi}$ would be the contraction of a tensor network with $J(\theta_i)$ on site $i$ and $J(\theta_j)$ on site $j$ and $T$ on all other sites.

\section{TDVP calculations}\label{appendix:TDVP}
Here we present the technical details of the variational dynamics calculations.  We first give a brief but self-contained review of the TDVP. 

\subsection{TDVP}
The TDVP (see Fig.~\ref{fig:manifold} for a visualization in the context of quantum scarred dynamics) is a way of calculating the approximate time evolution of a quantum state by projecting the true unitary time evolution onto a variational manifold, which in our case is the family of tensor network states introduced in Sec.~\ref{sec:variationalansatz}.
As explained in the main text, we may impose unit cell translation invariance, i.e. have all spins on each sublattice be in the same state, and due to energy conservation we may set $\Vec{\phi} = 0$, so that ultimately our variational manifold is parameterized by only two variables, $\theta_A$ and $\theta_B$.

Let the variational state be $\ket{\psi(\theta_A,\theta_B)}$.  Define the vector $\ket{\delta}$ as the difference between the time evolution within the manifold, and the true evolution within the full Hilbert space.  The latter is given by the Schrodinger equation, $i\frac{d\ket{\psi}}{dt} = H\ket{\psi}$. 
In the following we use Greek letters as placeholders for the sub-lattices $A$ and $B$, and use the notation  $\ket{\partial_\mu\psi}\equiv\frac{\partial\ket{\psi}}{\partial\theta_\mu}$. The leakage rate vector is given by
\begin{align}
    \ket{\delta}  = \sum_\mu \ket{\partial_\mu\psi}\Dot{\theta}_\mu + iH\ket{\psi}
\end{align}
The leakage rate is defined as $\Gamma^2 = \langle\delta|\delta\rangle$.  To derive the TDVP equations of motion, we minimize the leakage rate with respect to $\dot{\theta}_A$ and $\dot{\theta}_B$.  That is, we have $\partial\Gamma^2/\partial\dot{\theta}_A = \partial\Gamma^2/\partial\dot{\theta}_B = 0$, which gives us the equations of motion
\begin{align}\label{eq:EOM}
   \dot{\theta}_A = -i\frac{\langle\partial_A\psi|H|\psi\rangle}{\langle\partial_A\psi|\partial_A\psi\rangle} \,,\quad\dot{\theta}_B = -i\frac{\langle\partial_B\psi|H|\psi\rangle}{\langle\partial_B\psi|\partial_B\psi\rangle}.
\end{align}

In deriving these equations, we used the fact that $\langle\partial_A\psi|\partial_B\psi\rangle=0$, i.e. the Gram matrix $G_{\mu\nu}\equiv\langle\partial_\mu\psi|\partial_\nu\psi\rangle$ is diagonal. This is nontrivial, but we will derive this below.

After performing the calculations for the numerators and denominators of Eq. \eqref{eq:EOM}, it turns out that we can write the equations of motion as $\dot{\theta}_A = f(\theta_A,\theta_B)$ and $\dot{\theta}_B = f(\theta_B,\theta_A)$, where 

\begin{align}
    f(\theta_A,\theta_B) = \frac{\mathbb{K}(\theta_A,\theta_B)+D\mathbb{S}(\theta_A,\theta_B)}{\mathbb{G}(\theta_A,\theta_B)},
\end{align}
where $D$ is the dimension of space, and $\mathbb{K}$, $\mathbb{S}$, and $\mathbb{G}$ are each the contraction of a single (infinite) tensor network, the technical details of which are described below.

\subsection{Gram matrix calculation}
Here we describe the calculation of the Gram matrix, i.e. quantities of the form $\langle\partial_\mu\psi|\partial_\nu\psi\rangle\equiv G_{\mu\nu}$. 
We have, from the product rule of calculus, that $\ket{\partial_A\psi} = \sum_{a\in A} \ket{\partial_a \psi}$, so that $\langle\partial_\mu\psi|\partial_\nu\psi\rangle = \sum_{i\in\mu,j\in\nu}\langle\partial_i\psi|\partial_j\psi\rangle$, where $\ket{\partial_i \psi}$ means that the derivative is only applied on the site $i$.  In other words, $\ket{\partial_i \psi}$ is a tensor network state with the tensor $\partial M(\theta_i)$ on site $i$, and the usual $M(\theta_j)$ tensor on all other sites $j\neq i$.  The tensor $\partial M$ is defined as 
\begin{align}
    \partial M \equiv \frac{dM}{d\theta} = \vcenter{\hbox{\includegraphics[width=1.7cm]{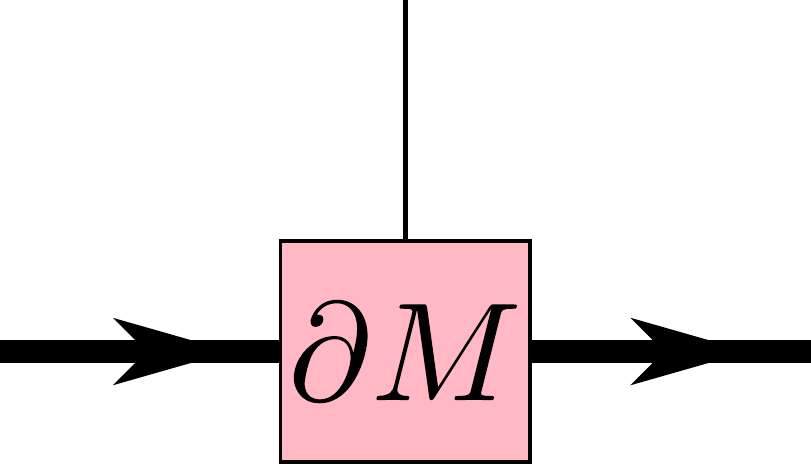}}}\qquad\quad\nonumber\\
    = -\frac{i}{2}\sin(\theta/2)\ket{\downarrow} |\vec{0}\rangle\langle\vec{0}|-\frac{1}{2}\cos(\theta/2)\ket{\uparrow}|\vec{0}\rangle\langle\vec{1}|
\end{align}
Define the tensor
\begin{align}
    \partial T = \vcenter{\hbox{\includegraphics[width=2cm]{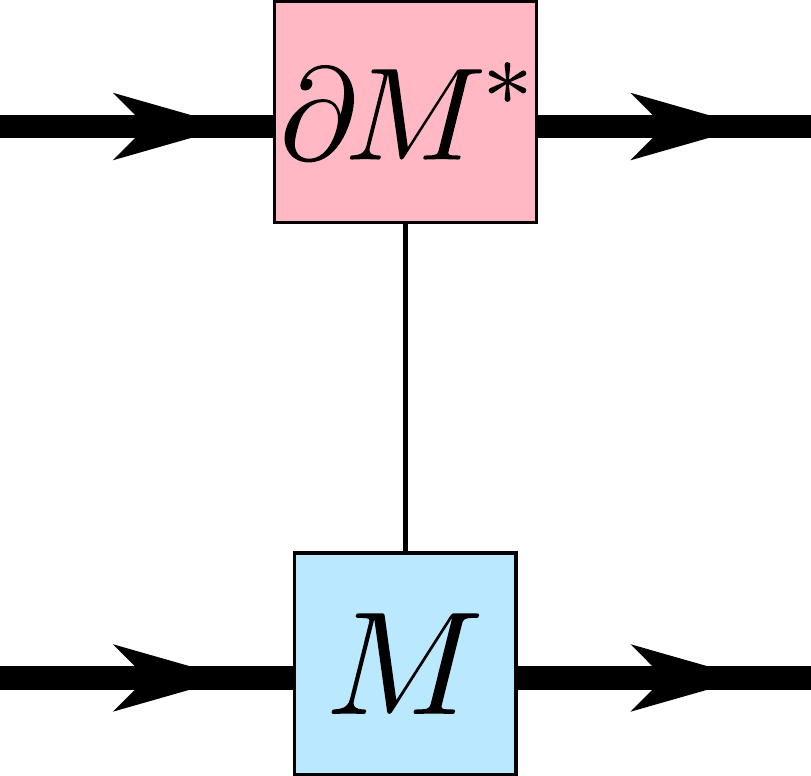}}} \;.
\end{align}
It turns out that, after performing the bond-dimension reduction, that $\partial T\propto q$.  This allows us to conclude that $\langle\partial_i\psi|\partial_j\psi\rangle=0$ for all pairs of sites $i\neq j$.
This leads to the conclusion that the Gram matrix is diagonal, and that $G_{AA}=(N/2)\langle\partial_a\psi|\partial_a\psi\rangle$, and the analogous equation for the $B$ sublattice. Here $a$ is any single site in $A$, and $N$ is the total number of sites ($N$ is infinite, but it always cancels out and never shows up in any final results).  

Let us define

\begin{align}
 g = \vcenter{\hbox{\includegraphics[width=2cm]{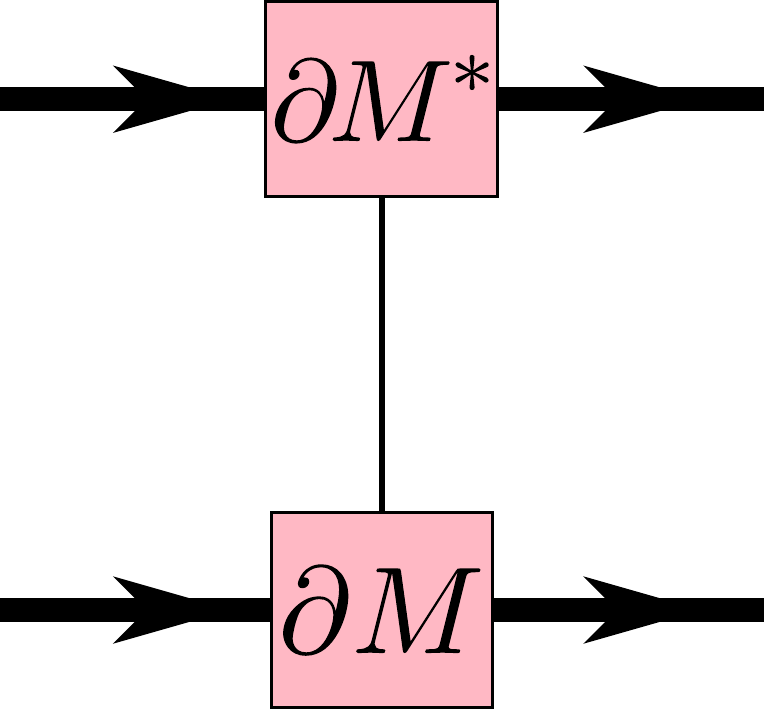}}}\longrightarrow\;\;\vcenter{\hbox{\includegraphics[width=1.8cm]{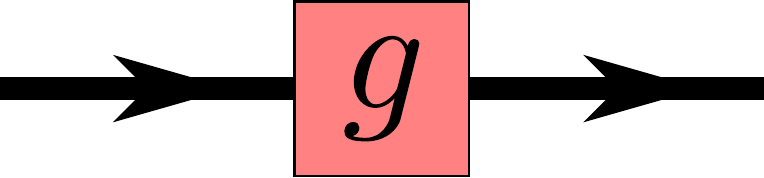}}}\;,
\end{align}
performing a bond dimension reduction as before.  The elements of $g$ are given by $g(\theta) = (1/4)\sin^2(\theta/2)|\vec{0}\rangle\langle\vec{0}|+(1/4)\cos^2(\theta/2)|\vec{0}\rangle\langle\vec{1}|$.

Thus $\braket{\partial_a\psi}$ is the contraction of a tensor network with $g(\theta_A)$ on a single site and $T$ on all other sites.   We define the aforementioned $\mathbb{G}$ as $\mathbb{G}(\theta_A,\theta_B) = \braket{\partial_a\psi} = G_{AA}/(N/2)$.  By symmetry, the contraction of a single $g(\theta_B)$ tensor on a $B$ lattice site is $\mathbb{G}(\theta_B,\theta_A)=\langle\partial_b\psi|\partial_b\psi\rangle=G_{BB}/(N/2)$.  Similar to the calculation of the energy, these tensor network calculations can either be performed explicitly or perturbatively, using the methods introduced above.

\subsection{Calculation of $\langle\partial_\mu\psi|H|\psi\rangle$}
Here we discuss the calculation of $\langle\partial_\mu\psi|H|\psi\rangle$, the numerator of the equations of motion, Eq.~\eqref{eq:EOM}.  Since $\ket{\psi}$ is inside the Rydberg-blockaded subspace we have 
\begin{align}
    \langle\partial_\mu\psi|H|\psi\rangle = \Omega\sum_{j\in\mu}\sum_i\langle\partial_j\psi|\sigma_i^x|\psi\rangle.
\end{align}
We will see that $\langle\partial_j\psi|\sigma_i^x|\psi\rangle$ is nonzero only if either $i=j$, or $j$ is immediately ``downstream" of $i$, i.e. in 1D, they are nearest neighbours with $i$ to the left of $j$ (and in 2D, either to the left or above, and so on). 

\subsubsection{Calculation of $\mathbb{K}$}

Let us first consider the $i=j$ contribution, which we call $\mathbb{K}$.
Define the tensor $K$ as 
\begin{align}
    K = \vcenter{\hbox{\includegraphics[width=2.2cm]{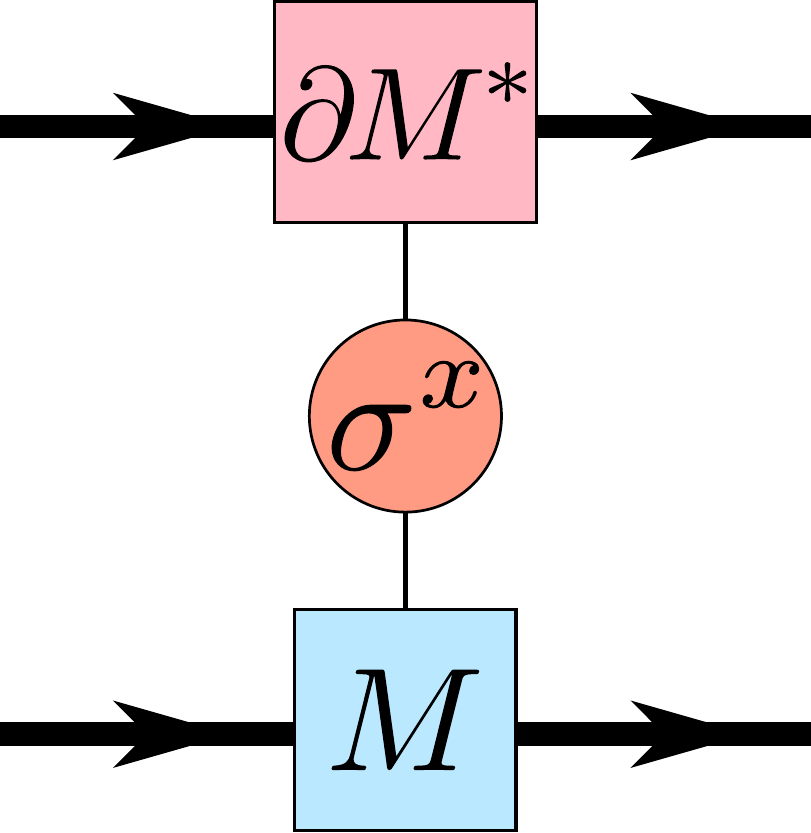}}},
\end{align}
where we find that the elements are given by $K(\theta) = (i/2)\cos^2(\theta/2)|\vec{0}\rangle\langle\vec{10}|+(i/2)\sin^2(\theta/2)|\vec{0}\rangle\langle\vec{01}|$.
The quantity $\mathbb{K}(\theta_A,\theta_B)=\langle\partial_a\psi|\sigma_a^x|\psi\rangle$ is the contraction of a network with $K(\theta_A)$ on one site $a\in A$ and $T$ on all other sites.  $A\leftrightarrow B$ symmetry gives the analogous quantity $\mathbb{K}(\theta_B,\theta_A)=\langle\partial_b\psi|\sigma_b^x|\psi\rangle$.

Similar to the calculation for $S$ for the energy in Appendix~\ref{sec:PXPenergy}, we cannot perform the bond dimension reduction on $K$ alone, so we consider the tensor 
 \begin{align}
  \vcenter{\hbox{\includegraphics[width=3.3cm]{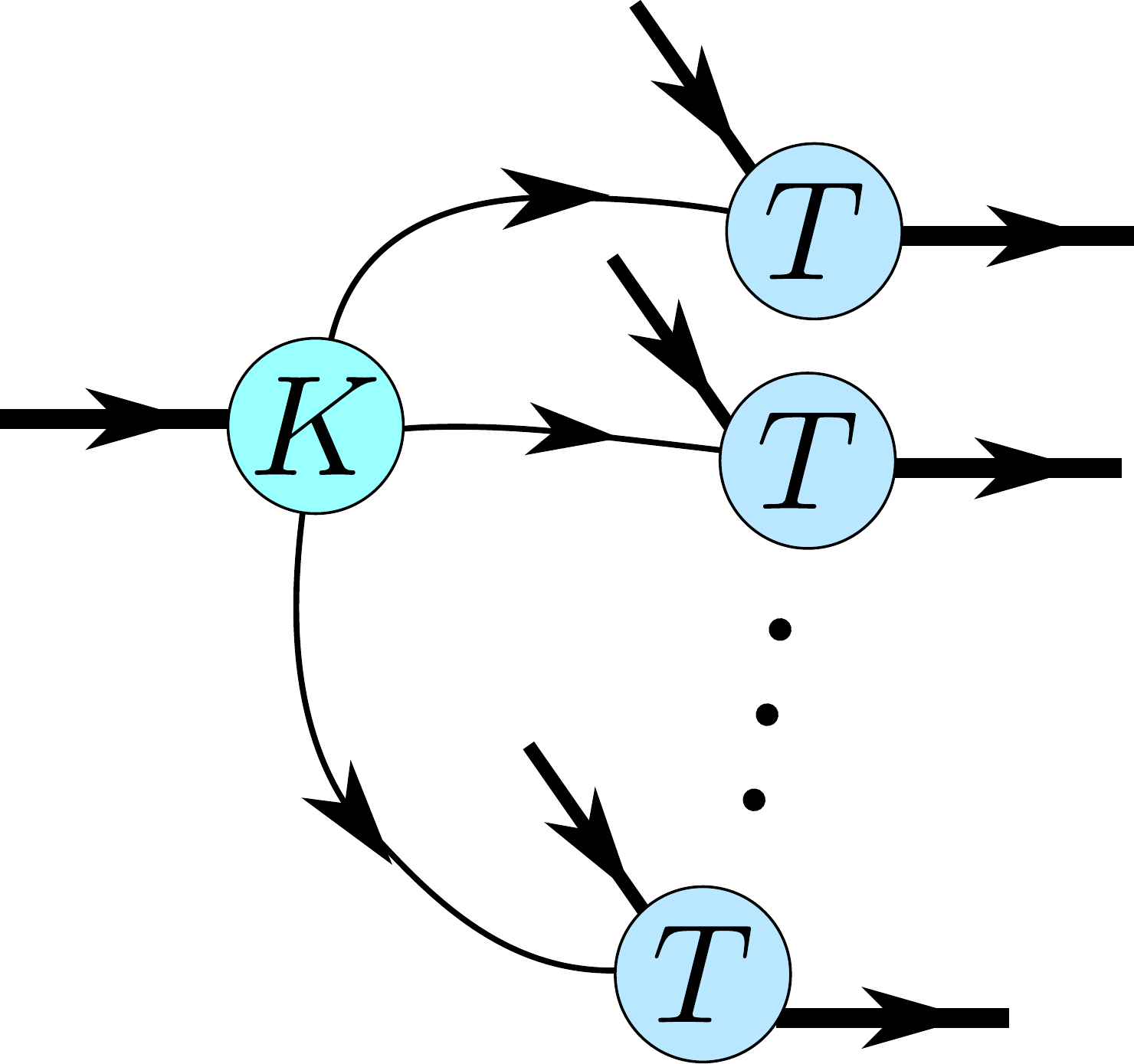}}}. 
 \end{align}

 We then contract this tensor with $T$ on all other sites using the same methods as for the other tensor network contractions above, to obtain $\mathbb{K}(\theta_A,\theta_B)$.

\subsubsection{Calculation of $\mathbb{S}$}\label{sec:Scalculation}
Next we consider $\langle\partial_j\psi|\sigma_i^x|\psi\rangle$ when $i\neq j$.  To start, consider the tensor $S$ as defined in Eq.~\eqref{eq:StensorDefinition}, but we now have $\phi_A=\phi_B=0$:
\begin{align}
    &\qquad\qquad\qquad S(\theta) \equiv \vcenter{\hbox{\includegraphics[width=2cm]{figs/S_tensor.pdf}}}\nonumber\\
    &=i\sin(\theta/2)\cos(\theta/2)(|\vec{0}\rangle\langle\vec{10}|-|\vec{0}\rangle\langle\vec{01}|).
\end{align}

In this case, we find that
 \begin{align}\label{eq:ST0}
  \vcenter{\hbox{\includegraphics[width=3.3cm]{figs/STcontraction.pdf}}}  =0.
 \end{align}

 Note that this implies that $\bra{\psi}\sigma^x_i\ket{\psi} = 0$ which implies that $E = \langle H\rangle = 0$ when $\ket{\psi}$ is a variational state with $\Vec{\phi} =0$. (Eq.~\eqref{eq:ST0} is true only if all the external legs are $0$ or $1$, i.e. if we perform bond dimension reduction.  This is sufficient for our purposes here, but the subtlety will matter when calculating the leakage rate.) 

Thus the only nonzero $i\neq j$ contributions are the cases where $i$ is next to $j$.  By symmetry, we only need to consider the diagram
 \begin{align}\label{eq:SdT}
  \vcenter{\hbox{\includegraphics[width=3.5cm]{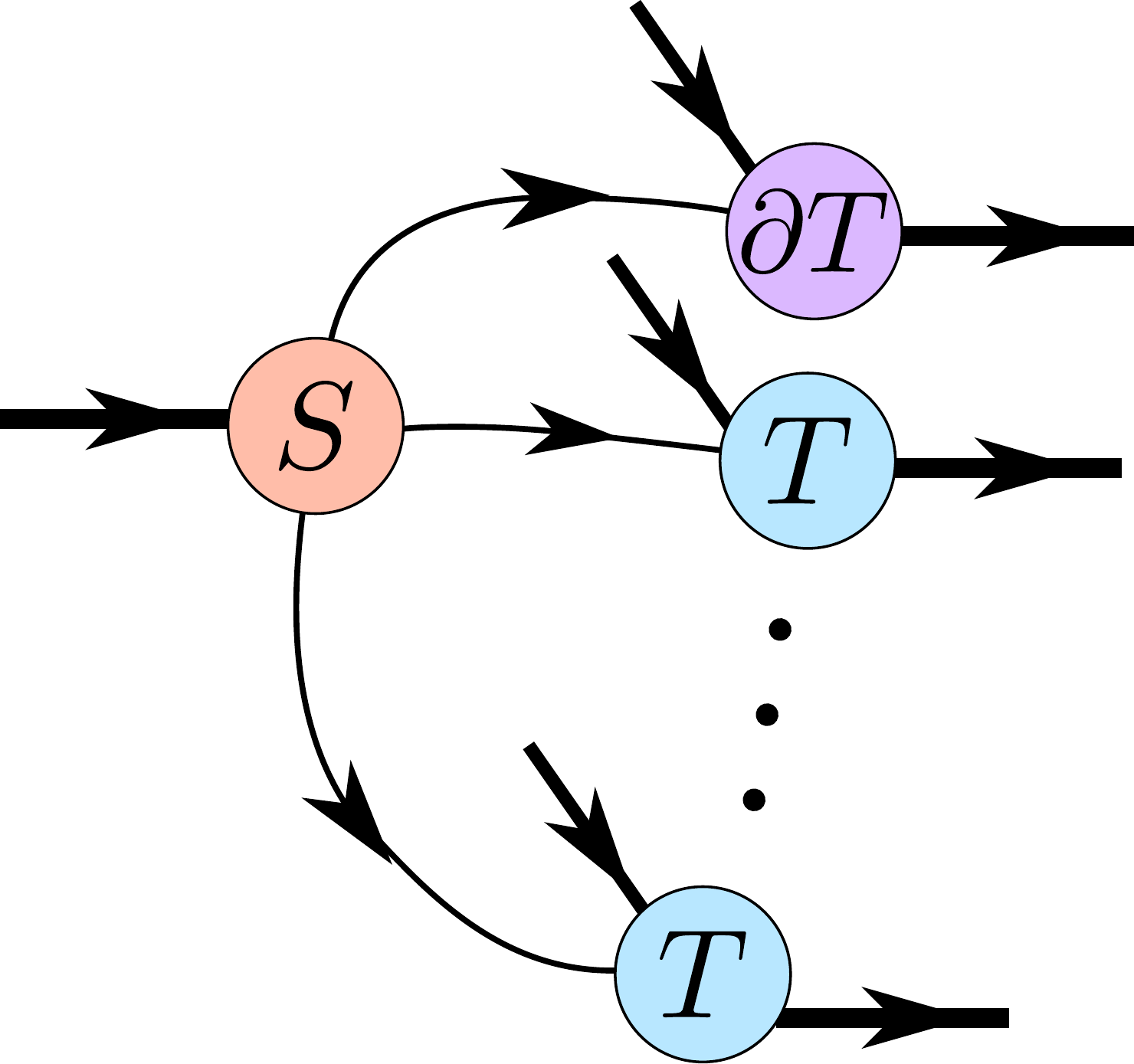}}}, 
 \end{align}
 multiplied by a factor of $D$. We contract this object with the $T$ tensor on all other sites, to get $\langle\partial_j\psi|\sigma_i^x|\psi\rangle$.  We define $\mathbb{S}(\theta_A,\theta_B)=\langle\partial_a\psi|\sigma_b^x|\psi\rangle$ and $\mathbb{S}(\theta_B,\theta_A)=\langle\partial_b\psi|\sigma_a^x|\psi\rangle$.

\section{Projected product states}\label{appendix:projprod}
Consider the set of projected (i.e. Rydberg blockaded) product states:  
\begin{align}
	|\psi(\Vec{\vartheta},\Vec{\varphi})\rangle = \mathcal{P}\bigotimes_i \ket{\vartheta_i,\varphi_i},
\end{align}
where the state of each site is a spin coherent state $\ket{\vartheta_i,\varphi_i} =  \cos(\vartheta_i/2)\ket{\downarrow} -i e^{i\varphi_i}\sin(\vartheta_i/2)\ket{\uparrow}$. These states resemble the variational ansatz, but are not normalized.  

As noted earlier, $\mathcal{P}$ is the application of the two-site operator $\mathcal{P}_{ij}=\ket{\downarrow\downarrow}\bra{\downarrow \downarrow}+\ket{\downarrow \uparrow}\bra{\downarrow \uparrow}+\ket{\uparrow \downarrow}\bra{\uparrow \downarrow}=\text{diag}(1,1,1,0)$ to all nearest-neighbour pairs of sites.  The two-site projector can be decomposed in a ``matrix product operator" manner: diagrammatically, we have

\begin{align}
	\vcenter{\hbox{\includegraphics[width=.7cm]{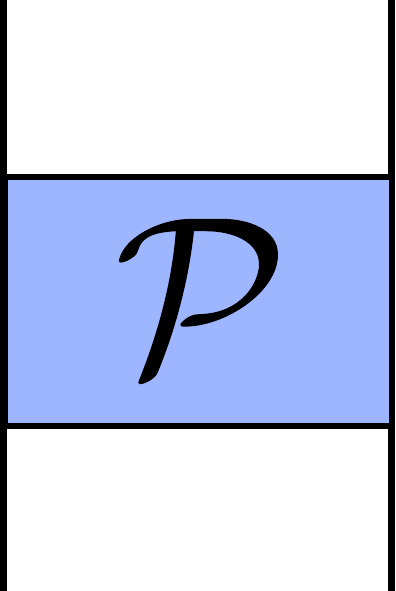}}} \;\;=\;\; \vcenter{\hbox{\includegraphics[width=2.1cm]{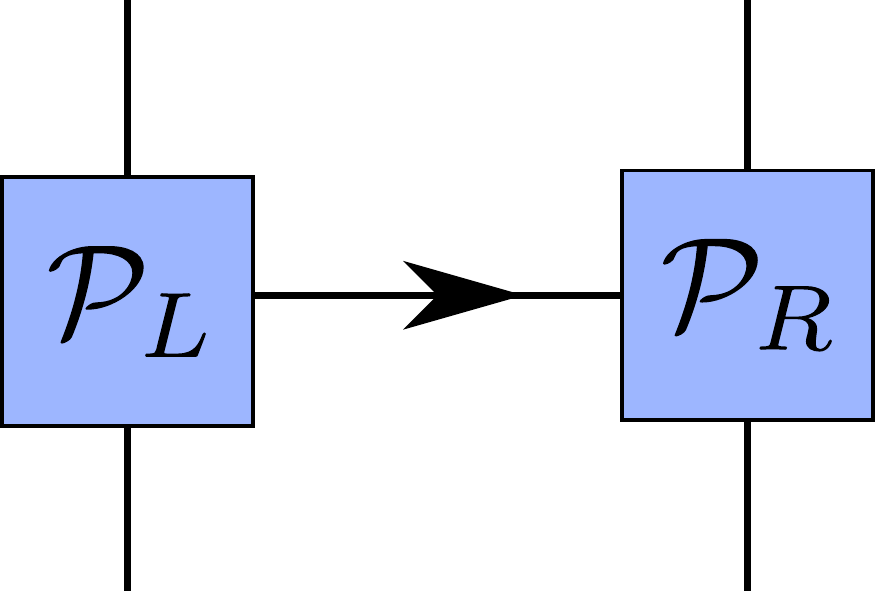}}}\;,
\end{align}
where the internal line has bond dimension 2.  Specifically, $\mathcal{P}_R$ and $\mathcal{P}_L$ are given by: $\mathcal{P}_R^0 = \begin{pmatrix}1&0\\ 0&1\end{pmatrix}
$, $\mathcal{P}_R^1=\mathcal{P}_L^0 = \begin{pmatrix}1&0\\ 0&0\end{pmatrix}
$, and $\mathcal{P}_L^1 = \begin{pmatrix}0&0\\ 0&1\end{pmatrix}$, where the superscript refers to the horizontal leg.

We can use this to write the state $ |\psi(\Vec{\vartheta},\Vec{\varphi})\rangle$ as a tensor network.  In one dimension, the MPS tensor is

\begin{align}
	\vcenter{\hbox{\includegraphics[width=2cm]{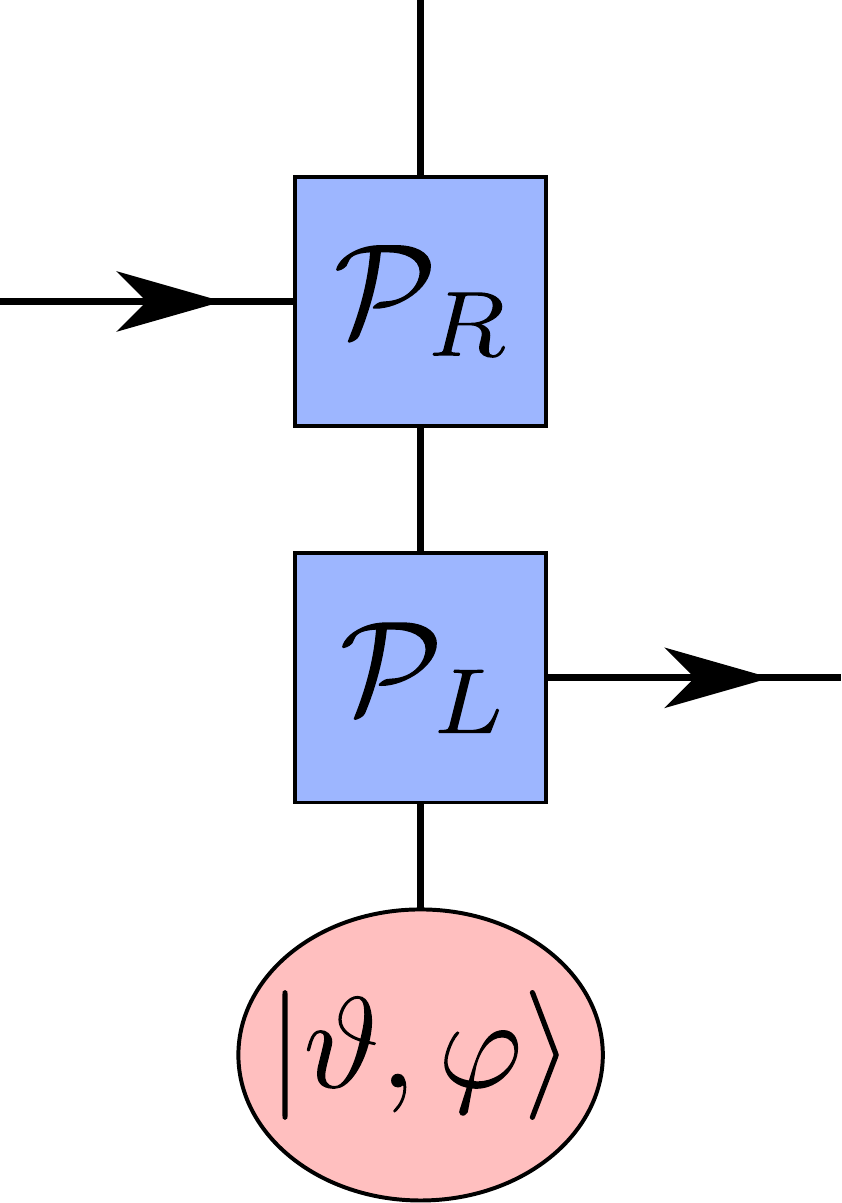}}} \;\;=\;\; \vcenter{\hbox{\includegraphics[width=3.1cm]{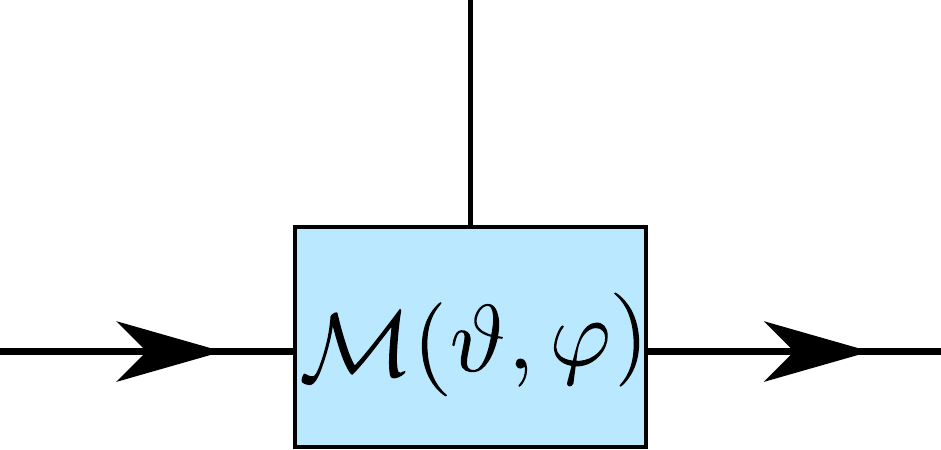}}}.
\end{align}

This can be easily generalized to PEPS in higher dimensions. We can read off all the nonzero elements of the PEPS tensors. Using the same notation as we used for the variational ansatz \eqref{eq:T_tensor}, the PEPS tensor at each site is given by

\begin{align}
    \mathcal{M}(\vartheta,\varphi) =\cos(\vartheta/2) \ket{\downarrow}\ket{\alpha}\langle\Vec{0}|
    -ie^{i\varphi}\sin(\vartheta/2)\ket{\uparrow}|\vec{0}\rangle\langle\vec{1}|.
    \label{eq:PPS}
\end{align}

Comparing this expression to Eq.~\eqref{eq:Mtensor} one can see that the only difference lies in the coefficient of the component $\ket{\beta}\langle\vec{0}|$. In 1D, a gauge transformation exists that maps the PEPS state resulting from the tensors contraction into each other~\cite{periodicOrbits}. We checked numerically that this is not the case in $D>1$. We did so by taking the exact contraction on periodic 2D systems of the PEPS in Eq.~\eqref{eq:PPS} and Eq.~\eqref{eq:Mtensor}. For a fixed value of the variational parameters in the former, we optimized the overlap between the two states over the variational parameters of the latter. The failure in finding unity overlap within numerical precision signals the non-equivalence of the two states. This can also be inferred from the fact that the projected product state 2D PEPS is known to exhibit a second-order phase transition along the line $\theta_A = \theta_B$~\cite{lesanovski_pps}, whereas this phase transition is absent in our ansatz~\eqref{eq:Mtensor}, as we verified numerically from infinite-cylinder calculations. We stress that the main advantage of our ansatz is the fact that is normalized in the thermodynamic limit. This property ensures a dramatic simplification of the perturbative expansion which allowed us to compute with high accuracy the overlaps and expectation values required for the ground state and TDVP analysis of the PXP model in two and three dimensions.

\section{Leakage rate}\label{appendix:error}
Here we discuss the calculation of the leakage rate for the variational dynamics, the rate at which the state evolving under the full unitary evolution leaves the variational manifold at a given instant.  Specifically, we will show how the evaluation of the leakage rate reduces to quantities of the form $\bra{\psi}\hat{\mathcal{O}}_{\text{local}}\ket{\psi}$, which can then be calculated via a tensor network contraction similar to the calculations described above. 

The leakage rate is the magnitude of the leakage rate vector $|\delta\rangle$:

\begin{align}
    \langle\delta|\delta\rangle = \sum_{\mu\nu}\Dot{\theta}_\mu\Dot{\theta}_\nu\langle\partial_\mu\psi|\partial_\nu\psi\rangle + i \sum_\mu \dot{\theta}_\mu \langle\partial_\mu\psi|H|\psi\rangle\\ - i\sum_\mu\dot{\theta}_\mu \langle\psi|H|\partial_\mu\psi\rangle + \langle\psi|H^2|\psi\rangle
    \\= \sum_\mu\Big[\dot{\theta}^2_\mu\langle\partial_\mu\psi|\partial_\mu\psi\rangle + 2i \dot{\theta}_\mu \langle\partial_\mu\psi|H|\psi\rangle \Big] \\+ \langle\psi|H^2|\psi\rangle \equiv \Gamma^2
\end{align}
where to get to the second line, we used the fact that (for the first term) the Gram matrix is diagonal, and (for the second term) the fact that $\langle\partial_\mu\psi|H|\psi\rangle$ is pure imaginary, so that $\langle\partial_\mu\psi|H|\psi\rangle = -\langle\psi|H|\partial_\mu\psi\rangle$. 

We can plug our derived expressions for $\dot{\theta}_A$ and $\dot{\theta}_B$ into the leakage rate expression to get
\begin{align}
    \Gamma^2 = \langle\psi|H^2|\psi\rangle - \sum_\mu\dot{\theta}^2_\mu\langle\partial_\mu\psi|\partial_\mu\psi\rangle
\end{align}
The new work that we have to do is to calculate $\langle\psi|H^2|\psi\rangle$. 

First, note that there are two equivalent ways of writing down the PXP Hamiltonian, which we call the ``global" and ``local" versions:
\begin{align}
    \text{global:}\quad H = \sum_i \mathcal{P}\sigma^x_i\mathcal{P} \\\text{local:}\quad H = \sum_i\bigg[\Big(\bigotimes_{\langle ji\rangle}P_j\Big)\sigma^x_i\bigg]
\end{align}
Here, $P= \ket{\downarrow}\bra{\downarrow}$, and $\mathcal{P}$ is the whole-system Rydberg-blockade projection, i.e. $\mathcal{P} = \prod_{\langle ij\rangle}\mathcal{P}_{ij}$, where $\mathcal{P}_{ij} = \text{diag}(1,1,1,0).$ In 1D, the local version can also be written as $H = \sum_i P_{i-1}\sigma^x_iP_{i+1}.$ 
Working with the global version, let us expand $\langle\psi|H^2|\psi\rangle$:
\begin{align}
   \langle\psi|H^2|\psi\rangle = \sum_{ij} \langle\psi|(\mathcal{P}\sigma^x_i\mathcal{P})(\mathcal{P}\sigma^x_j\mathcal{P})|\psi\rangle
\end{align}
At this point, it is convenient to use the local $H$ for the $i=j$ terms, and the global $H$ for the $i\neq j$ terms.    
For the $i=j$ terms, we have 
\begin{align}
   \sum_i \bra{\psi}\Big[\Big(\bigotimes_{\langle ji\rangle}P_j\Big)\sigma^x_i\Big]^2\ket{\psi} =  \sum_i \bra{\psi}\bigotimes_{\langle ji\rangle}P_j\ket{\psi} \\= \frac{N}{2}\bra{\psi}\bigotimes_{\langle ia\rangle}P_i\ket{\psi} + \frac{N}{2}\bra{\psi}\bigotimes_{\langle ib\rangle}P_i\ket{\psi}.
\end{align}
For the $i\neq j$ terms, making use of the fact that $\mathcal{P}\ket{\psi} = \ket{\psi}$, we have
\begin{align}
    \sum_{i\neq j} \langle\psi|(\mathcal{P}\sigma^x_i\mathcal{P})(\mathcal{P}\sigma^x_j\mathcal{P})|\psi\rangle=\sum_{i\neq j} \bra{\psi}\sigma^x_i\mathcal{P}\sigma^x_j\ket{\psi}
\end{align}
If $i$ and $j$ are not next to each other, all of the $\mathcal{P}$ operators in  $\mathcal{P}$ can be commuted out (this is easiest to see by drawing $\mathcal{P}$ as a brickwork-style ``quantum circuit", and noting that all the bricks are diagonal and therefore commute with each other, and that the bricks are annihilated upon hitting $\ket{\psi}$). So, for non-nearest-neighbour sites,  
\begin{align}
    \bra{\psi}\sigma^x_i\mathcal{P}\sigma^x_j\ket{\psi} = \langle\psi|\sigma_i^x\sigma_j^x|\psi\rangle = 0.
\end{align}
where in the last step we used the fact that we know from the explicit tensor network calculation that $\langle\psi|\sigma_i^x\sigma_j^x|\psi\rangle$ is nonzero only for adjacent sites.  Specifically, by nonadjacent, we mean neither directly adjacent nor diagonally adjacent. The quantity $\langle\psi|\sigma_a^x\sigma_{a^\prime}^x|\psi\rangle$ is nonzero if $a$ and $a^\prime$ are diagonally adjacent $A$ sites that are downstream of the same $B$ site. 

For quantities involving the projector, we only have to deal with the cases where $i$ and $j$ are next to each other.  In those cases, all of the $\mathcal{P}$ operators can be commuted out, except for the one acting on $i$ and $j$, so we have $\bra{\psi}\sigma^x_i\mathcal{P}\sigma^x_j\ket{\psi} = \bra{\psi}\sigma^x_i\mathcal{P}_{ij}\sigma^x_j\ket{\psi}$
Using symmetry, we can write the sum as 
\begin{widetext}
\begin{align}
    \sum_{\langle ij\rangle} \bra{\psi}\sigma^x_i\mathcal{P}\sigma^x_j\ket{\psi} = 2D\sum_{i} \bra{\psi}\sigma^x_i\mathcal{P}\sigma^x_{i+\hat{x}}\ket{\psi}
    +  \frac{N}{2} {D\choose 2}\langle\psi|\sigma_a^x\sigma_{a^\prime}^x|\psi\rangle + (a\leftrightarrow b)\\ =  \frac{N}{2} \bigg[ 2D \langle\psi|\sigma_a^x\mathcal{P}_{a,a+\hat{x}}\sigma_{a+\hat{x}}^x|\psi\rangle   +   {D\choose 2}\langle\psi|\sigma_a^x\sigma_{a^\prime}^x|\psi\rangle + (a\leftrightarrow b) \bigg]
\end{align}
In the first line, we gained a factor of $2D$: the factor of 2 comes from having considered each nearest-neighbour pair only once, and the factor of $D$ comes from the symmetry between the different directions (recall the fact for each PEPS tensor, all of the ``in" legs are interchangeable, likewise for the ``out" legs). In the second line, we used translation invariance.
Using $ \langle\partial_A\psi|\partial_A\psi\rangle = \frac{N}{2}\langle\partial_a\psi|\partial_a\psi\rangle$ (and the analogous equation for $B$) for the quantity $\sum_\mu\dot{\theta}^2_\mu\langle\partial_\mu\psi|\partial_\mu\psi\rangle$ , we now have the full leakage rate expression: 
\begin{align}\label{eq:leakage}
\begin{aligned}
    \frac{\Gamma^2}{N} \equiv \gamma^2 = \frac{1}{2}\bra{\psi}\bigotimes_{\langle ia\rangle}P_i\ket{\psi}  + D\Big(\langle\psi|\sigma_a^x\mathcal{P}_{a,a+\hat{x}}\sigma_{a+\hat{x}}^x|\psi\rangle\Big) - \frac{1}{2} \Big( \dot{\theta}_A^2\langle\partial_a\psi|\partial_a\psi\rangle   \Big)  +   \frac{1}{2}{D\choose 2}\langle\psi|\sigma_a^x\sigma_{a^\prime}^x|\psi\rangle + (a\leftrightarrow b), \end{aligned}
\end{align}
\end{widetext}
where $a$ and $a^\prime$ have a specific relationship as mentioned above, and where the $(a\leftrightarrow b)$ applies to all of the preceding terms.  At this point, each of the terms in the above equation can be calculated via a tensor network contraction.

\section{Explicit TDVP calculations for product states}\label{appendix:leakageprodstates}

Here we analytically perform the TDVP calculations for states along the $\theta_A$ and $\theta_B$ axes in Fig.~\ref{fig:flows1d2d} and Fig.~\ref{fig:flowsPert}. We derive the equations of motion Eq.~\eqref{eq:EOMaxes} and demonstrate that the leakage rate is zero, i.e. that $\gamma(\theta_A,0)=\gamma(0,\theta_B)=0$. 

Let us consider the case $\theta_B=0$; by symmetry, the case $\theta_A=0$ will be the same but with $\theta_A \leftrightarrow \theta_B$.
When doing the TDVP calculations, we work with the variational state $\ket{\psi(\theta_A,\theta_B)}$, and the states $\ket{\partial_a\psi(\theta_A,\theta_B)}$ and $\ket{\partial_b\psi(\theta_A,\theta_B)}$. Recall that the latter states are defined as applying a derivative $\partial\!/\!\partial\theta_A$ or $\partial\!/\!\partial\theta_B$ to $\ket{\psi(\theta_A,\theta_B)}$ at any one particular site $a\in A$ or $b\in B$.  In general, these are tensor network states; however, when $\theta_B=0$, these become product states, allowing one to perform the TDVP and leakage rate calculations by hand.

The state $\ket{\psi(\theta_A,0)}$ is the following product state:
\begin{align}
    \ket{\psi(\theta_A,0)}=\bigotimes_{i\in A, j\in B}\ket{\theta_A}_i\ket{\downarrow}_j
\end{align}
 where $\ket{\theta_A}\equiv \cos(\theta_A/2)\ket{\downarrow}-i\sin(\theta_A/2)\ket{\uparrow}$.  The state $ \ket{\partial_a\psi(\theta_A,0)}$ is obtained by taking the derivative at one site; it is given by

\begin{align}
    \ket{\partial_a\psi(\theta_A,0)} = \ket{\theta_A^\prime}_a\otimes\Big(\bigotimes_{\substack{i\in A\backslash\{a\}\\ j\in B}}\ket{\theta_A}_i\ket{\downarrow}_j\Big)
\end{align}
where $\ket{\theta_A^\prime}\equiv d\!\ket{\theta_A}\!\!/\!d\theta_A= -(1/2)\sin(\theta_A/2)\ket{\downarrow}-(i/2)\cos(\theta_A/2)\ket{\uparrow}$.

The state $\ket{\partial_b\psi(\theta_A,0)}$ is more tricky to write down, as one must start with $\ket{\partial_b\psi(\theta_A,\theta_B)}$ and then derive the product state that results when $\theta_B=0$.  A careful inspection yields the result:

\begin{widetext}

\begin{align}
    \ket{\partial_b\psi(\theta_A,0)} =\Big(\bigotimes_{a\in N_L(b)}\cos(\theta_A/2)\ket{\downarrow}_a\Big)\otimes\big(\frac{-i}{2}\ket{\uparrow}_b\big)\otimes\Big(\bigotimes_{a\in N_R(b)}\ket{\downarrow}_a\Big)\otimes\Big(\bigotimes_{\substack{i\in A\backslash N(b)\\ j\in B\backslash\{b\}}}\ket{\theta_A}_i\ket{\downarrow}_j\Big)\\
     =\frac{-i}{2}\cos^{D}(\theta_A/2)\Big(\bigotimes_{a\in N(b)}\ket{\downarrow}_a\Big)\otimes\ket{\uparrow}_b\otimes\Big(\bigotimes_{\substack{i\in A\backslash N(b)\\ j\in B\backslash\{b\}}}\ket{\theta_A}_i\ket{\downarrow}_j\Big)
\end{align}
\end{widetext}
where $N_L(i)$ is the ``left neighbourhood" of $i$, i.e. the set of sites immediately upstream of $i$, and $N_R(i)$ is the ``right neighbourhood" , i.e. the set of sites immediately downstream of $i$, and $N(i) = N_L(i)\cup N_R(i)$ is the full neighbourhood.  Note that $|N_L(i)|=|N_R(i)|=D$.

Note that $\bra{\theta_A}\sigma^x\ket{\theta_A} = 0$, $\langle\theta_A^\prime|\theta_A\rangle=0$, $\bra{\theta_A^\prime}\sigma^x\ket{\theta_A}=i/2$, and $\bra{\downarrow}\sigma^x\ket{\theta_A} = -i\sin(\theta_A/2)$.  This implies that $\bra{\partial_a\psi}\sigma^x_i\ket{\psi}=0$ for $i\neq a$ and $\bra{\partial_a\psi}\sigma^x_a\ket{\psi}=i/2$.  
Thus we find that 
\begin{align}
    \braket{\partial_a\psi(\theta_A,0)}=1/4\\
    \dot{\theta}_A(\theta_A,0) = -i\frac{\langle\partial_A\psi|H|\psi\rangle}{\langle\partial_A\psi|\partial_A\psi\rangle}= -i\frac{\langle\partial_a\psi|H|\psi\rangle}{\braket{\partial_a\psi}} = 2
\end{align}
Now let us consider $\dot{\theta}_B(\theta_A,0)$. Since $\ket{\partial_b\psi}$ has the site $b$ in the state $\ket{\uparrow}$, $\bra{\partial_b\psi}\sigma^x_i\ket{\psi}=0$ unless $i=b.$ Note that $\braket{0}{\theta_A} = \cos(\theta_A/2)$; thus when taking the scalar product in $\bra{\partial_b\psi}\sigma^x_b\ket{\psi}$ we gain a factor of $\cos^{2D}(\theta_A/2)$.  We find that 
\begin{align}
   \braket{\partial_b\psi(\theta_A,0)}=\frac{1}{4}\cos^{2D}(\theta_A/2)\\
\bra{\partial_b\psi}\sigma^x_b\ket{\psi}=\frac{i}{2}\cos^{3D}(\theta_A/2),
\end{align}
which yields 
\begin{align}
    \dot{\theta}_B(\theta_A,0) &= -i\frac{\langle\partial_B\psi|H|\psi\rangle}{\braket{\partial_B\psi}}\\&= -i\frac{\langle\partial_b\psi|H|\psi\rangle}{\braket{\partial_b\psi}} = 2\cos^D(\theta_A/2).
\end{align}
Now let us calculate the leakage rate using Eq.~\eqref{eq:leakage}.
For the case that we consider here where $\ket{\psi}$ is a product state, only the $\bra{\psi}\bigotimes P_i\ket{\psi}$ terms and the $\dot{\theta}^2_\mu\braket{\partial_\mu\psi}$ terms contribute to the leakage rate; the second and fourth terms in Eq.~\eqref{eq:leakage} are zero. 
Since $\bra{\theta_A}P\ket{\theta_A}=\cos^2(\theta_A/2)$ and $\bra{\downarrow}P\ket{\downarrow}=1$, we have   
\begin{align}
\bra{\psi(\theta_A,0)}\bigotimes_{i\in N(a)}P_i\ket{\psi(\theta_A,0)}&=1\\\bra{\psi(\theta_A,0)}\bigotimes_{i\in N(b)}P_i\ket{\psi(\theta_A,0)}&=\big(\cos^{2}(\theta_A/2)\big)^{2D}
\end{align}
Thus for the leakage rate at the state $\ket{\psi(\theta_A,0)}$, we have $2\gamma^2 = 1 + \cos^{4D}(\theta_A/2) - 1 - \cos^{4D}(\theta_A/2) = 0$.

\section{Details on exact diagonalization}\label{appendix:ED}
\begin{figure}
    \centering
    \includegraphics[scale=0.44]{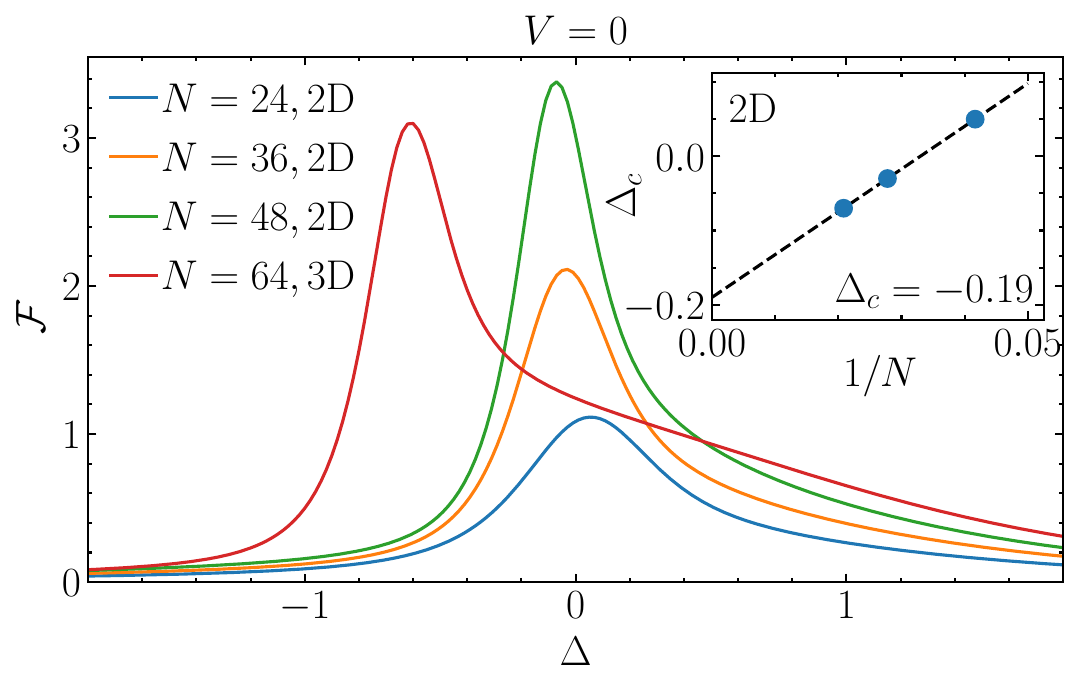}
    \caption{Ground state fidelity susceptibility for $V=0$ as function of $\Delta$ in 2D for $N=24,36,48$ sites and in 3D for 64 sites. The peak position signal the finite-size phase transition point from the disordered to the $\mathbb{Z}_2$-ordered phase. The inset shows the extrapolation of the thermodynamic limit critical value of $\Delta$ resulting in $\Delta_c \simeq -0.19$. }
    \label{fig:gs_ed}
\end{figure}

\begin{figure}
    \vspace*{2mm}
    \centering
    \includegraphics[scale=0.45]{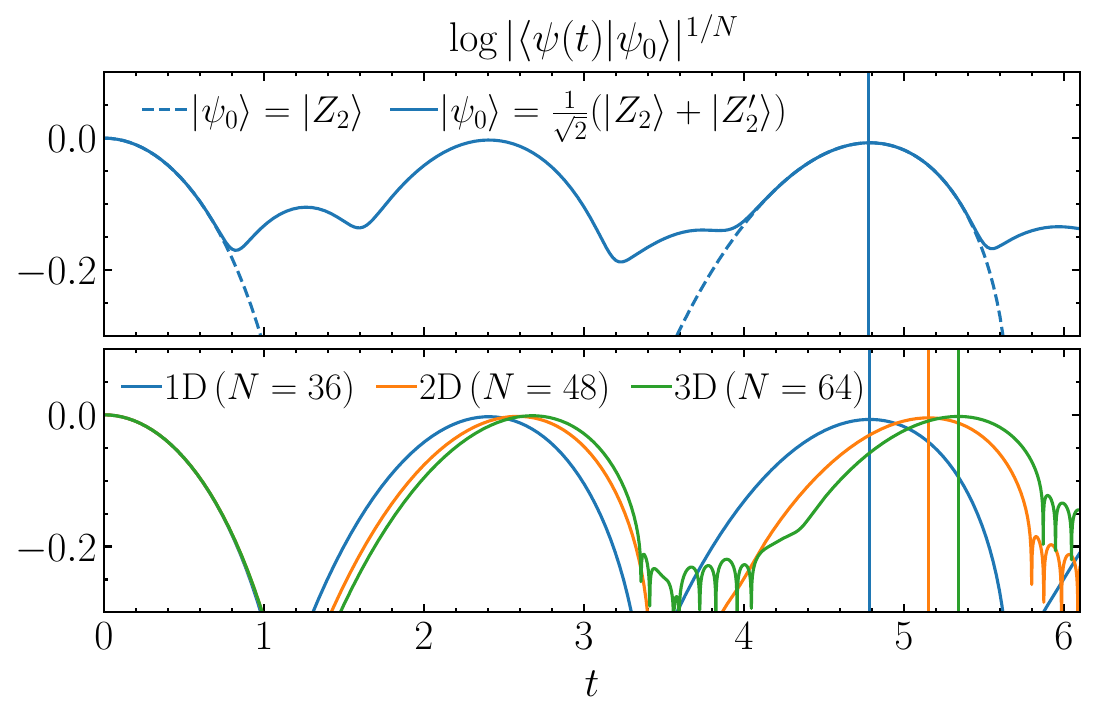}
    \caption{Revival fidelity employed to extract the first period of the approximately-periodic many-body dynamics generated by the $| \mathbb{Z}_2 \rangle $ initial state. The upper panel shows a comparison between the symmetric and non-symmetric dynamics starting from the cat state $(| \mathbb{Z}_2 \rangle + | \mathbb{Z}^{\prime}_2 \rangle)/\sqrt{2}$ and from the symmetry-breaking $| \mathbb{Z}_2 \rangle$ state, in 1D for $N=18$. The former case produces a revival at a time $t \simeq T/2$ due to the fact that the periodic orbit from the $| \mathbb{Z}_2 \rangle$ state back to itself goes through the $| \mathbb{Z}^{\prime}_2 \rangle$ state. The second revival in the symmetric time evolution exactly corresponds to the end of the orbit. The lower panel shows the symmetric dynamics for the largest systems diagonalized in all dimensions. Vertical lines correspond to the revival times.   }
    \label{fig:dyn_ed}
\end{figure}

In this appendix we provide details on the exact diagonalization techniques employed for the ground state and dynamics calculations presented in the main text.
For diagonalizing system sizes of up to 48 and 64 sites in 2D and 3D periodic lattices, respectively, we numerically computed the Hamiltonian directly in the sector which is invariant under the symmetry group of the lattice. In particular, we combined translations along and reflection with respect to the Cartesian axes. The dimensions of the neutral symmetry sectors for the largest system sizes considered in this work are 1682382 (48 sites in 2D) and 13292545 (64 sites in 3D), which can be easily tackled with sparse matrix techniques to obtain the lowest-energy eigenstates and to apply the exponential of the Hamiltonian for the time evolution of the $| \mathbb{Z}_2 \rangle$ state. \\
To extract the phase boundaries depicted in Fig.~\ref{fig:phasediagrams} we computed the ground state fidelity susceptibility 
\begin{equation}
    \mathcal{F} = \frac{1 - | \langle \mathrm{GS} (\Delta) | \mathrm{GS} ( \Delta + d \Delta ) \rangle | }{ d \Delta^2} ,
\end{equation}
which exhibits a peak at the finite-size phase transition point. We show in Fig.~\ref{fig:gs_ed} the finite size scaling of $\mathcal{F}$ for $V=0$ in 2D, from which we extrapolate the thermodynamic limit critical value $\Delta_c \simeq -0.19$, and for $V=0$ in 3D for the $4 \times 4 \times 4$ cube of 64 sites. \\
For the computation of the revival periods reported in Sec.~\ref{sec:dynamics}, we need to compute the fidelity $|\langle \mathbb{Z}_2 | e^{-i H t} | \mathbb{Z}_2 \rangle|$.
Although the $| \mathbb{Z}_2 \rangle$ state breaks the lattice symmetry of the Hamiltonian, the revival time can be extracted from the time evolution of the cat state $| \psi_0 \rangle = ( | \mathbb{Z}_2 \rangle + | \mathbb{Z}'_2 \rangle )/\sqrt{2}$, whose dynamics is symmetric and exhibits twice the revivals occurring in the non-symmetric case (see Fig.~\ref{fig:dyn_ed}). We also note, as can be seen by comparing the upper and bottom panels of Fig.~\ref{fig:dyn_ed}, that finite-size effects on revival times are negligible, since the position of fidelity maximum is unchanged from $L=18$ (upper panel) to $L=36$ (lower panel, blue line) in 1D. We verified that the same occurs in 2D for $N\geq 16$ ($4 \times 4$ square). For the 3D revival time a cube with 4 atoms in each direction ($N=64$) is required, otherwise the system is quasi-two-dimensional, and the resulting revival time is the same as in 2D. 

\bibliography{references}
\end{document}